\documentclass[iop]{emulateapj}

\usepackage[hyperfootnotes=false,naturalnames=true,pdfstartview=FitH,pdfpagemode=UseNone]{hyperref}

\newcommand\etal{et al.\ }

\newcommand\rmag{\ifmmode r_{625}\else$r_{625}$\fi}
\newcommand\imag{\ifmmode i_{775}\else$i_{775}$\fi}
\newcommand\zmag{\ifmmode z_{850}\else$z_{850}$\fi}
\newcommand\Vmag{\ifmmode V_{606}\else$V_{606}$\fi}
\newcommand\Imag{\ifmmode I_{814}\else$I_{814}$\fi}

\newcommand\rtwoh{\ifmmode {\rm r}_{200}\else r$_{200}$\fi}
\newcommand\ls{\mathrel{\hbox{\rlap{\hbox{\lower4pt\hbox{$\sim$}}}\hbox{$<$}}}}
\newcommand\gs{\mathrel{\hbox{\rlap{\hbox{\lower4pt\hbox{$\sim$}}}\hbox{$>$}}}}

\def\simgreat{\ifmmode{\mathrel{\mathpalette\@versim>}}
    \else{$\mathrel{\mathpalette\@versim>}$}\fi}
\def\simless{\ifmmode{\mathrel{\mathpalette\@versim<}}
    \else{$\mathrel{\mathpalette\@versim<}$}\fi}

\newcounter{thefigs}

\newcounter{thetabs}

\newcommand{\hGpc}{\ifmmode{h^{-1}{\rm Gpc}}\;\else${h^{-1}}${\rm Gpc}\fi}
\newcommand{\hkpc}{\ifmmode{h^{-1}_{70}\ {\rm kpc}}\;\else${h^{-1}_{70}}$ {\rm kpc}\fi}
\newcommand{\hMpc}{\ifmmode{h^{-1}_{70}\ {\rm Mpc}}\;\else${h^{-1}_{70}}$ {\rm Mpc}\fi}
\newcommand{\LCDM}{$\Lambda$CDM}

\def\simless{\mathbin{\lower 3pt\hbox
	{$\,\rlap{\raise 5pt\hbox{$\char'074$}}\mathchar"7218\,$}}} 
\def\simgreat{\mathbin{\lower 3pt\hbox
	{$\,\rlap{\raise 5pt\hbox{$\char'076$}}\mathchar"7218\,$}}} 

\newcommand{\HST}{{\em HST}}
\newcommand{\HSTACS}{{\em HST/ACS}}
\newcommand{\ACS}{{\em ACS}}
\newcommand{\WFCiii}{{\em WFC3}}

\shortauthors{Postman \etal 2011}
\shorttitle{CLASH Multi-Cycle Treasury Program}
\slugcomment{Accepted for Publication in the Astrophysical Journal Supplements, \today}

\begin{document}

\title{The Cluster Lensing And Supernova survey with Hubble: An Overview}

\author{Marc Postman\altaffilmark{1}}
\author{Dan Coe\altaffilmark{1}}
\author{Narciso Ben\'itez\altaffilmark{2}}
\author{Larry Bradley\altaffilmark{1}}         
\author{Tom Broadhurst\altaffilmark{3}}   
\author{Megan Donahue\altaffilmark{4}}
\author{Holland Ford\altaffilmark{5}}
\author{Or Graur\altaffilmark{6}}
\author{Genevieve Graves\altaffilmark{7}}
\author{Stephanie Jouvel\altaffilmark{13}}
\author{Anton Koekemoer\altaffilmark{1}} 
\author{Doron Lemze\altaffilmark{5}}
\author{Elinor Medezinski\altaffilmark{5}}
\author{Alberto Molino\altaffilmark{2}}
\author{Leonidas Moustakas\altaffilmark{8}}
\author{Sara Ogaz\altaffilmark{1}}
\author{Adam Riess\altaffilmark{1,5}}
\author{Steve Rodney\altaffilmark{5}}
\author{Piero Rosati\altaffilmark{9}}
\author{Keiichi Umetsu\altaffilmark{10}}
\author{Wei Zheng\altaffilmark{5}}
\author{Adi Zitrin\altaffilmark{6}}
\author{Matthias Bartelmann\altaffilmark{11}} 
\author{Rychard Bouwens\altaffilmark{12}}
\author{Nicole Czakon\altaffilmark{13}}
\author{Sunil Golwala\altaffilmark{13}}
\author{Ole Host\altaffilmark{14}}
\author{Leopoldo Infante\altaffilmark{15}}
\author{Saurabh Jha\altaffilmark{16}}
\author{Yolanda Jimenez-Teja\altaffilmark{2}}
\author{Daniel Kelson\altaffilmark{17}}
\author{Ofer Lahav\altaffilmark{14}} 
\author{Ruth Lazkoz\altaffilmark{3}}
\author{Dani Maoz\altaffilmark{6}}
\author{Curtis McCully\altaffilmark{16}}
\author{Peter Melchior\altaffilmark{18}}
\author{Massimo Meneghetti\altaffilmark{19}}
\author{Julian Merten\altaffilmark{11}}
\author{John Moustakas\altaffilmark{20}}
\author{Mario Nonino\altaffilmark{21}}
\author{Brandon Patel\altaffilmark{16}}
\author{Enik\"o Reg\"os\altaffilmark{22}}
\author{Jack Sayers\altaffilmark{13}}
\author{Stella Seitz\altaffilmark{23}}
\author{Arjen Van der Wel\altaffilmark{24}}

\altaffiltext{1}{Space Telescope Science Institute, 3700 San Martin Drive, Baltimore, MD 21208, U.S.A.}
\altaffiltext{2}{Instituto de Astrof\'isica de Andaluc\'ia}
\altaffiltext{3}{Department of Theoretical Physics, University of the Basque Country}
\altaffiltext{4}{Michigan State University}
\altaffiltext{5}{The Johns Hopkins University}
\altaffiltext{6}{Tel Aviv University}
\altaffiltext{7}{University of California, Berkeley}
\altaffiltext{8}{The Jet Propulsion Laboratory, California Institute of Technology}
\altaffiltext{9}{European Southern Observatory}
\altaffiltext{10}{ Academia Sinica, Institute of Astronomy \& Astrophysics}
\altaffiltext{11}{Universitat Heidelberg}
\altaffiltext{12}{Leiden Observatory, Leiden University}
\altaffiltext{13}{California Institute of Technology}
\altaffiltext{14}{University College London}
\altaffiltext{15}{Universidad Catolica de Chile}
\altaffiltext{16}{Rutgers University}
\altaffiltext{17}{The Carnegie Institute for Science; Carnegie Observatories}
\altaffiltext{18}{The Ohio State University}
\altaffiltext{19}{INAF, Osservatorio Astronomico di Bologna}
\altaffiltext{20}{University of California, San Diego}
\altaffiltext{21}{INAF, Osservatorio Astronomico di Trieste}
\altaffiltext{22}{European Laboratory for Particle Physics (CERN)}
\altaffiltext{23}{Universitas Sternwarte, M\"unchen}
\altaffiltext{24}{Max Planck Institut f\"ur Astronomie, Heidelberg}

\begin{abstract}
The Cluster Lensing And Supernova survey with Hubble (CLASH) is a 524-orbit multi-cycle treasury program 
to use the gravitational lensing properties of 25 galaxy clusters to accurately constrain their mass distributions. 
The survey, described in detail in this paper, will definitively establish the degree of concentration of dark matter in the 
cluster cores, a key prediction of structure formation models.
The CLASH cluster sample is larger and less biased than current samples of space-based imaging studies of
clusters to similar depth, as we have minimized lensing-based selection that favors systems with overly 
dense cores. Specifically, twenty CLASH clusters are solely X-ray selected. The X-ray selected clusters are massive ($kT > 5$ keV) and, in 
most cases, dynamically relaxed. Five additional clusters are included
for their lensing strength ($\theta_{Ein} > 35''$ at $z_s$ = 2) 
to optimize the likelihood of finding highly magnified high-$z$ ($z > 7$) galaxies.  
A total of 16 broadband filters, 
spanning  the near-UV to near-IR, are employed for each 20-orbit campaign on each cluster. These data 
are used to measure precise ($\sigma_z \sim 0.02(1+z)$) 
photometric redshifts for newly discovered arcs.
Observations of each cluster 
are spread over 8 epochs to enable a search for Type Ia 
supernovae at $z > 1$ to improve constraints on the time dependence of the dark energy equation of state and the evolution of supernovae.
We present newly re-derived X-ray luminosities, temperatures, and Fe abundances for the CLASH clusters as well as a representative
source list for MACS1149.6$+$2223 ($z = 0.544$). 
\end{abstract}


\section{Introduction}
\label{sec:intro}

The Universe has proven to be far more intriguing in its composition 
than we knew it to be even just 14 years ago. 
It is a ``dark'' Universe where $\sim 23\%$ of its mass-energy density is made up of 
weakly interacting (and, as yet, undetected) non-baryonic particles (a.k.a.~dark matter) 
and $\sim 73\%$ is as yet unknown physics (a.k.a.~dark energy) 
that is driving an accelerated expansion of the metric \citep[e.g.,][]{WMAP7,Riess11}.
The {\em Hubble Space Telescope} (\HST) has played a key role in providing evidence for and constraining the nature of
both of these mysterious dark components \citep[e.g.,][]{Riess98,Perlmutter99,Clowe06}.

Clusters of galaxies, by virtue of their position at the high end of the cosmic mass
power spectrum, provide a powerful way to constrain the frequency of 
high amplitude perturbations in the primordial density field. As such, they 
play a direct and fundamental role in testing cosmological models and in constraining
the properties of dark matter (DM), providing unique and independent tests of any viable cosmology and
structure formation scenario, and possible modifications of the laws of gravity. A key ingredient of such
cluster-based cosmological tests is the mass distribution of clusters, both on (sub) Mpc scales and across
the range of populations.
The best and highest resolution maps of DM distribution in massive galaxy clusters come
from observations of strong gravitational lensing
made by the Advanced Camera for Surveys \citep[ACS,][]{ACS} onboard \HST.
As part of a Guaranteed Time Observation (GTO) program,
deep (20-orbit) multiband (4-6 filter) observations were obtained for five galaxy clusters:
Abell 1689 \citep{Broadhurst05,Limousin07,Coe10};
Abell 1703 \citep{Limousin08,SahaRead09,Zitrin10A1703};
Abell 2218 \citep{Eliasdottir08};
CL0024+1654 \citep{Jee07,Zitrin09a,Umetsu10};
and MS1358+6245 \citep{Zitrin11MS1358}.
These results have contributed to a reported tension
between observed and simulated galaxy cluster dark matter halos.
Observed clusters appear to have denser cores
than simulated clusters of similar total (virial) mass 
\citep[e.g.,][]{Broadhurst08,Oguri09,Sereno10}.
Understanding the true constraints from observed concentration and mass profile measurements
on $\Lambda$ cold dark matter (\LCDM) structure formation models
is one of the important problems that can be tackled with a deep, high angular resolution
imaging survey of a significantly larger and more homogeneously selected sample of galaxy clusters.

In May 2009, NASA successfully executed the final planned Hubble Servicing mission, SM4, with
the installation of the Wide Field Camera 3 (\WFCiii\, \citealt{WFC3}) and the repair of \ACS.
Shortly thereafter, the Hubble Multi-Cycle Treasury (MCT) Program was conceived
to permit ambitious programs ($>500$ orbits) with broad scientific potential that could not be accomplished within
the constraints of a single {\em HST} observing cycle and that 
would take full advantage of the final era of a newly refurbished {\em Hubble Space Telescope}.

The Cluster Lensing and Supernova survey with Hubble (CLASH)
was one of three MCT programs selected. CLASH has four main science goals:
\begin{enumerate}
\item Measure the profiles and substructures of dark matter in galaxy clusters with unprecedented precision and resolution.
\item Detect Type Ia supernovae out to redshift $z \sim 2.5$ to measure the time dependence of
the dark energy equation of state and potential evolutionary effects in the supernovae themselves.
\item Detect and characterize some of the most distant galaxies yet discovered at $z > 7$.
\item Study the internal structure and evolution of the galaxies in and behind these clusters.
\end{enumerate}
To accomplish these objectives, the CLASH program targets 25 massive galaxy clusters and will
image each in 16 passbands using {\em WFC3/UVIS}, {\em WFC3/IR} and {\em ACS/WFC}. 
CLASH has been allocated 524 \HST\  orbits, spread out over Cycles 18, 19, and 20.
The majority of these orbits (474) are for cluster imaging and, simultaneously, for the parallel SN search program. An
additional 50 orbits were allocated as a reserve for supernova follow-up observations. Based on a current census of the 
\HST\  data archive,
CLASH will produce a six-fold increase the number of lensing clusters observed to a depth of 20 orbits and,
more importantly, will vastly increase the number lensing clusters with extensive multi-band \HST\  imaging.

Motivations for each of the main CLASH science goals are provided in
\S\ref{sec:DM}, \S\ref{sec:SNe}, \S\ref{sec:hiz}, and \S\ref{sec:gal}.
Subsequent sections describe the 
cluster sample (\S\ref{sec:sample}),
survey design (\S\ref{sec:survey}),
data pipeline (\S\ref{sec:pipeline}),
and supporting observations using other facilities (\S\ref{sec:supporting}).
Data products intended for public distribution to the community are briefly described in \S\ref{sec:summary}.
AB magnitudes are used throughout \citep{Oke74}. 
The cosmological parameters 
$H_o = 100h\ {\rm km}\  {\rm s}^{-1}\ {\rm Mpc}^{-1},\ \,h=0.7,\ \Omega_m = 0.30,\ \Lambda = 0.70$ 
are assumed in this paper.

\section{Scientific Motivation}
\label{sec:SciMotive}
\subsection{Galaxy Cluster Dark Matter Profiles and Formation Times}
\label{sec:DM}

Recent observations suggest that
galaxy clusters formed earlier in our universe than in simulated \LCDM\ universes. These
observations include the detection of perhaps unexpectedly massive galaxy clusters at $z > 1$
\citep{Stanford06,Eisenhardt08,Jee09,Huang09,Rosati09,Papovich10,Schwope10,Gobat11,Jee11,Foley11} and
the finding that some clusters at intermediate redshift ($z \sim 0.3$) have denser cores
than clusters of similar mass produced in simulations 
\citep{Broadhurst08,BroadhurstBarkana08,Oguri09,Richard10,Sereno10,Zitrin10}.
While the evidence to date for early cluster growth is suggestive,
possible explanations include
departures from Gaussian initial density fluctuation spectrum 
or higher levels of dark energy in the past,
so-called Early Dark Energy (EDE) \citep{FedeliBartelmann07,SadehRephaeli08,Francis09,GrossiSpringel09}.
If a significant quantity of EDE 
(for example $\Omega_{DE} \sim 0.1$ at $z = 6$)
suppressed structure growth in the early universe,
then clusters would have had to start forming sooner to yield the numbers we observe today.
These scenarios remain allowable within current observational constraints as described in the above papers,
although some non-Gaussian models can be ruled out by using the cosmic X-ray background measurements \citep{Lemze09b}.
The CLASH data permits significant advances to be made towards supporting or rejecting observational evidence for early cluster 
growth by measuring core densities for a larger, less biased sample of clusters.

\begin{figure}
\includegraphics[width=\columnwidth]{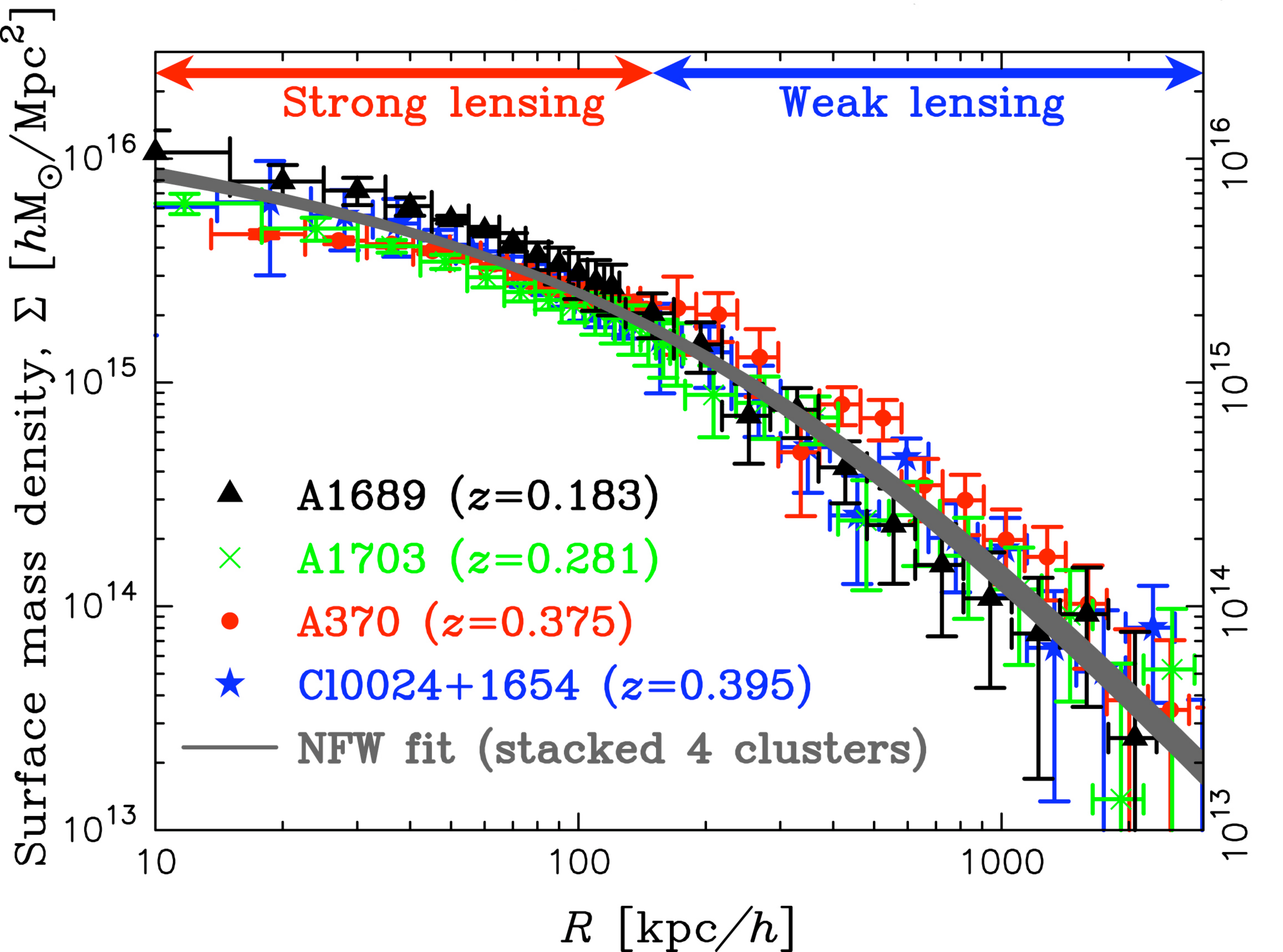}
\caption{\label{DMprofiles}
Mass profiles measured for four well-studied, strongly lensing galaxy clusters that are {\bf not} included in the CLASH sample.
All have similar mass profiles as measured from Hubble observations of strong-lensing
and Subaru observations of weak-lensing distortion and magnification
\citep[their Figure~6]{Umetsu11}.
The averaged mass profile is in remarkably good agreement with the standard NFW form \cite[their Figure~1]{Umetsu11b}
as shown by the gray area (2-$\sigma$ confidence interval of the NFW fit),
though with a higher concentration than predicted from cosmological simulations.
Both strong and weak lensing probes are required to map the continuously steepening mass profile
from the inner core ($\sim 10$ kpc$/h$) out to beyond the virial radius ($\sim 2$ Mpc$/h$).
}\end{figure}

\begin{deluxetable*}{llccccccc}
\tabletypesize{\scriptsize}
\tablewidth{0pt}
\tablecaption{\label{tab:cobs}Concentration Measurements for Previously Well-Studied Lensing-selected Clusters}
\tablehead{
\colhead{}&
\colhead{}&
\colhead{}&
\colhead{}&
\colhead{$M_{vir}$}&
\colhead{}&
\colhead{}\\
\colhead{Constraints\tablenotemark{a}}&
\colhead{Publication}&
\colhead{Cluster}&
\colhead{$z$}&
\colhead{[$10^{15} M_{\odot} h^{-1}$]}&
\colhead{$c_{vir}$}&
\colhead{$\chi^2/{\rm d.o.f.}$}&
}
\startdata
SL+WL+mag	& \citet{Umetsu11}\tablenotemark{b}	& A1689 & 0.187	& $1.34^{+0.20}_{-0.16}$	& $13.82^{+1.72}_{-1.62}$ & 4.73/17	\\
SL+WL+mag	& \citet{Umetsu11}\tablenotemark{b}	& A1703 & 0.281	& $1.29^{+0.22}_{-0.19}$	& $6.89^{+1.04}_{-0.91}$ & 7.14/19	\\
SL+WL+mag	& \citet{Umetsu11}\tablenotemark{b}	& A370 & 0.375	& $2.26^{+0.26}_{-0.23}$	& $4.56\pm0.33$ & 14.07/24	\\
SL+WL+mag	& \citet{Umetsu11}\tablenotemark{b}	& CL0024+17 & 0.395	& $1.37^{+0.20}_{-0.18}$	& $7.77^{+0.97}_{-0.87}$ & 11.47/20	\\
SL+WL+mag	& \citet{Umetsu11}\tablenotemark{b}	& RXJ1347.5-1145 & 0.451	& $1.73^{+0.12}_{-0.11}$	& $5.96^{+0.37}_{-0.35}$ & 45.06/25	\\
RE+WL	& \citet{Oguri09}	& SDSS J1531+3414 & 0.335	& $0.7^{+0.29}_{-0.24}$ & $7.9^{+3.0}_{-1.5}$	& 8.1/6 \\
RE+WL	& \citet{Oguri09}	& SDSS J1446+3032 & 0.464	& $0.8^{+0.3}_{-0.22}$ & $8.3^{+3.9}_{-3.1}$	& 6.4/6 \\
RE+WL	& \citet{Oguri09}	& SDSS J2111-0115 & 0.637	& $0.9^{+0.41}_{-0.32}$ & $14.1^{+25.9}_{-9.3}$	& 7.5/6 \\
\enddata
\vspace{-0.1in}
\tablecomments{Of these 8 clusters, only one -- RXJ1347.5-1145 -- is in the CLASH sample.}
\tablenotetext{a}{SL = strong lensing; WL = weak lensing; mag = magnification bias (number counts); RE = Einstein radius}
\tablenotetext{b}{The parameters here are from standard NFW fits to profiles in \citet{Umetsu11}, and not from the generalized NFW (gNFW) 
fits given in that paper.}
\end{deluxetable*}

In cosmological simulations, CDM-dominated halos of all masses consistently evolve 
to have a roughly ``universal'' density profile that steepens with radius. Functional forms
that fit such a profile well include the ``NFW'' profile \citep{NFW96,NFW97}
and the Einasto / S\'ersic profile \citep{Sersic63,Einasto65,Navarro04,Navarro10}. 
Furthermore, each simulated halo's core density is related to 
the background density of the universe at the halo's formation time.
Halos that form later, including the most massive galaxy clusters,
are found to have the least dense cores in a relative sense.
Determining the relationship between the shape and depth of a halo's gravitational potential
and its total mass as a function of time thus provides fundamental constraints on structure formation.

In practice, the relative core densities, or ``concentrations'',
are measured (both in simulated and observed halos) 
as $c_{vir} = r_{vir} / r_{-2}$,
a ratio between the virial radius and the inner radius at which the density slope 
of the fitted profile is isothermal ($\rho \propto r^{-2}$).
Analyses of gravitational lensing spanning such a large range of radius
allows one to map the (primarily dark) matter profiles of observed halos
and measure their concentrations.

Dark matter profiles are best mapped in cluster cores using strong lensing analysis of multiband \HST\  imaging. 
The lensing-based mass profile mapping is extended to the virial radii of clusters
using weak lensing analysis of wider field ground-based multiband imaging, such as from Subaru.
Results from \citet{Umetsu11,Umetsu11b}
for four of the currently best-studied (non-CLASH) clusters are shown in Figure~\ref{DMprofiles}.

\begin{figure}
\plotone{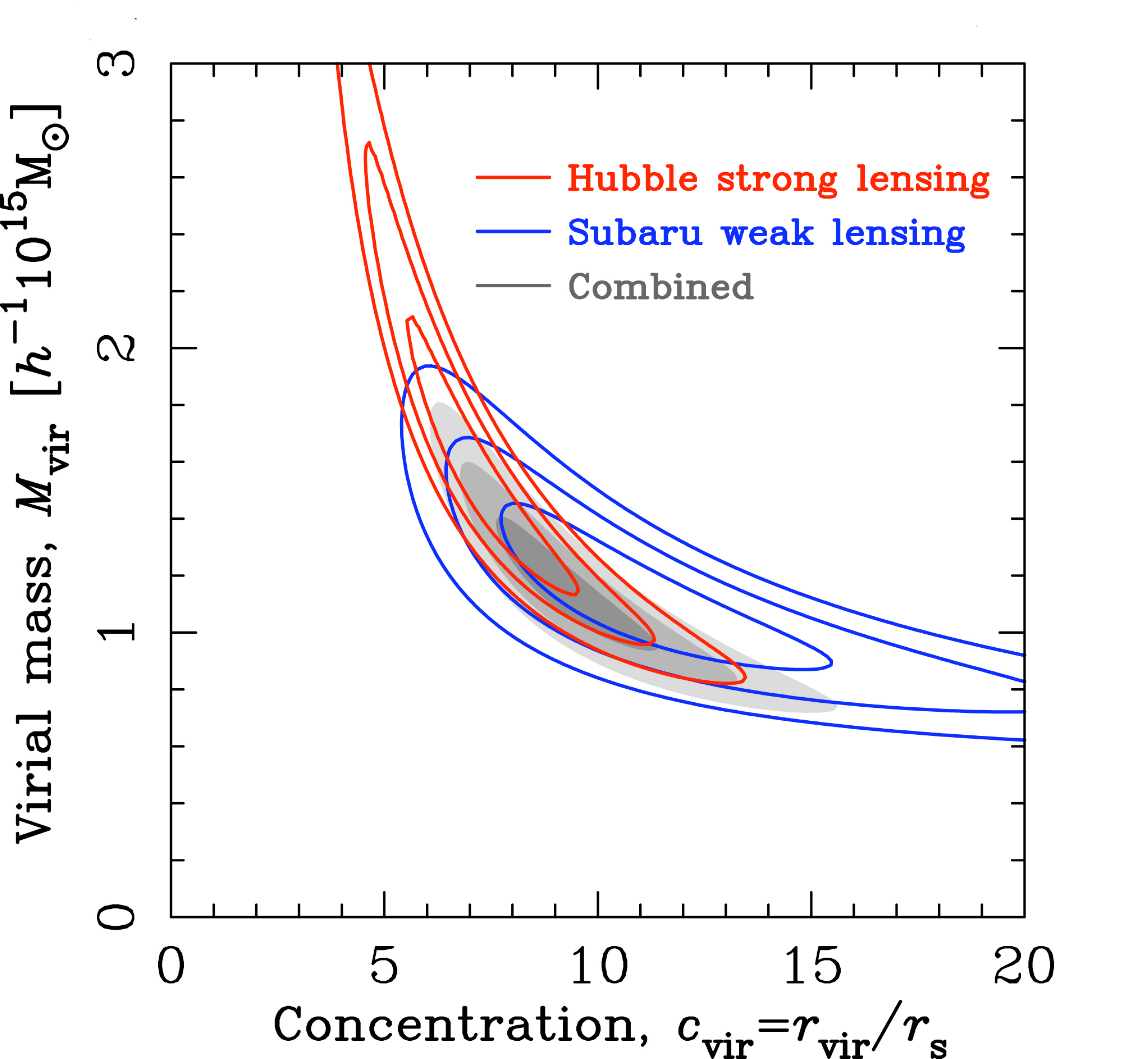}
\caption{\label{CMfit}
Joint strong and weak lensing analyses are required to obtain tight
constraints on cluster concentrations as shown here for CL0024+17, a non-CLASH cluster
\cite[from their Figures 15 and 17]{Umetsu10}. 
Confidence levels of 68.3\%, 95.4\%, and 99.7\%
are plotted in the $c_{\rm vir}$--$M_{\rm vir}$ plane.
}\end{figure}

Joint modeling of strong and weak lensing (SL+WL)
yields significantly better constraints on concentrations than either probe alone.
Quantitatively, \cite{Meneghetti10} found their joint SL+WL analyses of simulated clusters
yield concentrations to $\sim 11\%$ accuracy,
while WL-only and SL-only analyses yielded $\sim 33\%$ and $\sim 59\%$ scatters, respectively.
Figure~\ref{CMfit} demonstrates how SL and WL analyses combine
to yield robust constraints on the mass and concentration of CL0024+17 \citep{Umetsu10}.

\begin{figure}
\includegraphics[width=\columnwidth]{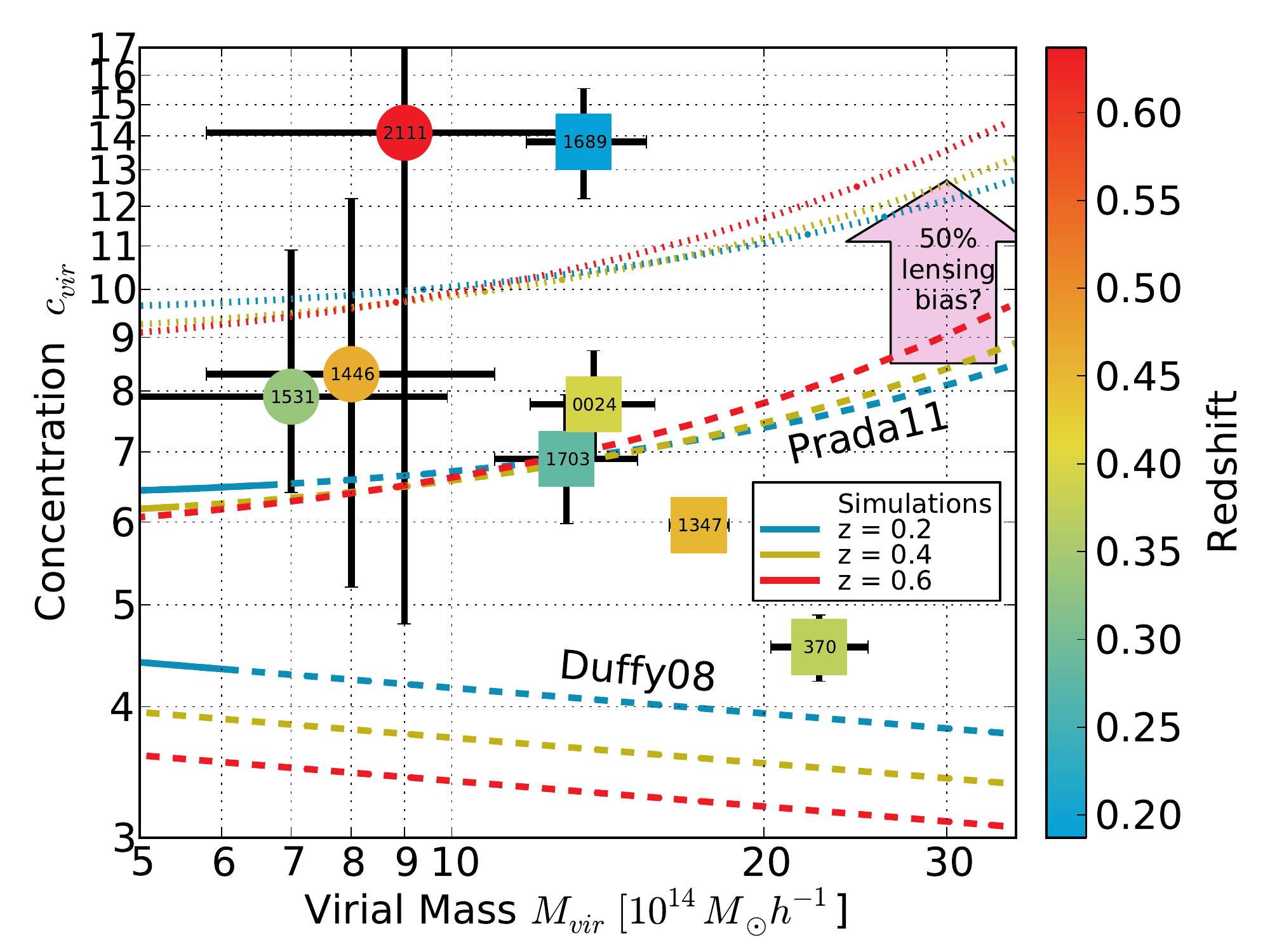}
\caption{\label{cobs}
Mass profiles of the best studied clusters to date
are revealed to have higher central density concentrations
than simulated clusters of similar mass and redshift.
Reconciliation may be within reach
given results from the latest simulations
and an estimated lensing bias which CLASH will avoid.
The plotted lines are mean concentrations at three different redshifts
for clusters in these simulations as calculated from the fitting formulae provided in those papers.
However note that halos of this great mass are rare (or even non-existent at these redshifts)
in these simulations, so these results are mainly extrapolations, as designated by the dashed lines.
The thinner dashed lines above illustrate a 50\% observational bias applied to the \citet{Prada11} results.
This bias has been roughly estimated for non-CLASH clusters such as these 
which were selected for study based on exceptional lensing strength
\citep{Hennawi07,OguriBlandford09,Meneghetti10,Meneghetti11}.
Results from the currently best-studied clusters are plotted here as
squares \citep{Umetsu11} and circles \citep{Oguri09}
labeled with abbreviated names and described further in Table \ref{tab:cobs}.
}\end{figure}

The best studied (SL+WL) galaxy clusters to date have been found to have overly high concentrations (dense cores)
compared to haloes in N-body simulations with similar masses, as shown in Figure~\ref{cobs} \citep{Broadhurst08,Oguri09,Sereno10}
and detailed in Table~\ref{tab:cobs}. Recent simulations \citep{Prada11}
yield cluster concentrations that are over 50\% higher
than previous simulations \citep[e.g.,][]{Duffy08}.
This is a product of upturns as a function of both mass and redshift
found in these newer simulations, which are not yet understood.
Similarly, clusters have also been found to have somewhat larger than expected Einstein radii,
a direct and particularly accurate measure of the projected mass in a halo's core \citep{BroadhurstBarkana08,Richard10,Zitrin10,Zitrin11RE}.

Unfortunately, the best studied clusters to date have also been among the strongest gravitational lenses known.
Such lensing-selected clusters are highly biased toward halos with high concentrations,
both intrinsically and as projected on the sky due to halo elongation along the line of sight
\citep{Hennawi07,OguriBlandford09,Meneghetti10a,Meneghetti11}.
These biases are estimated to lead to systematically higher concentrations by as much as 50\% 
or more. However, such a bias is insufficient to account for the discrepancy between the observations and the predictions
that is until a very recent analysis of new simulations was performed
(see Figure~\ref{cobs} and discussion below).

More robust conclusions require analysis of a larger, unbiased cluster sample.
Progress toward this goal has been made by LoCuSS, the Local Cluster Substructure Survey \citep{Smith05}.
A large sample of 165 clusters between $0.15 < z < 0.30$ was selected based on X-ray brightness.
Strong lensing analyses of 20 of these based on \HST\ imaging (mostly single-band ``snapshots'')
were presented by \cite{Richard10}.
\cite{Okabe10} published weak lensing analyses of 30 LoCuSS clusters,
22 of these being more ``secure'' based on multiband Subaru imaging,
and 9 of these 22 overlapping with the \cite{Richard10} subset.
Subaru images are especially desirable for WL studies as they enable excellent galaxy shape measurements
to be performed over a wide area.
Multiband imaging is also critical to properly select background galaxies
and avoid significantly diluting the weak lensing signal
(and thus the virial mass and concentration) with unlensed foreground galaxies \citep{Medezinski07,Medezinski10}.
Stacked WL-only analyses have been performed on large cluster samples \citep{Johnston07,Mandelbaum08},
but we re-emphasize the need for combining strong + weak lensing analyses in addition to 
cross-comparisons with mass profile estimates from other techniques. For example,
mass concentrations can be measured from X-ray profiles \citep{Buote07,Ettori10},
although these are subject to uncertainties due to assumptions about hydrostatic equilibrium.
Importantly, cluster elongation along the line of sight (a potential bias in concentration measurements) can 
be measured by the combination of lensing and X-ray analysis \citep{Morandi10,Newman11}.
The caustic technique \citep{DiaferioGeller97,Rines06} is, like lensing-based methods, independent 
of the dynamical state of the cluster and also provides an important cross check on mass estimates from
lensing and gas kinematics.
Further discussion of some of the previous results from various methods are given in
\citet{ComerfordNatarajan07, Coe10DMprofiles, Rines10, KingMead11}.

Reconciliation of high observed concentrations with results from simulations
may ultimately come from
significantly reducing the observational sample bias along with finding higher concentrations in simulated clusters.
Baryons, for example, are currently absent from those cosmological simulations large enough to produce massive clusters.
Baryons can result in significant ``adiabatic contraction'' on galaxy scales,
however they constitute a much smaller fraction of the mass on cluster scales.
Simulations including baryons show that
cluster halo concentrations are likely only varied by $\sim 10\%$ or so relative to DM-only halos,
and that the direction of variation (increase or decrease) is not even clear, 
depending on the gas physics assumed \citep{Duffy10,Mead10,KingMead11}. Nonetheless,
measuring the velocity dispersion of the Brightest Cluster Galaxy (BCG)  as an additional constraint on the inner ($\lesssim 80$ kpc)
mass profile, where its stellar mass is a non-negligible component of the matter distribution, provides for a more thorough mapping of the total mass 
profile. Some clusters have been shown to have inner mass profiles that are shallower than NFW \citep{Sand04,Sand08,Newman09,Newman11}. 
Deviations of cluster mass profiles from NFW or Einasto forms at small and/or large radii may 
slightly bias concentration measurements \citep{OguriHamana11}. The use of multiple probes of the matter distribution, as CLASH is designed to do, will 
enable such biases to be measured and the significance of deviations determined.

More recently, an analysis by \cite{Prada11} 
found that clusters in the Bolshoi and MultiDark simulations
have concentrations $\sim 50\%$ higher than clusters in previous simulations.
While the robustness of this new result is still being assessed,
it raises the possibility that the combination of new observations of an unbiased sample of clusters and 
new simulations may be able to bridge the concentration gap.
We stress here that estimates of the observational bias from previous cluster studies have large uncertainties and
likely varies for each cluster. CLASH will determine mass profiles and 
concentrations for a new cluster sample free of lensing selection bias.
As we demonstrate in \S\ref{sec:samplesize},
CLASH is designed to detect (or rule out) with 99\% statistical confidence
average deviations of 15\% or more from predicted concentrations.
However, given the outstanding uncertainties in the expected concentrations, 
we prefer to recast the problem as follows: CLASH will deliver robust observational concentration measurements
for a sample of clusters that simulations will be tasked to reproduce.
Ultimately this will lead to a better calibration of many mass estimation techniques and,
consequently, to a better understanding of structure formation on cluster scales
and perhaps of our cosmological model as well.

\subsection{Improved Constraints on the Dark Energy Equation of State and SNe Evolution}
\label{sec:SNe}

The biggest cosmological surprise in decades came from observations of high-redshift type-Ia supernovae 
(SNe Ia), providing the first evidence that the expansion of the Universe now 
appears to be accelerating \citep{Riess98, Perlmutter99}, and
indicating the Universe is dominated by ``dark energy.'' 
The presence of dark energy has galvanized cosmologists as they seek to understand it.  
Observations of high-redshift SNe Ia have continued to lead the way 
in measuring the properties of dark energy \citep[e.g.,][]{Riess11}. 
The goal for cosmologists now is to measure the equation of state of dark energy, $w = P/(\rho c^2)$, 
and its time variation in the hope of discriminating between viable explanations.  
A departure of the present equation of state, $w_0$, from $-1$ 
or a detection of its variation, $\partial w/\partial z$, 
would invalidate an innate vacuum energy (i.e., the cosmological constant) 
as the source of dark energy and would point towards a present epoch of ``weak inflation.'' 
A difference between the expansion history and the growth history of structure expected for $w(z)$ 
would point towards a breakdown in General Relativity as the cosmic scale factor approaches unity.  

\HST\ paired with \ACS\ is a unique tool in this investigation, 
providing the only means to collect SNe Ia at $1 < z < 1.5$,
which, in turn, provide the only constraints we have to date on the time variation of $w$. 
From the 23 SNe Ia at $z>1$ with \HST\ data \citep{Riess04,Riess07} we have learned: 
1) that cosmic expansion was once decelerating before it recently began accelerating, 
2) that dark energy, i.e., an energy density with $w<0$, was already present during this prior decelerating phase, 
3) that SNe Ia at a look-back time of 10 Gyr appear both spectroscopically and photometrically 
similar to those seen locally 
and 4) no rapid change is seen in $w(z)$ and thus no departure is yet seen from the cosmological constant, 
though the constraint on the time variation remains an order of magnitude worse than on the $w_0$.

SNe Ia play a central role not only as distance indicators for cosmography, 
but also as major contributors to cosmic metal production and distribution. 
The measurement of high-redshift SN Ia rates is therefore integral to understanding the history of chemical enrichment. 
The high-redshift rate cannot easily be predicted from the star formation rate 
because the nature and, hence, the timescales of the process behind the growth of the white dwarf
toward the Chandrasekhar mass is not known. 
The two leading competing scenarios are accretion from a close binary companion 
-- the single-degenerate scenario \citep{WhelanIben73,Nomoto82}
or merger with another white dwarf, 
following loss of orbital energy and angular momentum by emission of gravitational waves
-- the double-degenerate scenario \citep{IbenTutukov84,Webbink84}.
One way to constrain the different progenitor scenarios is to 
measure the delay-time distribution (DTD) of SNe Ia. This is the distribution of times that elapse between a brief burst 
of star formation, and the subsequent SN Ia explosions. Observations have suggested various different forms for the DTD 
(e.g., \citep{Dahlen04,Dahlen08,Mannucci06,Pritchet08}). However, a number of more recent 
measurements and analyses point to a DTD that is a power law of index $\approx -1$ \citep{Totani08,
Maoz10a,Maoz10b,Brandt10,Horiuchi10,Maoz11,Graur11}. 
Specifically, \cite{Graur11} have shown that such a DTD fits well the measured SN Ia rate out to $z \approx 2$.
Small sample sizes are the major limiting factor at high redshifts.  
Large high-$z$ SN samples are therefore needed to resolve the issue, and 
control possible biases in cosmological studies (due to the evolving SN Ia channel mix).

\begin{figure}
\plotone{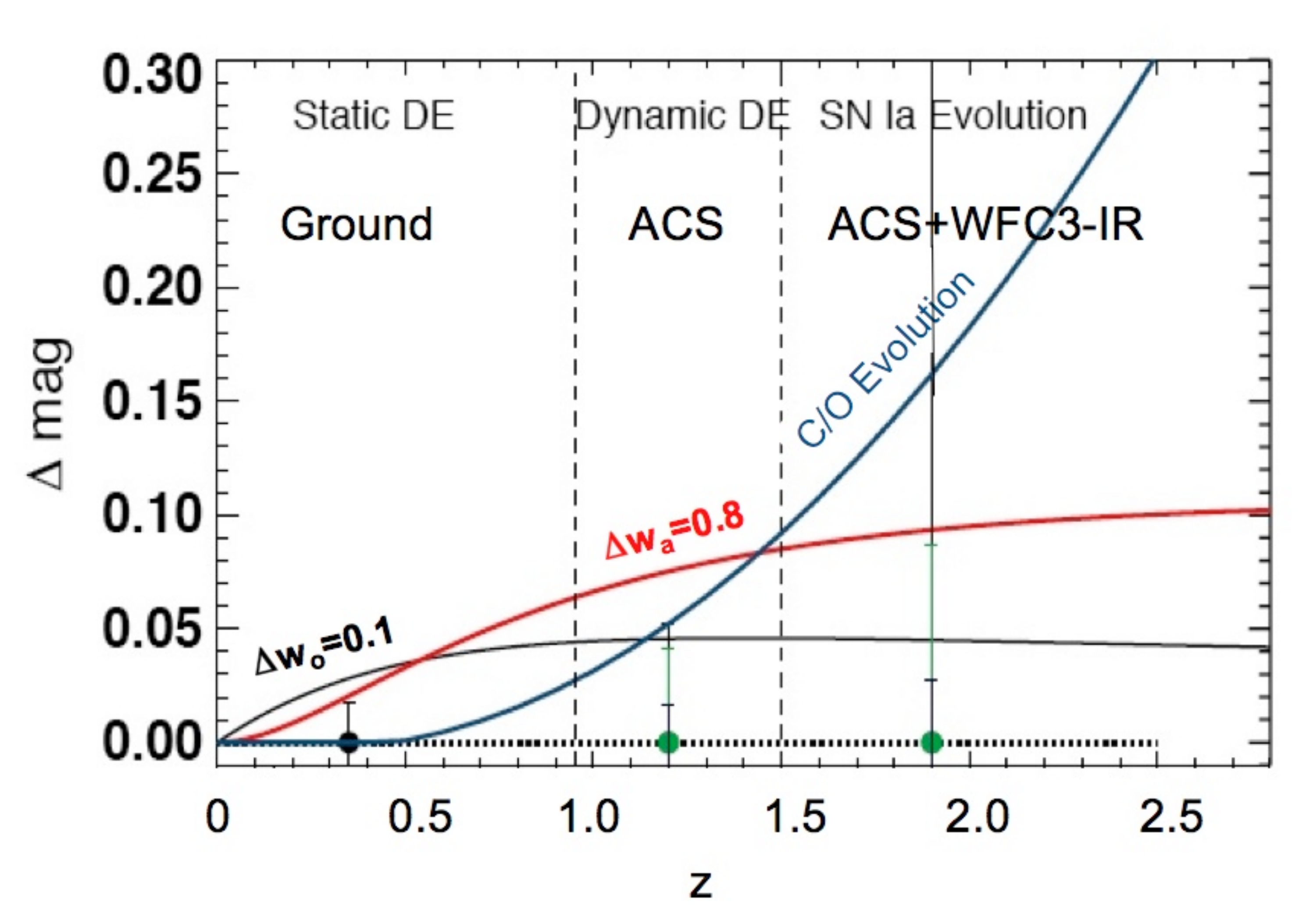}
\caption{\label{SNconstraints}
Dark Energy and Evolution Sensitivity. 
Three models consistent with current data are shown:
$w_{0} = Ð0.9$, $w_{a}= 0$ (black);
$w_{0} = Ð1$, $w_{a}= 0.8$ (red);
and 
$w_{0} = Ð1, w_{a} = 0$ (dashed).
Also plotted blue is the SNe Ia evolution model of \cite{Dominguez01},
where the peak luminosity changes 3\% per solar mass change in the donor star.
The error bars show the constraints at present, as projected after CLASH, 
and if \HST\ continues to collect SNe Ia at the present rate for 7 years.
}\end{figure}

The CLASH survey uses \ACS\ in parallel with the cluster program to continue the discovery of SNe 
Ia at $1 < z < 1.5$, the objects which tell us about the variation in $w$.  With {\em WFC3} in parallel, CLASH will 
yield SNe Ia at $1.5 < z < 2.5$ \citep{Rodney11}.  Observations in this fully matter dominated epoch provide the unique chance 
to test SN Ia distance measurements for the deleterious effects of evolution independent of our ignorance of 
dark energy \citep{RiessLivio06}. Because the SNe Ia are detected when these cameras are in parallel, 
they are far from the cluster core ($\sim2$ Mpc at the median cluster redshift of $z = 0.4$)  
and, hence, the effects of lensing are small (and correctable), making the SNe 
usable for improving the limits on the redshift variation of the dark energy equation of state.  
At $z < 1$, SN Ia distance measurements are most sensitive 
to the static component of dark energy, $w_{0}$.
At $1 < z < 1.5$, the measurements are most sensitive to the dynamic component, $w_{a}$. 
By $z > 1.5$, the measurements are most sensitive to evolution if present 
(e.g., the changing C/O ratio of the donor star),
providing the means to diagnose and calibrate the degree of SN Ia evolution in dark energy measurements. 
Figure~\ref{SNconstraints}
shows how variations in the DE equation of state or an evolution in the white dwarf C/O ratio can change the
SN Ia distance modulus as a function of redshift.

Accounting for the systematic uncertainty introduced by the cosmic star formation history, \cite{Graur11} 
predicted the SN Ia rate out to higher redshifts (their figure 13), which we use here, in conjunction with the CLASH
observational parameters, to estimate the number of $z \ge 1$
SNe Ia that will be discovered over the course of this program. These estimates 
are presented in Table~\ref{tab:sne} and 
are the 68\% confidence intervals, accounting for both statistical and systematic uncertainties. 

\begin{deluxetable}{cc}
\tablewidth{2in}
\tablecaption{\label{tab:sne}Estimated Number of $z \ge 1$ SNe Ia to be Discovered During the CLASH Program}
\tablehead{
\colhead{Redshift}&
\colhead{Estimated Number}\\
\colhead{Range}&
\colhead{of SNe Ia Found}
}
\startdata
$1.0 \le z < 1.5$ & 7 -- 11 \\
$1.5 \le z < 2.0$ & 4 -- 13 \\
$2.0 \le z \le 2.7$ & 0 -- ~~4 \\
\\
$1.0 \le z \le 2.7$ & 11 -- 28\\
\enddata
\vspace{-0.1in}
\tablecomments{These estimates are 
the 68\% confidence intervals, accounting for both statistical and systematic uncertainties. The
assumed SN Ia rate is from \cite{Graur11}.}
\end{deluxetable}

\subsection{Detection and Characterization of $z > 7$ Galaxies}
\label{sec:hiz}

One of the most important goals in observational cosmology is to find the first generation of galaxies 
and understand their roles in the reionization of the intergalactic medium (IGM) 
between $z \sim 6$ to $z \sim 12$.
Substantial progress has been made in 
recent years in finding objects at $z \sim 6 - 7$ \citep{Bouwens06, Bouwens09, Bunker10, Oesch10a, Oesch10, McLure10, Capak11},
which are, at most, $\sim 700 - 900$ Myr of age. We wish to find objects at even earlier epochs, in order to understand 
(1) how the first galaxies were formed through a hierarchical merging process; 
(2) how the chemical elements were generated and redistributed through the galaxies; 
(3) how the central black holes exerted influence over the galaxy formation; 
and 
(4) how these objects contributed to the end of the ``dark ages.''

The majority of galaxies at $z = 6 - 7$ have been discovered via two methods: 
(1) deep pencil-beam surveys, such as the Hubble Ultra Deep Field (HUDF; \cite{HUDF06}) 
and the Great Observatories Origins Deep Survey (GOODS; \cite{GOODS}); and 
(2) degree-size surveys with 10m-class ground-based telescopes, such as the Subaru Deep Field. 
Both use a color selection to search for ``dropout'' candidates -- galaxies  
with deep IGM absorption at the wavelengths shortward of the redshifted Ly$\alpha$ break. 
These efforts have proven successful at finding objects at $z<7$, 
but there is growing evidence for a rapid decrease in the number of candidates at higher redshifts
\citep{Iye06,Bouwens06,Castellano10}.
{\em HST/WFC3} data yields 73 $z \sim 7$ and 59 $z \sim 8$ candidates \citep{Bouwens10b}.
These results are based on $\sim500$ orbits of HUDF observations. 
In comparison, $>600$ galaxies are found at $z \sim 6$ \citep{Bouwens07,Su2011}. 
To date, there have been only two or three spectroscopically confirmed galaxies at $z>7$ \citep{Lehnert10,Ono11,Schenker11}
and one spectrum for a Gamma Ray Burst at $z \sim 8.2$ \citep{Salvaterra09,Tanvir09}.
The reason, in part, is a lack of photons: 
at redshift $z=8$ and 10, an $L^*$ galaxy would have an apparent magnitude in 
the first NIR detection band of 28.2 and 29.6, respectively. 

Gravitational lensing by clusters amplifies the flux of background sources considerably. 
These cosmic telescopes improve the efficiency
of searches for relatively bright high redshift galaxies \citep{Bouwens09,Maizy10}.
In fact, the majority of $m< 25.5$ AB, $z \simgreat 6.5$ galaxy candidates have been found in cluster fields
\citep[e.g.,][]{Kneib04,Bradley08,Zheng09,Bradley11,Schenker11}.
Interestingly, most of these candidates are found in regions with amplification of $\sim10$. 
Furthermore, lens models can be used to help discriminate 
between highly-reddened objects and truly distant, high redshift objects,
as the projected positions of the lensed images are strong functions of the source redshifts.

\begin{figure}
\plotone{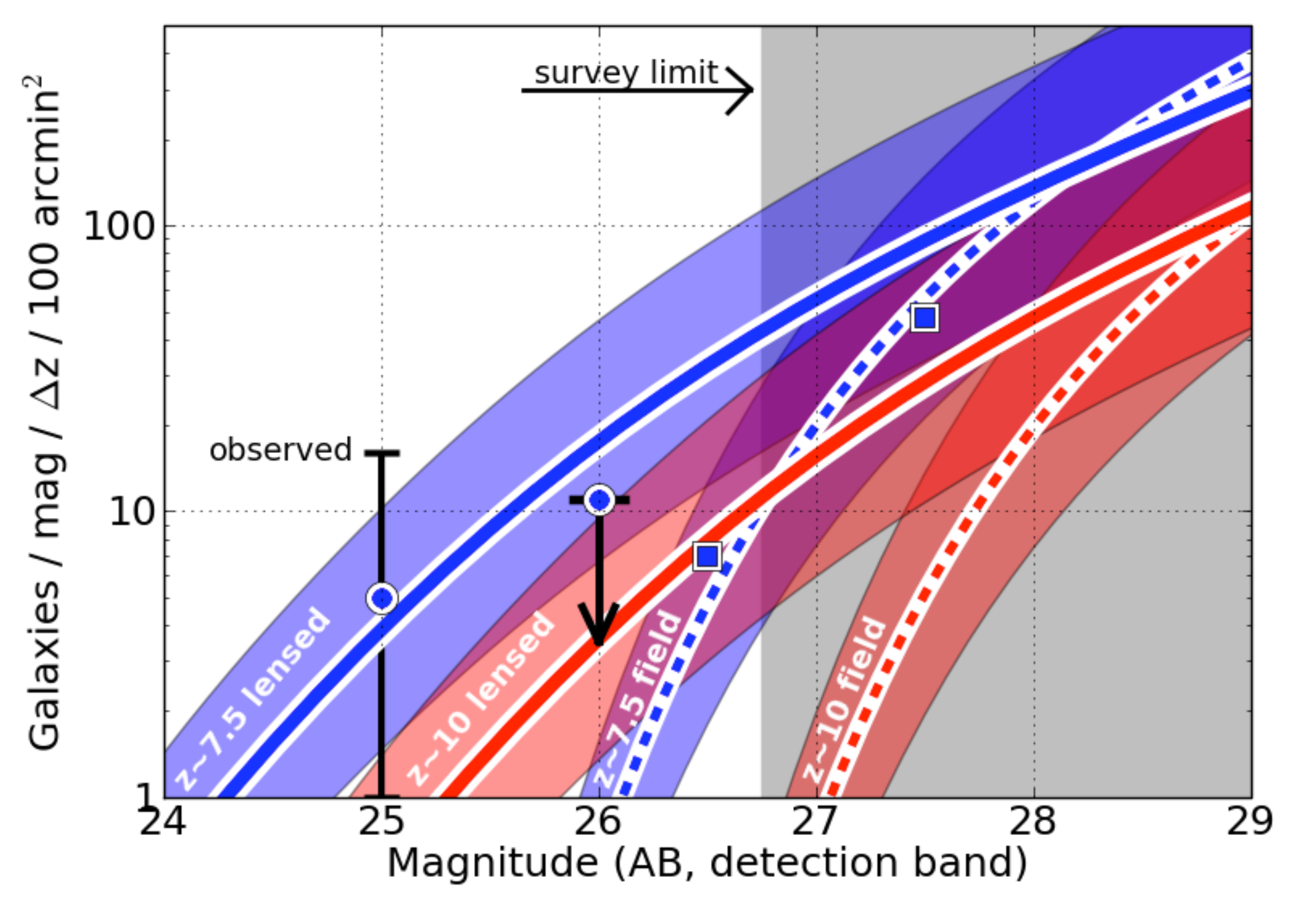}
\caption{\label{highz}
The estimated number counts in the total CLASH survey at $z \sim 7.5$ and 10 
are plotted as the solid ``lensed'' blue and red lines, respectively,
as functions of magnitudes in the detection bands F110W and F160W.
Note the large uncertainties (shaded regions).
A NIR survey of a comparable area (100 square arcmin) in unlensed fields
would have a $\sim 10\times$ lower efficiency at these magnitudes
(in agreement with \citealt{Maizy10}).
We have assumed an evolving Schechter luminosity function 
with $dM^{\star}/dz = 0.36$ as derived by \citep{Bouwens06,Bouwens08} 
and plotted as the dashed lines with shaded uncertainties.
Observed number counts from \cite{Bouwens09} are plotted as squares.
Lensing estimates were derived using the \cite{Zitrin09a} CL0024+17 mass model 
to simulate the magnifications of sources and reduction of source area.
These estimates are in agreement with observed lensed counts in cluster fields
plotted as circles \citep{Bouwens09}.
}\end{figure}

Luminous $z>7$ galaxies are extremely valuable as their spectra can
be used to determine the epoch of the IGM reionization. This is because only a tiny fraction of
neutral hydrogen is needed to produce the high opacity of Ly$\alpha$
observed at $z \sim 6$.  The damped Ly$\alpha$ absorption profile that results from a
completely neutral IGM \citep{Miralda-Escude98} can be measured even at
low-spectral resolution.  Furthermore, direct measurements of the
early star-formation rate (via Ly$\alpha$ and H$\alpha$ emission, \citealt{Iye06}) 
can be derived from the spectra of bright high-$z$ galaxies.
CLASH may detect dozens of relatively bright (magnified to $m < 26.7$ AB)
$z > 7$ galaxies, including some bright enough for spectroscopic follow-up.
A NIR survey of a comparable area (100 square arcmin) in unlensed fields
would have a $\sim 10\times$ lower efficiency. An estimate, albeit a highly uncertain one, of the 
CLASH high-$z$ detection efficiency enhancement is shown in Figure~\ref{highz}.

\subsection{Galaxy Evolution}
\label{sec:gal}

\LCDM\ provides a robust theoretical framework for the evolution
of dark matter haloes.  However, the formation of galaxies within
these haloes is governed by complicated baryonic interactions that are
difficult to simulate on cosmological scales.  The gap in our understanding of the
connection between the formation
of galaxies and that of their parent haloes is illustrated by
observations of galaxy ``downsizing'' \citep{Cowie96}.  \LCDM\ 
predicts that structures form hierarchically, with small haloes
forming early and later assembling into larger haloes.  In contrast,
massive galaxies appear to have formed at early times, as evidenced by
the redshift at which their star formation peaks
\citep[e.g.,][]{Cowie96, Guzman97, Brinchmann00, Juneau05} and the old ages
of their stellar populations at $z=0$ 
(e.g., \citealt{faber95,  worthey96, proctor02, thomas05}).  
The stellar mass density of the universe is also increasing with cosmic time, 
as galaxies form stars, then shut down their star formation 
and, ultimately, accumulate on the red sequence\citep{Bell04, faber07}.

Several open questions remain about the origin and nature of this stellar mass growth. 
Is it dominated by {\it in situ} star formation, or through mergers of
existing stellar systems? How does the balance of these two processes
change with galaxy mass and cosmic time?  In a system dominated by
hierarchical assembly, this is akin to asking whether most of the mass
accreted by galaxies comes in as gas (e.g., the ``cold flows'' of
\citealt{keres05, dekel_birnboim06}) or as stars (e.g., ``dry
mergers'', \citealt{khochfar03, Bell04}).  Theoretical
expectations predict that the {\it in situ} population should be very
centrally concentrated, while an accreted stellar component should
extend to larger radius \citep{oser10}.  Thus massive galaxies are
predicted to have substantial gradients in the origin of their stars,
with the innermost stars having formed {\it in situ} and the outer
stars largely accreted through merging.

Recent work has demonstrated that massive, passive galaxies already exist
at $z \sim 2$ (e.g., \citealt{trujillo06, kriek09}) and that
these systems are much more compact than local massive galaxies
\citep{trujillo06, longhetti07, van_dokkum08}, suggesting that they
must grow significantly in size between $z \sim 2$ and $z \sim 0$.
This process likely happens through ``dry" (dissipationless) merging, since such massive
galaxies are observed to have old stellar populations locally. 
There are also substantial populations of massive
star-forming galaxies over a similar redshift range. These galaxies tend to exhibit large star forming
``clumps'' \citep{cowie95, van_den_bergh96, elmegreen04a,
  elmegreen04b}, although at least some of these galaxies may form disks with
coherent rotation \citep{genzel10}.

There are several observational challenges to measuring and
characterizing these two modes of growth.  It is difficult to resolve
substructure in high-redshift galaxies, even from space.  Both
resolution and signal-to-noise considerations limit observations to
the upper end of the mass function.  Searches that use a small number
of imaging filters struggle to identify the redshifts (and therefore
the masses) of the targets and can only poorly constrain their
spectral energy distributions (SEDs) and stellar populations.  Studies
that rely on emission lines probe only the assembly of star-forming
galaxies.

The CLASH program is well suited for studying galaxy assembly at
$z < 3$.  The high magnification
and spatial stretching of strongly lensed distant galaxies, coupled with 
the broad 16-band photometry and resulting photometric redshifts, enables the measurement
of star formation rates and stellar ages in each lensed galaxy over
many resolution elements.  The magnification makes it possible to resolve
substructures in high-redshift lensed galaxies further down the
galaxy mass function than is possible in unlensed galaxies at similar
redshifts.  In addition, with CLASH, we can contrast the relative
properties of galaxy cores versus their outer regions, and (where they
exist) the properties of substructures such as clumps, spiral arms,
and ongoing mergers.
In the full CLASH sample of 25 clusters, we expect to find
several strongly lensed background galaxies per cluster that are
suitable for such analyses, providing a sample of 50 to 100 galaxies
with $1 < z < 3$.     

\section{CLASH Cluster Sample}
\label{sec:sample}

The CLASH program is robustly measuring galaxy cluster dark matter profiles and concentrations
for a systematic comparison with those realized in cosmological simulations. 
Specifically, our cluster sample size and selection criteria were chosen to allow the robust
measurement of deviations from the predicted cluster concentration 
distribution of $\sim 15\%$ or more at high statistical
confidence ($\sim 99$\% C.L.) given a relatively unbiased ensemble of clusters (\S\ref{sec:DM}). 

\subsection{Cluster Sample Selection}
\label{sec:clselection}

To date, robust joint SL+WL analyses have only been performed on a small, highly-biased sample of five to
ten clusters (Table~\ref{tab:cobs} and Figure~\ref{cobs}).
These clusters were selected for study primarily based on their gravitational lensing strength, which
tends to preferentially select systems with higher central matter concentrations.
To establish a sample that is largely free of lensing bias, we selected
20 massive clusters from X-ray-based compilations of dynamically relaxed systems. Sixteen of these
20 clusters were taken from the \cite{Allen08} compilation of massive relaxed clusters.  
Clusters in our X-ray selected subsample all have $T_x \ge 5$ keV and exhibit a high degree of dynamical
relaxation as evidenced by {\em Chandra X-ray Observatory} images that show 
well-defined central surface brightness peaks and nearly 
concentric isophotes. The clusters, in general, also show minimal evidence for 
departures from hydrostatic equilibrium in X-ray pressure maps.
Most of the 20 clusters have smooth and only mildly elliptical ($\left<\epsilon\right> = 0.19$) X-ray emission and 
a BCG within a projected distance of  23 kpc of the X-ray centroid. No lensing information was 
used a priori to select them in order to ensure that they are not preferentially aligned along the line of sight, 
in contrast with a purely lensing-selected sample, where the 
surface mass density is, on average, biased upwards along the line of sight by 
intrinsic triaxiality \citep{Hennawi07,Oguri09,Meneghetti10,Meneghetti11}.
A handful of the clusters in the CLASH X-ray selected subset have some evidence for departures from
symmetric X-ray surface brightness distributions. These systems are briefly discussed in
\S\ref{sec:cnotes}.

\begin{figure*}
\plotone{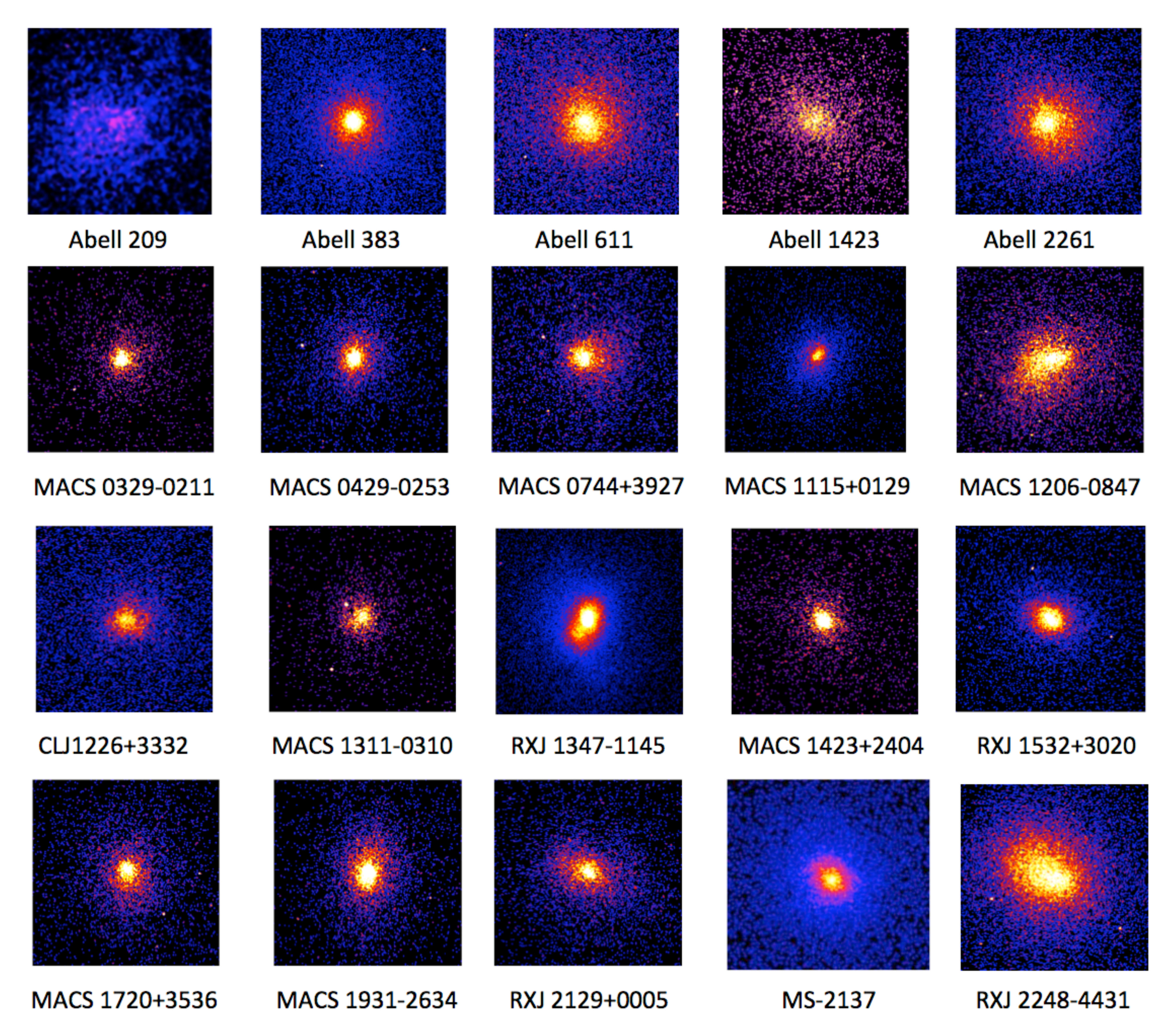}
\caption{\label{xrayim}
Cutouts of Chandra X-ray images centered on the 20 CLASH clusters in
the X-ray selected subsample. These images are taken from the Archive
of Chandra Cluster Entropy Profile Tables (ACCEPT). Each cutout subtends
$\sim$3.45 arcminutes, which is nearly the same as the {\em ACS/WFC} field of view on \HST.
}\end{figure*}

These 20 clusters comprise the 
primary sample for our DM distribution studies. Existing X-ray data will be used to measure the 
gas density profile for subtraction from the total mass profile (as 
applied by \cite{Lemze08} for A1689), which is necessary when making detailed comparison 
with the predictions of DM-only simulations. Although these 
clusters are X-ray selected, one or more lensed arcs are indeed visible in all 
of the clusters for which sufficiently high-quality imaging was available. 
This indicates the X-ray selected clusters in our sample have Einstein radii in the range 15$''$ to 35$''$.
X-ray images from the
ACCEPT\footnote{ACCEPT = Archive of Chandra Cluster Entropy Profile Tables. See http://www.pa.msu.edu/astro/MC2/accept/ for more information}
database for these 20 CLASH clusters are shown in
Figure~\ref{xrayim}.

In addition, CLASH includes five clusters 
selected based solely on their exceptional strength as gravitational lenses 
(large Einstein radii, $\theta_{Ein} > 35''$ for $z_s = 2$).
Analysis of these clusters will allow us to further quantify the lensing selection bias toward high concentrations. 
The excellent strong lensing data yielded by these clusters will enable derivation of 
some of the highest resolution dark matter maps which may be obtained \citep[as in][]{Coe10}.
The primary motivation for selecting these five ``high-magnification" clusters, however, was to significantly
increase the likelihood of discovering very highly magnified high redshift galaxies (\S \ref{sec:hiz}).
While most of the CLASH clusters lie behind regions of low Galactic extinction (median E(B-V) = 0.026), 
there is one cluster, MACS2129.4-0741, with
significant amounts of Galactic cirrus located northeast of the cluster core. Because this cluster has one of the larger Einstein
radii known, its inclusion in the ``high-magnification" sample was deemed worthwhile in spite of the presence of this cirrus.

\begin{figure*}
\plottwo{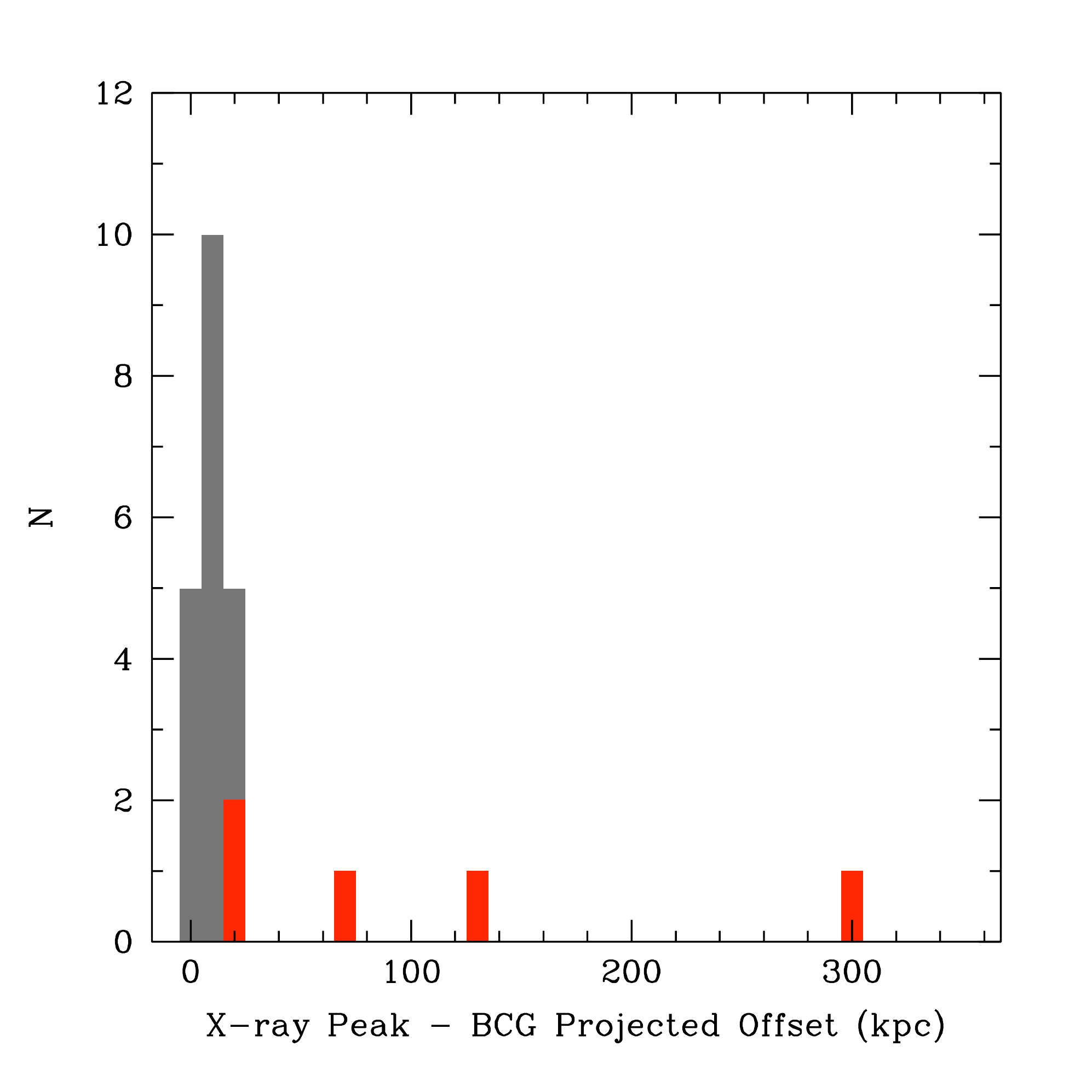}{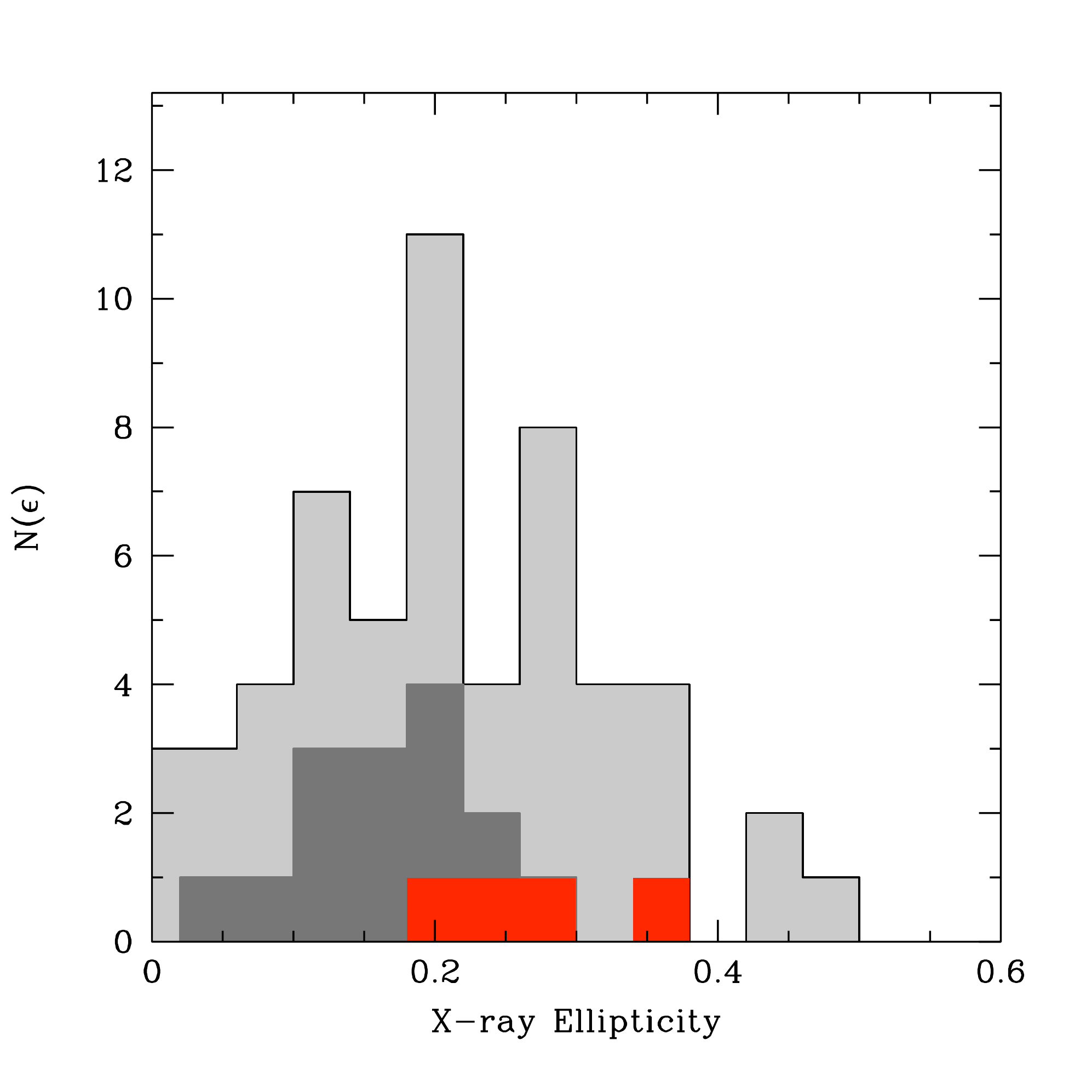}
\caption{\label{xoffset}
{\bf Left:} The distribution of the projected separation between the Brightest Cluster Galaxy (BCG) and the peak of the
X-ray surface brightness for the X-ray selected (dark grey) and Einstein-radii selected (red) cluster subsamples.
The average projected BCG - X-ray peak separation for the X-ray selected sample is 10.4 kpc.
{\bf Right:} The distribution of the ellipticity of the X-ray emitting intracluster gas 
for the X-ray selected (dark grey) and Einstein-radii selected (red) cluster subsamples.
The distribution of the ellipticities of the 96 clusters in the \cite{Maughan08} sample that
are {\bf not} in common with CLASH is shown in the light grey histogram.
The mean X-ray emission ellipticities for the X-ray and Einstein-radii selected cluster
samples are 0.19 and 0.28, respectively.
}\end{figure*}

Figure~\ref{xoffset} shows, respectively, the distribution
of the projected separation between the BCG and the peak of the X-ray surface brightness and the
distribution of the ellipticity of the X-ray emission for the CLASH sample. Ellipticity measurements
are from \cite{Maughan08}. The X-ray selected cluster
sub-sample exhibits, on average, a very small offset between the BCG and peak X-ray flux. The X-ray
selected cluster sub-sample also exhibits a lower average ellipticity than our high-magnification 
(Einstein-radii selected) sample but is consistent with the mean ellipticity of the larger \cite{Maughan08} sample.

The 25 clusters in the CLASH program are presented in Table~\ref{tab:MCT} and some of
their key X-ray properties are given in Table~\ref{tab:ClusProp}, including the bolometric luminosity
(defined for convenience to be between 0.1 and 100 keV), temperature,
and cluster-to-solar [Fe/H] ratio, where the solar abundance reference is from \cite{Anders89}. Table~\ref{tab:ClusProp} also lists the
source of the X-ray selection in column 6. 
The CLASH sample is drawn heavily from the Abell and MACS cluster catalogs 
\citep{Abell58,Abell89,Ebeling01,Ebeling07,Ebeling10}. The CLASH clusters
span almost an order of magnitude in mass ($\sim5$ to $\sim30 \times 10^{14}$ M$_{\odot}$).
These clusters were also selected to cover a wide redshift range ($0.18 < z < 0.90$ with a median $z_{med} = 0.40$)
allowing us to probe the full $c(M,z)$ relations expected from simulations.

\begin{figure}
\plotone{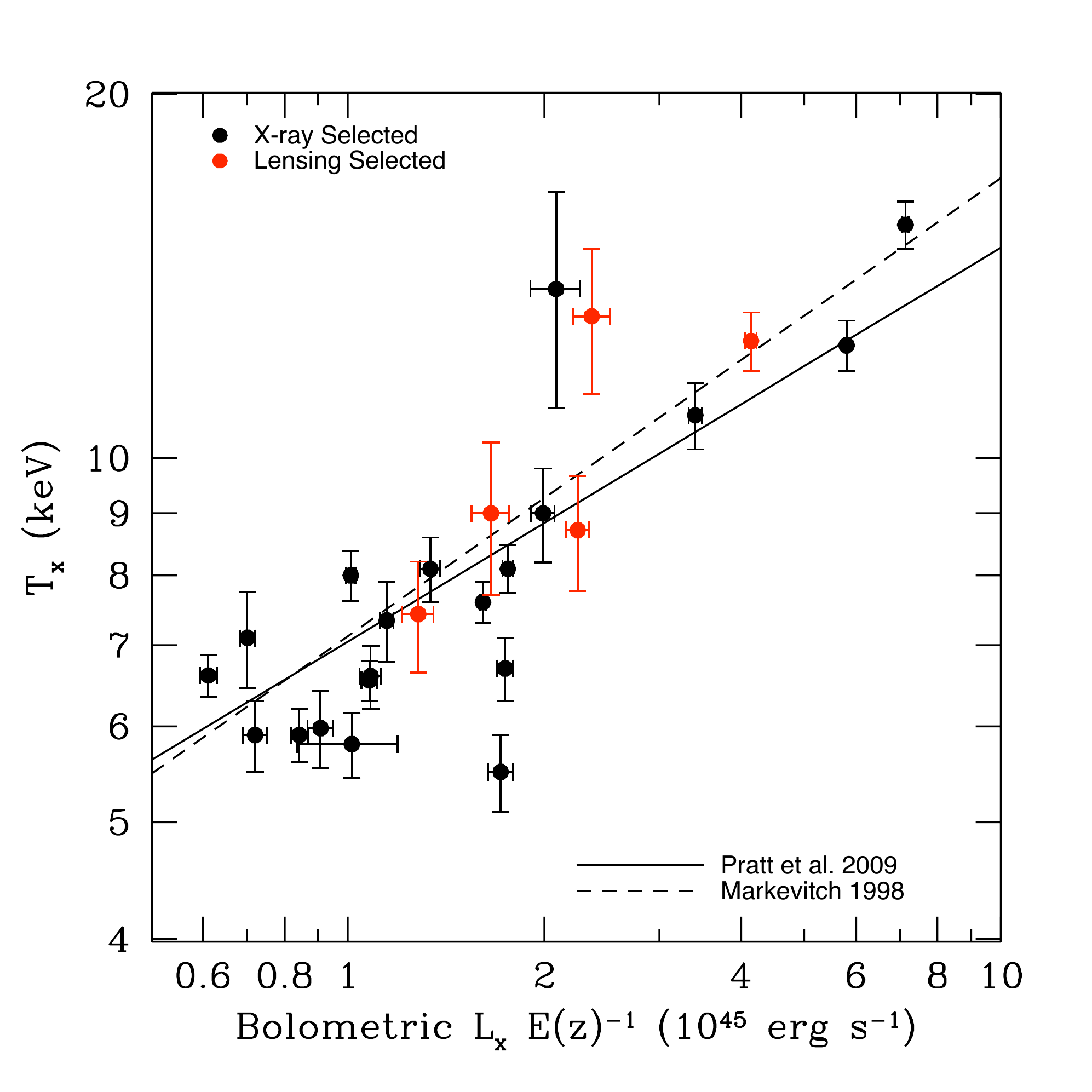}
\caption{\label{lxtx}
The X-ray temperature, T$_x$, as a function of the scaled bolometric luminosity, L$_x$/E(z), for the CLASH cluster sample.
Data points are from Table~\ref{tab:ClusProp}. The X-ray luminosities are scaled assuming self-similar evolution, to allow direct
comparison to the low-redshift luminosity temperature relationships from \cite{Markevitch98b}
 and \cite{Pratt09}. The $L_x - T_x$ relations from \cite{Markevitch98b} and \cite{Pratt09} (L$_2$ (measured within 0.15-1.00 R
 $_{500}$) vs. T$_3$  (0.15-0.75 R$_{500}$)) are shown. CLASH clusters are representative of these larger X-ray cluster
samples.
}\end{figure}

The X-ray parameters in Table~\ref{tab:ClusProp} are derived by us
using {\tt CIAO v4.3} and {\tt CALDB v4.4.3}. We filtered
the Chandra data for flares and reprojected the deep Chandra background fields to match the cluster 
observations. The background data were also filtered for ``status$=0$" events to be suitable for use with the 
{\tt VFAINT} mode data. To obtain X-ray luminosities and temperatures uniformly across the sample, we extracted 
spectra from apertures with radii of 714 kpc (500 $h_{100}^{-1}$ kpc) and excluded the 
central 71.4 kpc (50 $h_{100}^{-1}$ kpc).
These ``core-excised'' spectra provide X-ray temperature estimates that are relatively unaffected by the
presence or absence of a cool core (e.g., \cite{Markevitch98b}). Figure~\ref{lxtx} shows 
the X-ray temperature, T$_x$, as a function of the scaled bolometric luminosity, L$_x/E(z)$, for CLASH clusters using the data in Table~\ref{tab:ClusProp}.
The X-ray luminosities are scaled assuming self-similar evolution, $E(z) =  \sqrt{\Omega_{m}(1+z)^3 + \Lambda}$, to allow direct
comparison to the low-redshift luminosity temperature relationships from \cite{Markevitch98b}
 and \cite{Pratt09} (L$_2$ (measured within 0.15-1.00 R$_{500}$) vs. T$_3$  (0.15-0.75 R$_{500}$), to be specific.)
We verified that the deep background particle
event rates matched that seen in the cluster spectra between $9 - 12$ keV. The X-ray counts were binned to a minimum
of 20 counts per energy bin. Since the extracted spectra were
dominated by source counts, the results were not very sensitive to the background scaling. To estimate
the cluster temperature, we used {\tt XSPEC 12.6.0q} to fit the X-ray spectra between $0.7 - 7.0$ keV.
We assumed a single temperature plasma model (the {\tt XSPEC} apec model) absorbed
by a Galactic hydrogen column fixed to the value obtained from the 
Bell survey \citep{DickeyLockman90}. The
metallcity, temperature, and normalization were allowed to be free. The best fit temperatures and $1\sigma$
error bars, and the bolometric X-ray luminosities  
are reported in Table~\ref{tab:ClusProp}.

\begin{deluxetable*}{llllcccc}
\tablewidth{0pt}
\tablecaption{\label{tab:MCT}The CLASH Cluster Sample and \HST\ Observing Plan}
\tablehead{
\colhead{}&
\colhead{}&
\colhead{}&
\colhead{}&
\colhead{HST}&
\colhead{CLASH}&
\colhead{Program}&
\colhead{Archival}\\
\colhead{Cluster}&
\colhead{$\alpha_{\rm J2000}$}&
\colhead{$\delta_{\rm J2000}$}&
\colhead{$z_{\rm Clus}$}&
\colhead{Cycle}&
\colhead{Orbits}&
\colhead{ID}&
\colhead{Orbits$^c$}
}
\startdata
X-ray Selected Clusters:  \\
   ~~Abell 209                              & 01:31:52.57 & $-$13:36:38.8                &  ~0.206        & ~  19 & 20 & 12451 & (3)   \\
   ~~Abell 383                              & 02:48:03.36 & $-$03:31:44.7                &  ~0.187        & ~  18 & 20 & 12065 & (3)  \\
   ~~MACS0329.7-0211                 & 03:29:41.68 & $-$02:11:47.7                &  ~0.450         & ~  19 & 20 & 12452 & (2.5) \\
   ~~MACS0429.6-0253                 & 04:29:36.10 & $-$02:53:08.0                &  ~0.399         & ~  20 & 20 & 12788 & (0.5) \\
   ~~MACS0744.9+3927                & 07:44:52.80 & +39:27:24.4                   &   ~0.686        & ~  18 &17 & 12067 & 6  \\
   ~~Abell 611                              & 08:00:56.83 & +36:03:24.1                   &   ~0.288        & ~  19 & 18 & 12460 & 2 \\
   ~~MACS1115.9+0129                & 11:15:52.05  &+01:29:56.6                    &   ~0.352        & ~  19 & 20 & 12453 & 0.5 \\
   ~~Abell 1423                               & 11:57:17.26 &+33:36:37.4                     & ~0.213          & ~ 20  & 20 & 12787 & (0.5) \\
   ~~MACS1206.2-0847                 & 12:06:12.28 & $-$08:48:02.4                &   ~0.440         & ~  18 & 20 & 12069 & 0.5 + (0.5)  \\
   ~~CLJ1226.9+3332                     & 12:26:58.37 & +33:32:47.4                   &   ~0.890         & ~  20 & 18 & 12791 & 16 \\ 
   ~~MACS1311.0-0310                  & 13:11:01.67 & $-$03:10:39.5               &   ~0.494          & ~  20 & 20 & 12789 & 0 \\
   ~~RXJ1347.5-1145                      & 13:47:30.59 & $-$11:45:10.1               &   ~0.451          & ~  18 & 15  & 12104 & 6 + (0.5) \\
   ~~MACS1423.8+2404                 & 14:23:47.76 & +24:04:40.5                  &   ~0.545          & ~  20 & 17  & 12790 & 5 \\
   ~~RXJ1532.9+3021                     & 15:32:53.78 & +30:20:58.7                  &  ~0.345           & ~  19 & 20  & 12454 & (0.5) \\
   ~~MACS1720.3+3536                 & 17:20:16.95 & +35:36:23.6                  &  ~0.391           & ~  19 & 20  & 12455 & (0.5) \\
   ~~Abell 2261                             & 17:22:27.25 & +32:07:58.6                  &  ~0.224           & ~  18 & 20 & 12066 & (0.5) \\
   ~~MACS1931.8-2635                  & 19:31:49.66 & $-$26:34:34.0                & ~0.352           & ~  19 & 20  & 12456  & 0 \\
   ~~RXJ2129.7+0005                     & 21:29:39.94 & +00:05:18.8                   &  ~0.234           & ~  19 & 20 & 12457 & (1)  \\
   ~~MS2137-2353                        & 21:40:15.18 & $-$23:39:40.7               &  ~0.313           & ~  18 & 18 & 12102 & 5 + (7) \\
   ~~RXJ2248.7-4431 (Abell 1063S) & 22:48:44.29 & $-$44:31:48.4             &  ~0.348           & ~  19 & 20 & 12458 & (0.5) \\
 ~~\\
   High Magnification Clusters:  \\
   ~~MACS0416.1-2403                 & 04:16:09.39 & $-$24:04:03.9  		    & ~0.42$^b$    & ~  19 & 20 & 12459 & (1)  \\
   ~~MACS0647.8+7015                & 06:47:50.03 & +70:14:49.7                     & ~0.584           & ~  18 & 18 & 12101 & 9  \\
   ~~MACS0717.5+3745                & 07:17:31.65 & +37:45:18.5                     & ~0.548           & ~  18 & 17 & 12103 & 7 \\
   ~~MACS1149.6+2223                & 11:49:35.86  & +22:23:55.0                    & ~0.544           & ~  18 & 18 & 12068 & 5  \\
   ~~MACS2129.4-0741                & 21:29:26.06$^a$ & $-$07:41:28.8$^a$   & ~0.570           & ~  18 & 18 & 12100 & 5  \\
\enddata
\tablenotetext{a}{Central cluster coordinates derived from optical image instead of X-ray image}
\tablenotetext{b}{Cluster redshift for MACS0416.1$-$2403 is 
based on {\em Chandra} X-ray spectrum (this work). Uncertainty on this value is $\pm0.02$. For this fit, the core
was not excluded. To maximize the counts, an aperture of 714 kpc was used, binned to achieve a minimum of 20 counts per energy bin. The background in
the 0.7-7.0 keV range was fit from the {\em Chandra} deep fields. }
\tablenotetext{c}{Archival \ACS\ or {\em WFPC2} imaging data only; {\em WFPC2} orbits shown in parentheses. Archival \ACS\ images, when available, are
used in conjunction with new CLASH data to achieve the desired depths in all filters.}
\end{deluxetable*}

\begin{deluxetable*}{lrrcccc}
\tablewidth{0pt}
\tablecaption{\label{tab:ClusProp}X-ray Properties of the CLASH Cluster Sample}
\tablehead{
\colhead{}&
\colhead{kT}&
\colhead{(L$_{Bol}$)$^a$}&
\colhead{[Fe/H] Ratio}&
\colhead{Ellipticity$^b$}&
\colhead{Centroid Shift$^b$}&
\colhead{X-ray Morphology}\\
\colhead{Cluster}&
\colhead{(keV)}&
\colhead{(10$^{44}$ erg s$^{-1}$)}&
\colhead{(Cluster/Solar)$^c$}&
\colhead{$\epsilon$}&
\colhead{$\left< w \right>$(10$^{-2}R_{500}$)}&
\colhead{Reference}
}
\startdata
X-ray Selected Clusters:  \\
   ~~Abell 209                                & $7.3\pm 0.54$      &$12.7\pm 0.3$ 	&$0.18\pm0.09$&  $0.21 \pm 0.01$ & $0.55 \pm 0.05$  &     1 \\
   ~~Abell 383                                & $6.5\pm 0.24$      &   $6.7\pm 0.2$ 	&$0.59\pm0.10$& $0.04 \pm 0.01$ & $0.18 \pm 0.02$  &  2 \\
   ~~MACS0329.7-0211                    & $8.0\pm 0.50$   &$17.0\pm 0.6$	&$0.48\pm0.10$ & $0.15 \pm 0.03$ & $1.38 \pm 0.13$  &  2,3 \\
   ~~MACS0429.6-0253                    & $6.0\pm 0.44$      &$11.2\pm 0.5$ 	&$0.34\pm0.11$& $0.21 \pm 0.02$ & $0.38 \pm 0.03$ &  2,3 \\
   ~~MACS0744.9+3927                   & $8.9\pm 0.80$    &$29.1\pm 1.2$	&$0.37\pm0.10$& $0.11 \pm 0.02$ & $1.41 \pm 0.13$  &   2,3 \\
   ~~Abell 611                                 & $7.9\pm 0.35$    &$11.7\pm 0.2$	&$0.30\pm0.07$&\nodata  		     &\nodata    		&  2 \\
   ~~MACS1115.9+0129                    & $8.0\pm 0.40$    &$21.1\pm 0.4$	&$0.32\pm0.06$&\nodata  		     &\nodata    		&  2,3 \\
  ~~Abell 1423                                   & $7.1\pm 0.65$   &$7.8\pm 0.2$               &$0.23\pm0.13$&\nodata                       &\nodata                  &   4\\
   ~~MACS1206.2-0847                    & $10.8\pm0.60$     &$43.0\pm 1.0$	&$0.25\pm0.09$&\nodata  		     &\nodata     		&  3,4\\
   ~~CLJ1226.9+3332                        & $13.8\pm2.80$     &$34.4\pm 3.0$ 	&$0.36\pm0.25$& $0.10 \pm 0.03$  & $1.30 \pm 0.11$   & 2 \\ 
   ~~MACS1311.0-0310                     & $5.9\pm0.40$      &$9.4\pm 0.4$	&$0.42\pm0.10$&  $0.08 \pm 0.02$ & $0.39 \pm 0.03$   &  2,3 \\
   ~~RXJ1347.5-1145                         & $15.5\pm0.60$    &$90.8\pm 1.0$	&$0.20\pm0.06$& $0.26 \pm 0.01$ & $0.63 \pm 0.05$    &  2 \\
   ~~MACS1423.8+2404                   & $6.5\pm0.24$       &$14.5\pm 0.4$	&$0.35\pm0.06$& $0.17 \pm 0.03$ & $0.25 \pm 0.02$    &  2,3 \\
   ~~RXJ1532.9+3021                       & $5.5\pm 0.40$      &$20.5\pm 0.9$	&$0.52\pm0.12$& $0.18 \pm 0.02$ & $0.07 \pm 0.01$    &  2 \\
   ~~MACS1720.3+3536                   & $6.6\pm0.40$       &$13.3\pm 0.5$	&$0.29\pm0.09$& $0.17 \pm 0.02$ & $0.57 \pm 0.05$    &  2,3 \\
   ~~Abell 2261                               & $7.6\pm  0.30$     &$18.0\pm 0.2$	&$0.31\pm0.06$& $0.10 \pm 0.01$ & $0.71 \pm 0.06$    &  5 \\
   ~~MACS1931.8-2635                    & $6.7\pm 0.40$      &$20.9\pm 0.6$	&$0.26\pm0.08$& $0.30 \pm 0.01$ & $0.28 \pm 0.02$     &  2,3 \\
   ~~RXJ2129.7+0005                       & $5.8\pm0.40$       &$11.4\pm 2.0$	&$0.33\pm0.10$& $0.26 \pm 0.02$ & $0.55 \pm 0.05$     &   2 \\
   ~~MS2137-2353                          & $5.9\pm 0.30$      &$9.9\pm 0.3$	&$0.41\pm0.08$& \nodata 		    &\nodata      		&  2 \\
   ~~RXJ2248.7-4431 (Abell 1063S) & $12.4\pm0.60$    &$69.5\pm 0.1$	&$0.39\pm0.06$& $0.20 \pm 0.01$ & $0.74 \pm 0.06$     & 6\\
 ~~\\
   High Magnification Clusters:  \\
   ~~MACS0416.1-2403                 & $7.5\pm0.80$          &$16.0\pm 0.9$	&$0.40\pm0.14$&\nodata  		    &\nodata    		&   3 \\
   ~~MACS0647.8+7015               & $13.3\pm 1.80$        &$32.5\pm 2.1$ 	&$0.30\pm0.19$& $0.36 \pm 0.02$ & $0.62 \pm 0.06$   &   3 \\
   ~~MACS0717.5+3745                & $12.5\pm 0.70$       &$55.8 \pm 1.1$    &$0.22\pm0.07$& $0.30 \pm 0.01$ & $2.11 \pm 0.18$   &  3 \\
   ~~MACS1149.6+2223                & $8.7\pm 0.90$         &$30.2 \pm 1.2$ 	&$0.24\pm0.11$& $0.25 \pm 0.02$ & $1.21 \pm 0.10$   &  3 \\
   ~~MACS2129.4-0741                & $9.0\pm 1.20$          &$22.6\pm 1.5$ 	&$0.40\pm0.17$& $0.19 \pm 0.03$ & $1.56 \pm 0.14$   &   3 \\
\enddata
\tablenotetext{a}{The X-ray bolometric luminosity covers the energy range 0.1 $-$ 100 keV.}
\tablenotetext{b}{The X-ray ellipticity, $\epsilon$, and centroid shift, $\left< w \right>$, values are from \cite{Maughan08}.}
\tablenotetext{c}{The solar abundance reference values used here are from \cite{Anders89}.}
\tablerefs{(1) \cite{Maughan08}; (2) \cite{Allen08}; (3) \cite{Ebeling07}; (4) \cite{Cavagnolo08}; (5) \cite{Mantz10}; (6) CLASH team selected cluster.}
\end{deluxetable*}

\subsection{Cluster Sample Size Requirements}
\label{sec:samplesize}

The required size of our ``relaxed'' cluster sample is derived from 
the goal to measure ``average'' cluster concentrations to $\sim 10\%$ 
(after accounting for variations in mass and redshift)
and to detect a $\sim 15\%$ deviation from the concentrations of simulated clusters at 99\% confidence.

\begin{figure}
\plotone{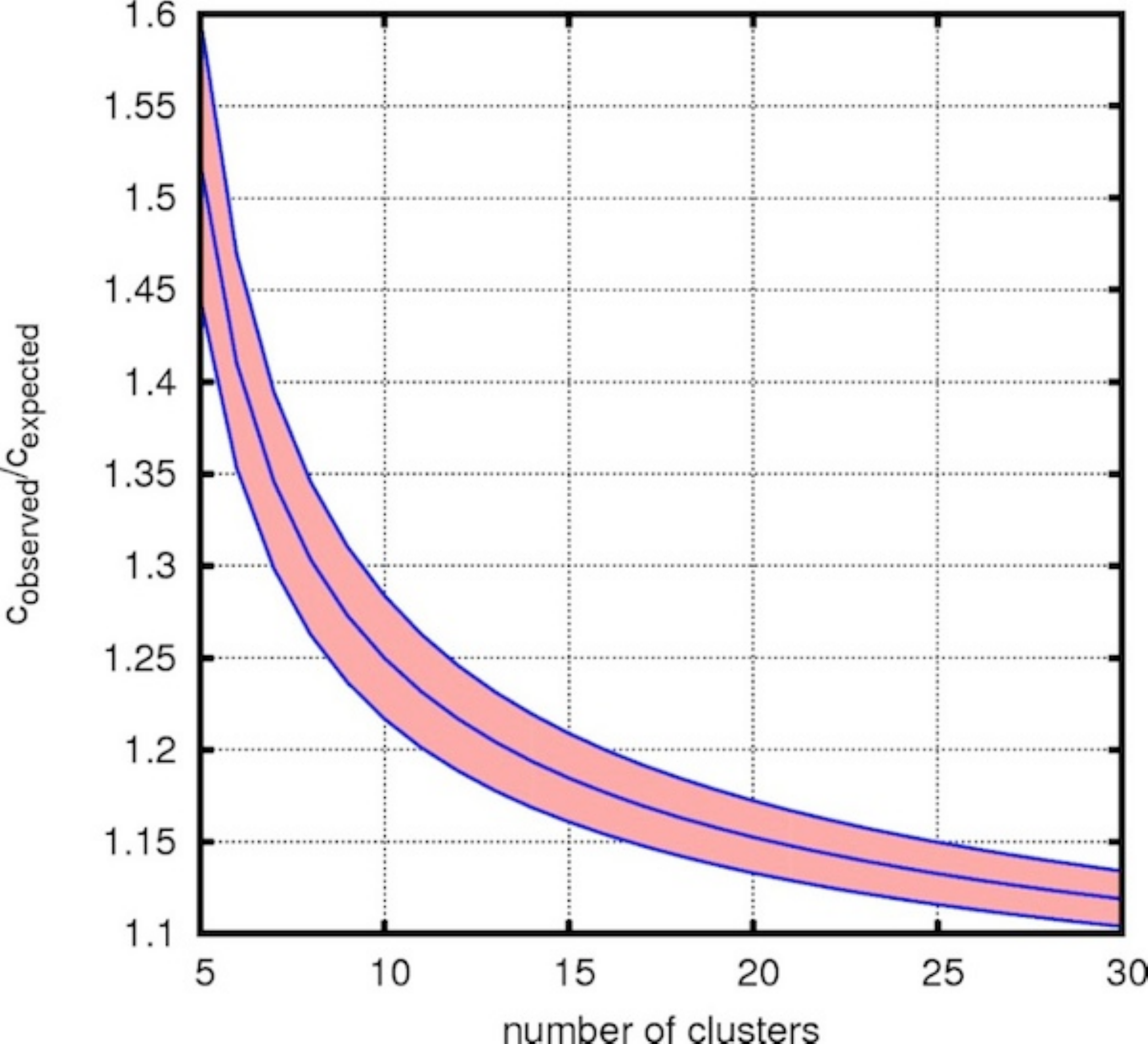}
\caption{\label{samplesize}
Mean ``over-concentration'' ratio 
necessary to reject with 99\% confidence
the hypothesis that measured concentrations 
are drawn from a sample with the expected values.
Average deviations of $\sim 15\%$ from expectations
will be detected with 99\% confidence
based on our relaxed sample of 20 clusters.
We have adopted the finding from \LCDM\ simulations that 
DM halo concentrations are log-normally distributed 
with a standard deviation of $\sigma \approx 0.25 \pm 0.03$ \citep[e.g.,][]{Duffy08}.
The shaded band includes this uncertainty.
}\end{figure}

Statistically, if we assume measurement errors are independent and normally distributed then the number of clusters, $N_{clus}$, 
needed to measure the average concentration to a fractional accuracy of $f$ is just
\begin{equation}
N_{clus} = (\sigma_{tot}/f)^2
\end{equation}
where the total scatter in an individual measurement of the concentration, $\sigma_{tot}$, is 
\begin{equation}
\sigma_{tot}^2 = \sigma_{int}^2 + \sigma_{LSS}^2 + \sigma_{meas}^2
\end{equation}
and includes contributions from
intrinsic variation (including the effects of orientation averaging of triaxial halo shapes), variations
due to intervening large scale structure (LSS), and measurement uncertainties.
Concentrations of relaxed simulated cluster halos 
have intrinsic scatters of $\sim 30\%$ \citep[e.g.,][]{Neto07,Maccio08,Duffy08}.
LSS can scatter concentration measurements by $\sim 13\%$ \citep{Hoekstra02}.
And we conservatively estimate measurement uncertainties to be $\sim 37\%$
as follows based on empirical data
(though analysis of simulated data suggests $\sim 11\%$ accuracy is possible \citealt{Meneghetti10}).

Analysis of CL0024+17 yielded a concentration measurement uncertainty of $\sim 22\%$ \citep{Umetsu10},
which we will use here as a baseline.
We assume here that our precisions will be dominated by the quality of our SL data
and that our measurement uncertainty will 
scale roughly as $\sigma \propto 1 / \sqrt{N_{arc}} \propto 1 / R_{E}$,
where for CL0024+17 with $R_{E} \sim 35\arcsec$, $N_{arc} = 33$ arcs were detected.
Note we have also assumed $N_{arc} \propto R_{E}^{2}$.
Based on archival {\em WFPC2} and \ACS\ images,
we measured a mean $R_{E} \approx 20\arcsec$ for our sample of 20 relaxed clusters,
which should conservatively yield $N_{arc} \sim 12$ arcs per cluster.
This scales the $\sim 22\%$ uncertainty for CL0024+17 with 33 arcs
up to a conservative $\sim 37\%$ for an average cluster with just 12 arcs.

Adopting $\sigma_{int} = 0.30,\ \sigma_{LSS} = 0.13$ and $\sigma_{meas} = 0.37$, we find $\sigma_{tot}= 0.49$,
and hence, if $f = 0.10$ then $N_{clus} \sim 24$.
This empirical sample size estimate is consistent with one derived from numerical simulations of strong 
lensing, as shown in Figure~\ref{samplesize}. 
The approximately log-normal distribution of DM halo concentrations 
seen in such simulations \citep{Meneghetti10a}
indicates that $\sim$20 massive clusters (M$_{vir} > 5 \times 10^{14}$ M$_{\odot}$) 
are required to detect 15\% deviations from expectations in the standard cosmological model 
at 99\% confidence.

Note that we may expect deviations to be greater than $15\%$ 
if results from lensing-selected clusters are any indication.
Simulated $10^{15} M_{\odot} h^{-1}$ clusters at $z = 0$ typically have $c_{vir} \sim 5$ \citep[e.g.,][]{Duffy08}.
For this fiducial cluster, \cite{Oguri09} instead find $c_{vir} \sim 12$ for observed clusters
(based on a fit to values observed for 10 lensing clusters accounting for the various masses and redshifts).
Even after including a 50\% lensing bias,
these observed concentrations are still $\sim 70\%$ greater than expectations.
Being more conservative and assuming a factor of 2 (100\%) lensing bias \citep{Meneghetti10},
the observed concentrations would still be $\sim 20\%$ greater than expectations.
Although as discussed in \S\ref{sec:DM}, more recent simulations \citep{Prada11} may alleviate these discrepancies.

\subsection{Notes on Clusters with Possible Substructure}
\label{sec:cnotes}

While our X-ray selection criteria favors the inclusion of highly relaxed clusters in the CLASH sample,
for 8 of our clusters the dynamical state is somewhat ambiguous. 
Some researchers have reported evidence for substructure in the X-ray surface brightness profiles
of these 8 clusters. The presence of substructure may suggest that a cluster is not fully dynamically relaxed. 
We summarize this evidence below. We note that the presence of a minor degree of substructure does not inhibit our ability to
determine a cluster's mass profile characteristics. In simulations, relaxed clusters have DM concentrations that are, on average,
$\sim20$\% higher than the general cluster population \citep{Duffy08,Prada11}.
\begin{itemize}
\item Abell 209: This system is marginally unrelaxed according to \cite{Smith05}. 
They based their relaxed/unrelaxed characterization mainly on a mass ratio between the cluster main 
component and the total mass (both at $r<500$ kpc). Clusters with mass ratio values below $0.95$ are considered unrelaxed. 
Their value for Abell 209 is $0.87 \pm 0.06$. \cite{Maughan08}, hereafter M08, derived 
an ellipticity of $\epsilon =0.21\pm  0.01$ for this
cluster. \cite{Gilmour09}, hereafter G09, find it to be relaxed based on a visual examination of its X-ray morphology.
\item MACSJ0329.7-0211 and CLJ1226.9+3332: these two clusters are included in the A08 compilation (as well as in \cite{SchmidtAllen07}, hereafter SA07,
 and were classified in these works as dynamically relaxed. 
However, M08 report that
both exhibit evidence for substructure. 
\item MACSJ0744.9+3927: This cluster shows evidence for substructure (SA07; M08). On the other 
hand G09 find it to be relaxed based on a visual examination of its X-ray 
morphology.
\item MACSJ1206.2-0848: In both optical and X-ray images the cluster appears close to relaxed in projection, with a pronounced X-ray peak at 
the location of the BCG. 
However, some evidence of merger activity along the line of sight may be suggested by the very high velocity dispersion of 
1580 km s$^{-1}$. G09 visually classify this cluster as relaxed.
\item RXJ1347.5-1145: Significant substructure is observed in this system (SA07; M08). M08 measure
an average ellipticity of the X-ray isophotes of $\epsilon = 0.26 \pm  0.01$.
\item Abell 2261: M08 found a small level of substructure and G09 classified it as disturbed. 
\item RXJ2248.7-4431: M08 found this cluster to be slightly elliptical $\epsilon = 0.2  \pm 0.01$ and with some level of substructure, 
while G09 found it to be relaxed. 
\end{itemize}

\section{Survey Design and Implementation}
\label{sec:survey}

\begin{figure*}
\plotone{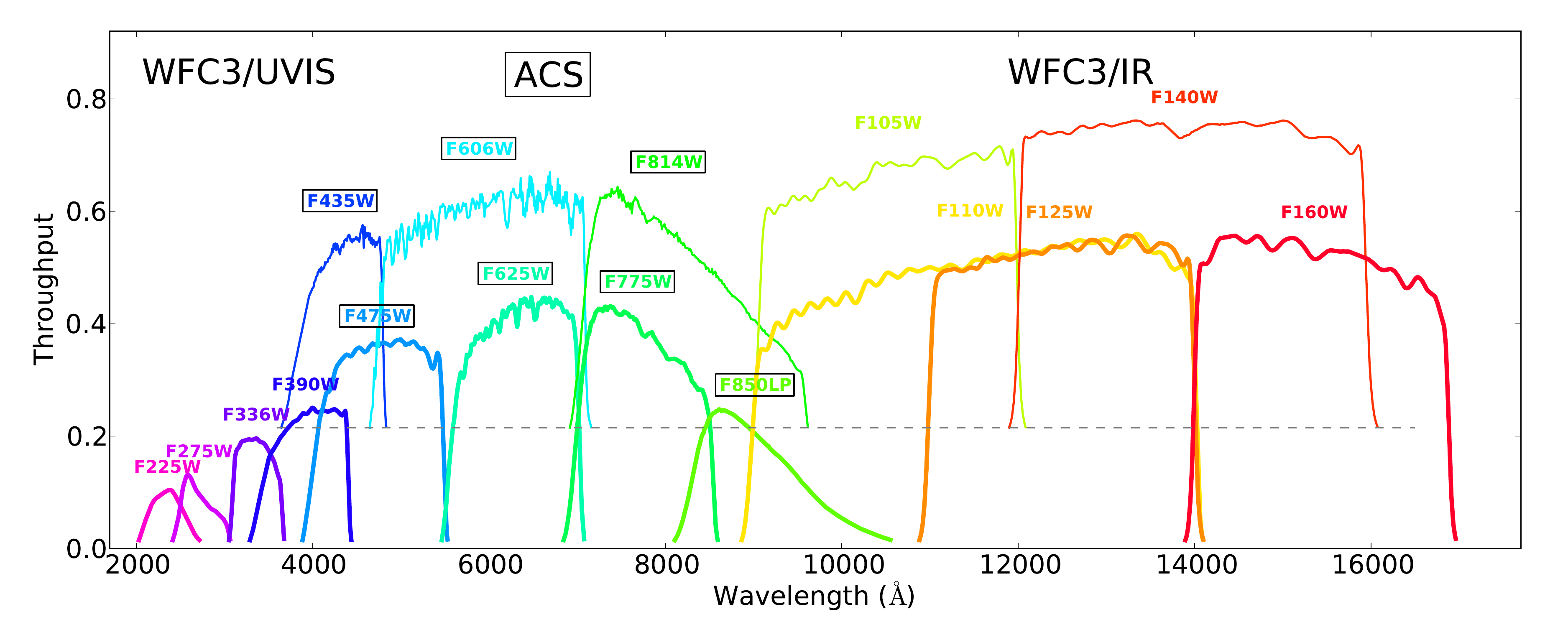}
\caption{\label{filters}
Each CLASH cluster is observed in 16 \HST\ filters 
spanning $\sim$2,000--17,000\AA\ with 
{\em WFC3/UVIS} in the near-ultraviolet,
\ACS\ in the optical (extending into the near-infrared),
and {\em WFC3/IR} in the near-infrared (see Table \ref{tab:filters}).
Total throughput curves are plotted for each filter.
For clarity, some curves are offset vertically by 0.2 (dashed line).
}\end{figure*}

The CLASH program consists of 524 \HST\ orbits,
including 50 for supernova follow-up.
The bulk of the program (474 orbits) will be used to image 25 galaxy clusters
each to a depth of 20 orbits divided among 16 \HSTACS\ and \WFCiii\ filters
as shown in Table \ref{tab:filters} and Figure~\ref{filters}.
The number of required cluster orbits is reduced from 500 to 474,
as some of these filter depths have already been achieved in existing data.
Columns 6, 7, and 8 in Table~\ref{tab:MCT} provide the number of new orbits allocated, the
\HST\ Program ID (where available), and the number of
existing archival orbits available, respectively, for each cluster.
The multi-band observations span
the near-ultraviolet to near-infrared (2,000 -- 17,000\AA).
In the wavelength range $\sim$4,000--9,000\AA,
we use \ACS\ for its greater throughput efficiency (especially toward the red end)
and larger observing area.

\begin{figure}
\plotone{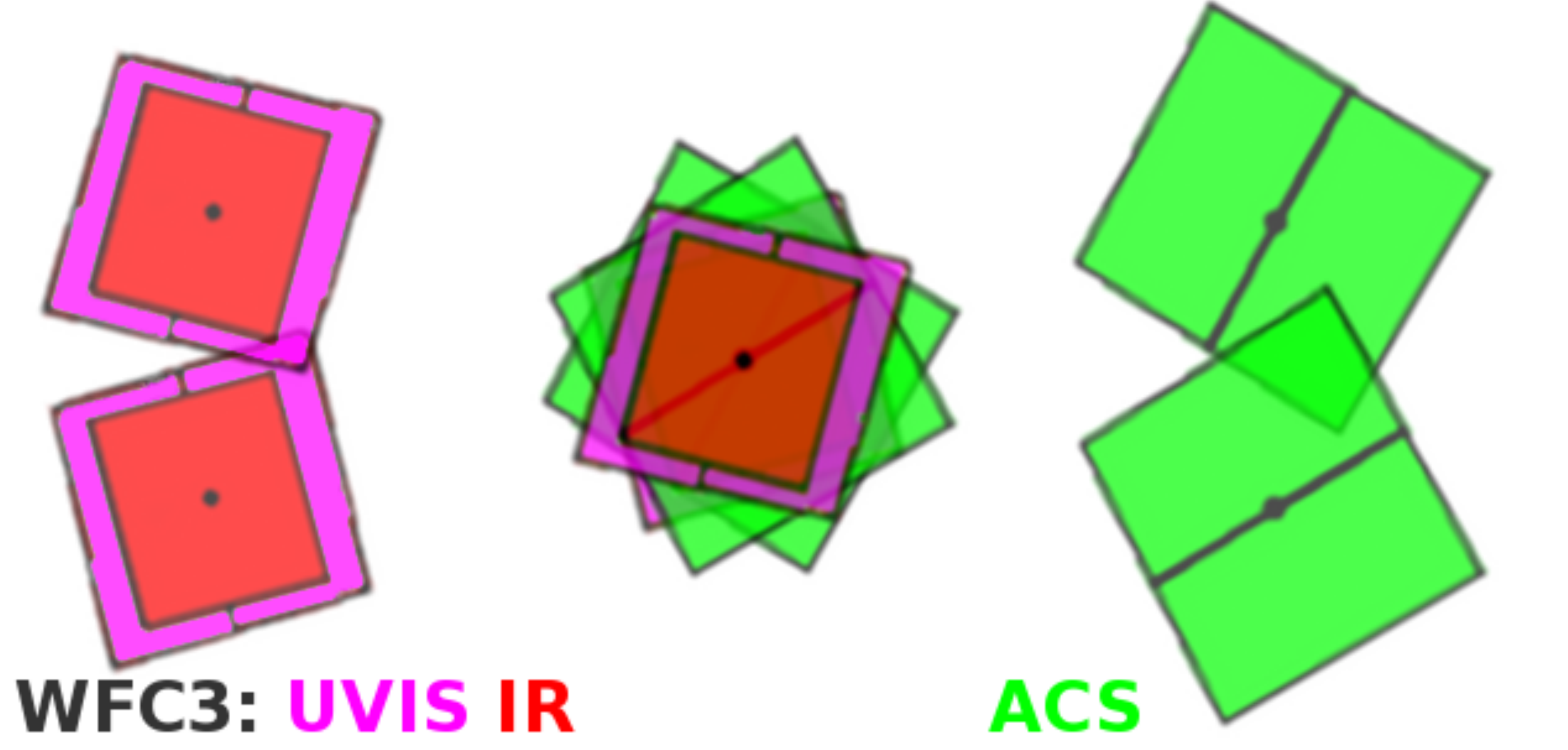}
\caption{\label{pointings}
Each cluster is observed at two orientations
approximately 30 degrees apart
to increase the supernova search area in the parallel fields.
{\em WFC3} parallel observations are obtained while the cluster core is being observed with \ACS\ 
and vice versa. At the median cluster redshift of $z_{med} = 0.4$, the 6 arcminute separation 
between the center of the parallel field and the cluster center corresponds to a projected distance of
1.9 Mpc.
}\end{figure}

\begin{deluxetable*}{ccccccc}
\tablewidth{0pt}
\tablecaption{\label{tab:filters}CLASH Exposure Times, Limiting Magnitudes, and Extinction Coefficients}
\tablehead{
\colhead{Camera /}&
\colhead{Filter}&
\colhead{}&
\colhead{Average Exposure} &
\colhead{$10\sigma$ Limit}&
\colhead{$5\sigma$ Limit}&
\colhead{Galactic Ext.}\\
\colhead{Channel}&
\colhead{Element}&
\colhead{Orbits}&
\colhead{Time (sec)$^a$}&
\colhead{(AB mag)$^b$}&
\colhead{(AB mag)$^b$}&
\colhead{(AB mag/E(B-V))}
}
\startdata
WFC3/UVIS & F225W & 1.5 & 3558 & 25.7 & 26.4& 7.474 \\
WFC3/UVIS & F275W & 1.5 & 3653 & 25.7 & 26.5& 6.140\\
WFC3/UVIS & F336W & 1.0 & 2348 & 25.9 & 26.6& 5.090\\
WFC3/UVIS & F390W & 1.0 & 2350 & 26.5 & 27.2& 4.514\\
ACS/WFC    & F435W & 1.0 & 1984 & 26.4 & 27.2& 4.117\\
ACS/WFC    & F475W & 1.0 & 1994 & 26.8 & 27.6& 3.724\\
ACS/WFC    & F606W & 1.0 & 1975 & 26.9 & 27.6& 2.929\\
ACS/WFC    & F625W & 1.0 & 2008 & 26.4 & 27.2& 2.671\\
ACS/WFC    & F775W & 1.0 & 2022 & 26.2 & 27.0& 2.018\\
ACS/WFC    & F814W & 2.0 & 4103 & 27.0 & 27.7& 1.822\\
ACS/WFC    & F850LP & 2.0 & 4045 & 26.0 & 26.7& 1.473\\
WFC3/IR      & F105W & 1.0 & 2645 & 26.6& 27.3& 1.015\\
WFC3/IR      & F110W & 1.0 & 2415 & 27.0 & 27.8& 0.876\\
WFC3/IR      & F125W & 1.0 & 2425 & 26.5 & 27.2& 0.757\\
WFC3/IR      & F140W & 1.0 & 2342 & 26.7 & 27.4& 0.609\\
WFC3/IR      & F160W & 2.0 & 4920 & 26.7 & 27.5& 0.470\\
\vspace{-0.1in}
\enddata
\tablenotetext{a}{Exposure times are the average time for each filter for all CLASH cycle 18 observations. 
Exposure times for cycles 19 and 20 may differ slightly due to
scheduling considerations.}
\tablenotetext{b}{~Limiting magnitudes are for a circular aperture that is 0.4 arcsec in diameter.}
\end{deluxetable*}

\begin{figure*}
\plottwo{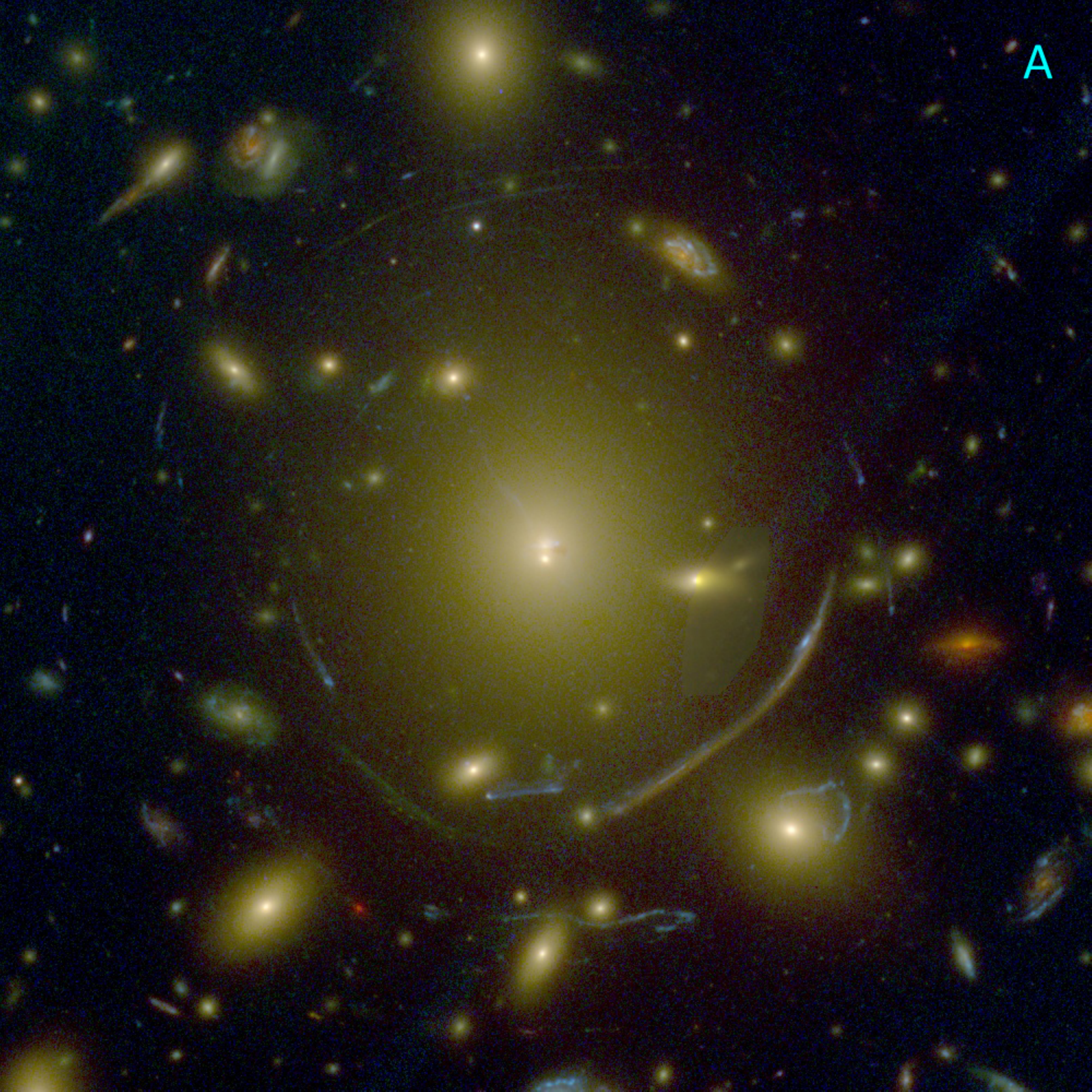}{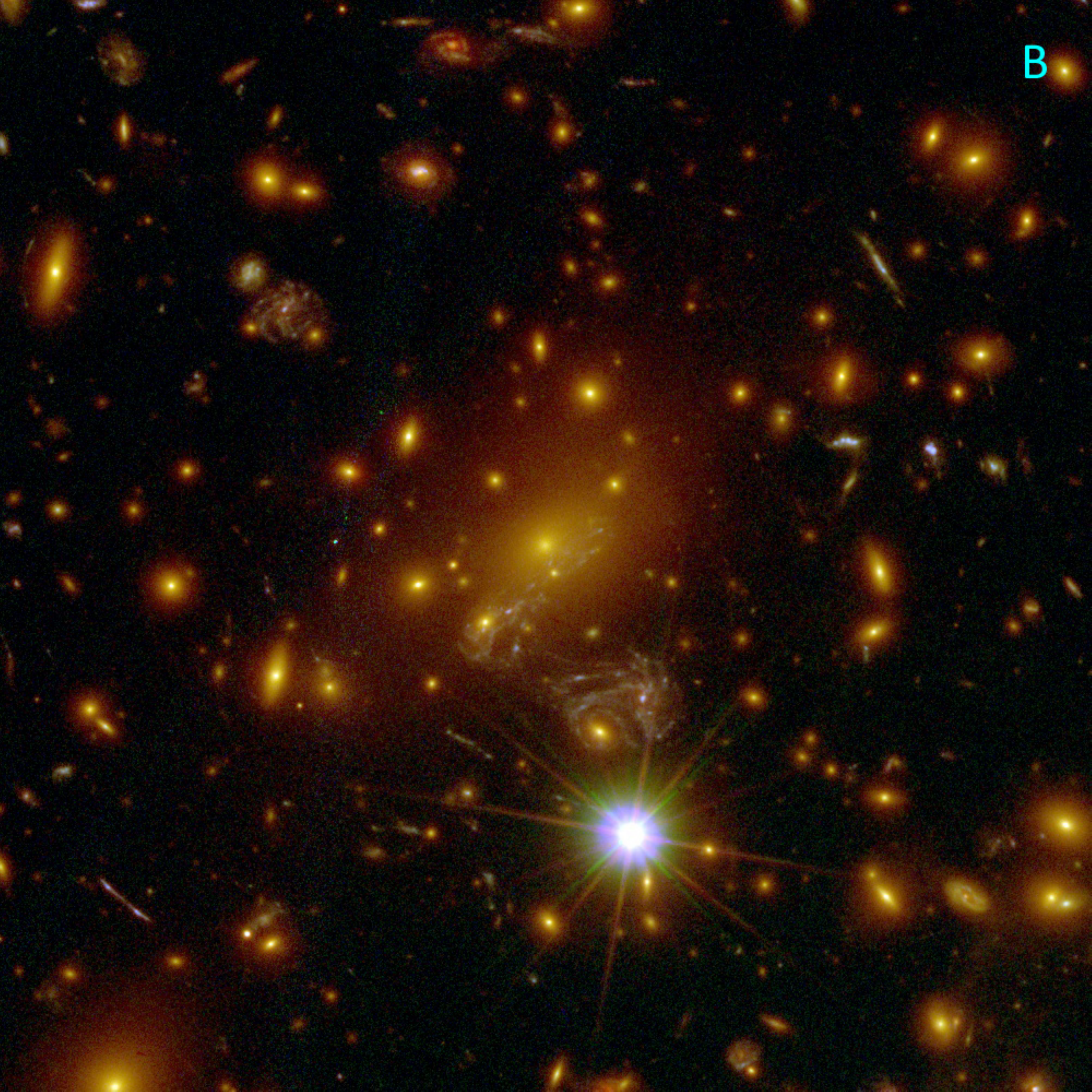}
\vskip 25pt
\plottwo{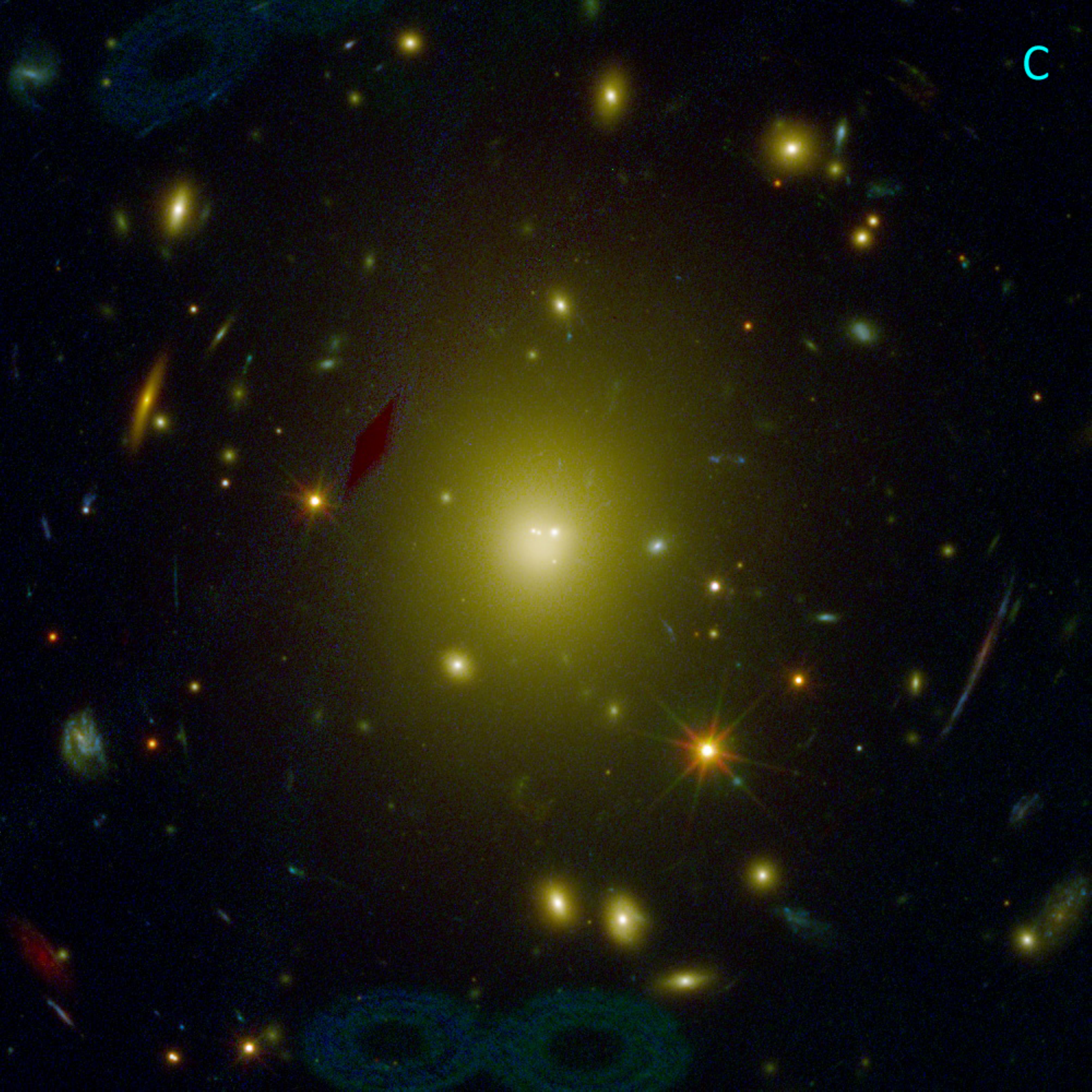}{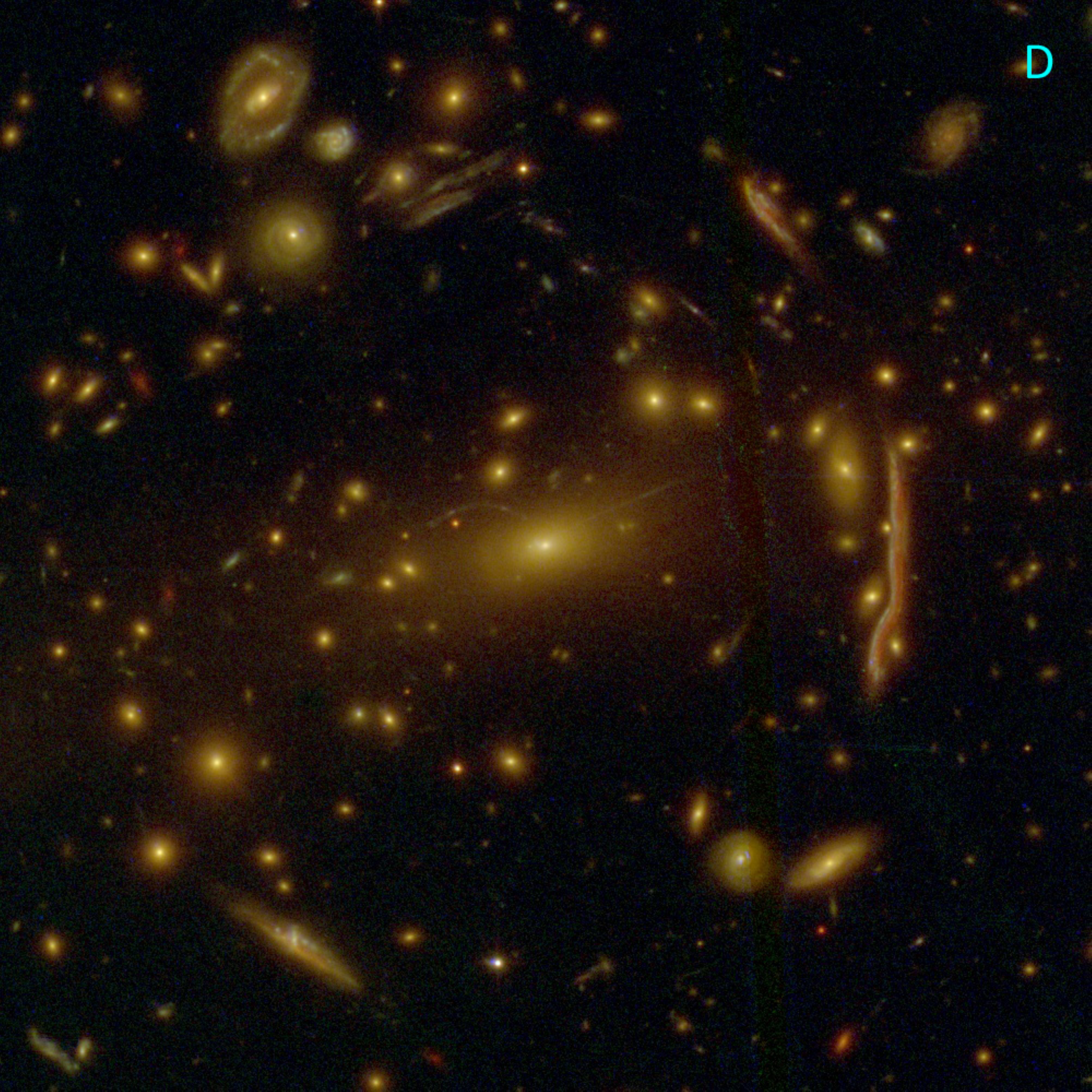}
\caption{\label{climage1}
Cutouts of the CLASH images of the central regions of (A) Abell 383, (B) MACS1149.6$+$2223, (C) Abell 2261, and (D) MACS1206.2$-$0847. 
The field of view shown in each image 
is 65 $\times$ 65 arcseconds and is centered on the brightest cluster galaxy. These color 
composites are made from a combination of 12 of the 16 filters available for
each cluster. The blue channel is a sum of the \ACS\ F435W and F475W filters, 
the green channel is a sum of the \ACS\ F606W, F625W, F775W, F814W, and F850LP filters, 
and the red channel is a sum of all five of the {\em WFC3/IR} broad band filters.  Each cutout
shown here represents $\sim28$\% of the area covered by all 16 filters and 
just $\sim10$\% of the total area imaged in the cluster core.
}\end{figure*}

A typical CLASH observing sequence is presented in Table~\ref{tab:visits}.
Each cluster is observed at two orientations to increase the area covered in the 
parallel field SN search. The two orientation angles are typically $\sim 30^{o}$
apart to minimize the overlap between the \ACS\ parallel pointings (see \S\ref{sec:cadence} 
for more details). 
Although we label the visits as ``A" and ``B" in Table~\ref{tab:visits},
either orientation may be executed first. The choice depends solely on scheduling
constraints. Indeed, there are often overlaps in time when the ``A" and ``B"
orientations are both being executed. When the entire sequence of exposures for a cluster is
completed, the region covered by all 16 filters subtends an area of 4.08 square arcminutes.
Larger areas about the cluster center are covered by \ACS\ in 7 filters. The 
survey footprint and exposure map are shown in Figure~\ref{pointings}. Color images
of the co-added \HST\  data from the first 4 CLASH cluster are shown in Figure~\ref{climage1}.

\begin{deluxetable*}{cccccc}
\tablewidth{0pt}
\tablecaption{\label{tab:visits}Typical CLASH Observing Sequence}
\tablehead{
\colhead{}&
\colhead{}&
\colhead{Primary}&
\colhead{Primary Cluster}&
\colhead{Parallel}&
\colhead{Parallel Field}\\
\colhead{Visit ID}&
\colhead{Epoch$^b$}&
\colhead{Camera}&
\colhead{Target Filters}&
\colhead{Camera}&
\colhead{Filters}
}
\startdata
Orientation A:\\
A0, A1 & 1A &  ACS & F625W, F850LP & WFC3 & F350LP, F125W, F160W \\
     &     &      & F475W, F775W &           &                \\
A2 & 1A & WFC3 & F110W, F160W & ACS & F775W, F850LP \\
A3 & 2A & ACS & F606W, F814W & WFC3 & F350LP, F125W, F160W \\
A4 & 2A & WFC3 & F110W, F160W & ACS & F775W, F850LP \\
A5 & 3A & ACS & F435W, F814W & WFC3 & F350LP, F125W, F160W \\
A6$^{a}$ & 3A & WFC3 & F225W, F390W & ACS & F775W, F850LP \\
A8 & 4A & ACS & F814W, F850LP & WFC3 & F350LP, F125W, F160W \\
A9 & 4A & WFC3 & F110W, F160W & ACS & F775W, F850LP \\
 \\
 Orientation B:\\
 B0 & 1B &  ACS & F625W, F850LP & WFC3 & F350LP, F125W, F160W \\
 B1 & 1B & WFC3 & F125W, F160W & ACS & F775W, F850LP \\
 B2 & 2B & ACS & F606W, F775W & WFC3 & F350LP, F125W, F160W \\
 B3$^{a}$ & 2B & WFC3 & F275W, F390W & ACS & F775W, F850LP \\
 B5 & 3B & ACS & F435W, F814W & WFC3 & F350LP, F125W, F160W \\
 B6 & 3B & WFC3 & F336W, F105W, F140W & ACS & F775W, F850LP \\
 B7 & 4B & ACS & F475W, F850LP & WFC3 & F350LP, F125W, F160W \\
 B8 & 4B & WFC3 & F125W, F160W & ACS & F775W, F850LP \\
\vspace{-0.1in}
\enddata
\tablenotetext{a}{Visits A6 and B3 are each executed as a single 2-orbit visit using the same guide 
star acquisition. This is essential to facilitate alignment of the NUV exposures.}
\tablenotetext{b}{Epochs 1A -- 4A are sequential in time, as are epochs 1B -- 4B. However, depending
on scheduling constraints, exposures for orientation B may precede or be interlaced with those for orientation A.}
\end{deluxetable*}

CLASH clusters will be observed over the course of three annual \HST\ observing cycles,
with 10, 10, and 5 clusters to be observed in cycles 18, 19, and 20, respectively. The
current assignments of the cluster observations to the \HST\ cycle number are shown in 
column 5 of Table~\ref{tab:MCT}.
The duration of this MCT program is driven, in part, by the cadence required by 
supernova search (see \S\ref{sec:cadence} below).

\subsection{Filter Selection and Exposure Times}
\label{sec:filters}

\begin{figure}
\plotone{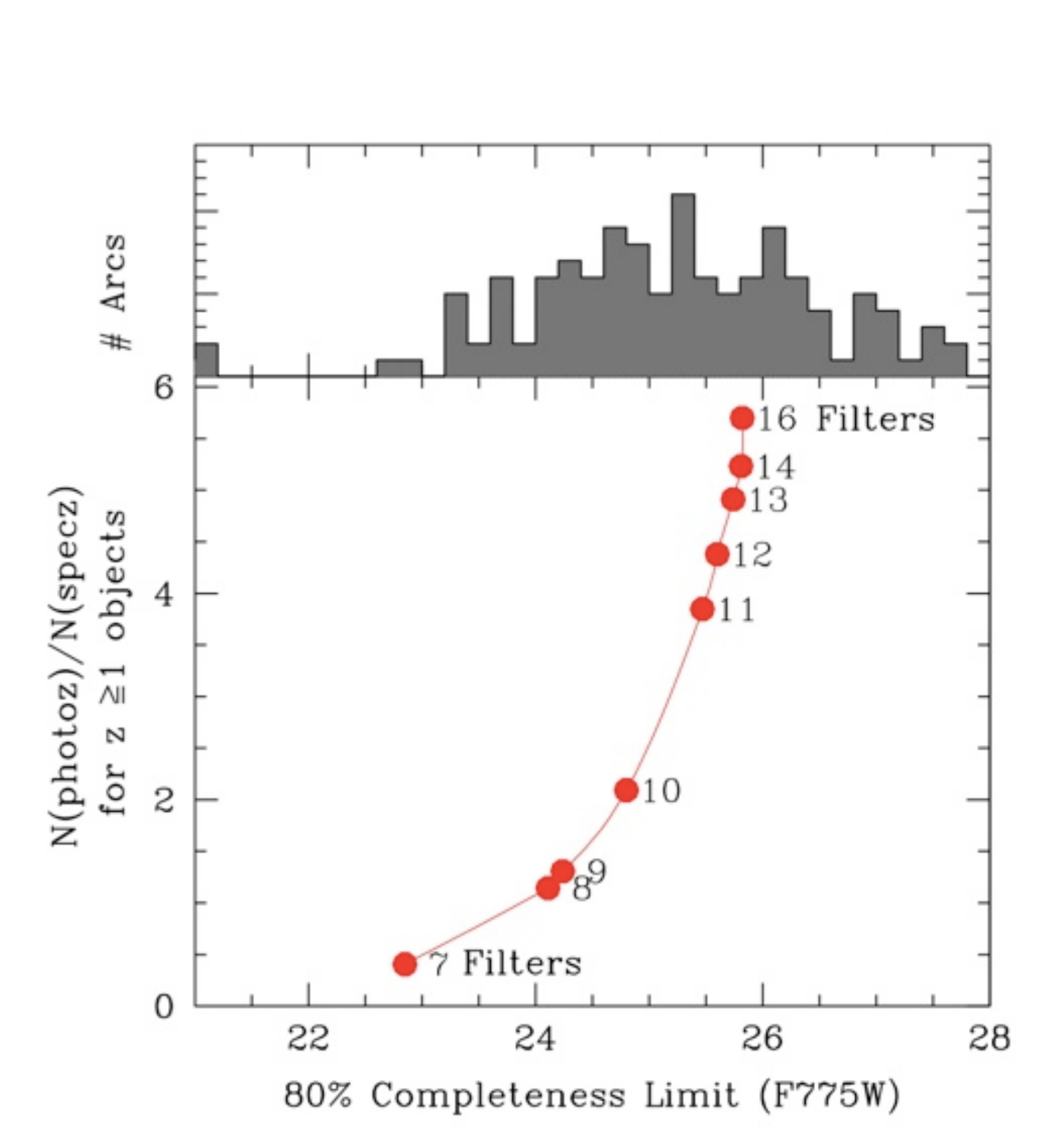}
\caption{\label{photoz}
Top: Magnitude distribution of 132 multiply lensed images
detected in A1689 and CL0024+17 \citep{Broadhurst05,Zitrin09a}.
Most are too faint for spectroscopic follow-up.
Bottom: By increasing the number of filters for a fixed observing time,
we show how photometric redshifts are improved using two metrics.
The horizontal axis shows, for example, 
that 16 filters yield very accurate ($\Delta z \sim 0.02 (1+z)$) photo-z's
for 80\% of F775W mag $< 26$ objects.
This allows us to obtain precise redshifts for $\sim 6$ times more objects
than attainable using spectroscopy, as shown on the vertical axis.
These estimates are based on simulated photometric and photo-z catalogs
derived from galaxies observed in the UDF \citep{Coe06}.
}\end{figure}

Redshift estimates for multiply-lensed images are crucial for breaking lensing degeneracies 
and tightening constraints on mass profiles \citep[e.g.,][]{Broadhurst05,Zitrin09a,SahaRead09}.
However, most of the useful lensed images are much too faint for spectroscopy. 
Typical lensed source magnitudes are $23 < I < 28$ (Figure~\ref{photoz}),
so that only the brightest arcs yield spectroscopic redshifts
even when observed with the largest ground-based facilities.
With continuous sampling of the broad wavelength range
from the NUV to NIR ($\sim$ 2,000--17,000 \AA) that is enabled with 
\WFCiii\ and \ACS\ we can now obtain very 
accurate photometric redshifts (photo-z's) for most of the lensed objects down
to an apparent F775W AB magnitude limit of 26. 

For a fixed total observing time,
splitting observations into multiple (ideally overlapping) filters
significantly improves photo-z precision \citep{Benitez09}.
We performed simulations to inform our filter selection
and estimate our eventual photo-z precision.
Galaxy magnitudes, redshifts, and SEDs (spectral energy distributions)
were drawn from the UDF \citep{Coe06}
and ``re-observed'' with our filter set,
adding noise as appropriate given our proposed depths in each filter.
Photo-z's were then re-estimated using {\tt BPZ} \citep{Benitez00,Benitez04,Coe06}.
In this simulation, we find that for 16 filters, we obtain very accurate ($\Delta z \sim 0.02 (1+z)$) 
photo-z's for 80\% of objects with F775W mag $< 26$. Most importantly, we find that
we will be able to acquire $\sim 6$ times as many reliable photometric redshifts than
spectroscopic redshifts for objects at $z > 1$, enabling a very substantial improvement in the number
of unique constraints on the DM mass distributions (see Figure~\ref{photoz}). 

The coverage provided by the 16 \WFCiii\ and \ACS\ filters 
allows the Lyman-limit feature (rest frame 912\AA)
to be photometrically traced to redshifts as low as $z \sim 1.5$
and Ly$\alpha$ to be detected out to $z \sim 10$. The inclusion of
NUV photometry, for example, resolves one of the most common photo-z degeneracies
between the Balmer break in $z \sim 0.2$ galaxies
and the Lyman break in $z \sim 3$ galaxies \citep{Rafelski09}.

The exposure times for the primary camera (cluster center position) 
are set primarily by the need to achieve 80\% photometric 
redshift completeness down to F775W = 26 AB mag. 
This requires $\sim$18 orbits per cluster as we need to achieve a $10\sigma$ limiting AB magnitude of 26 in each filter. 
We augment this by two orbits per cluster 
to extend the NIR depth to a 10$\sigma$ limiting AB mag of F160W = 26.7
and F814W = 27.0 as needed for our lensed high-$z$ galaxy search (\S\ref{sec:hiz} and see Table~\ref{tab:filters}). 
The limiting magnitudes in Table~\ref{tab:filters} are for a 0.4$''$ diameter circular aperture and a point source with a flat $F_{\nu}$ spectrum.
The minimum exposure in any given filter is one orbit and
observations in each filter are distributed over at least four exposures
to ensure robust cosmic ray rejection.

\begin{figure*}
\plottwo{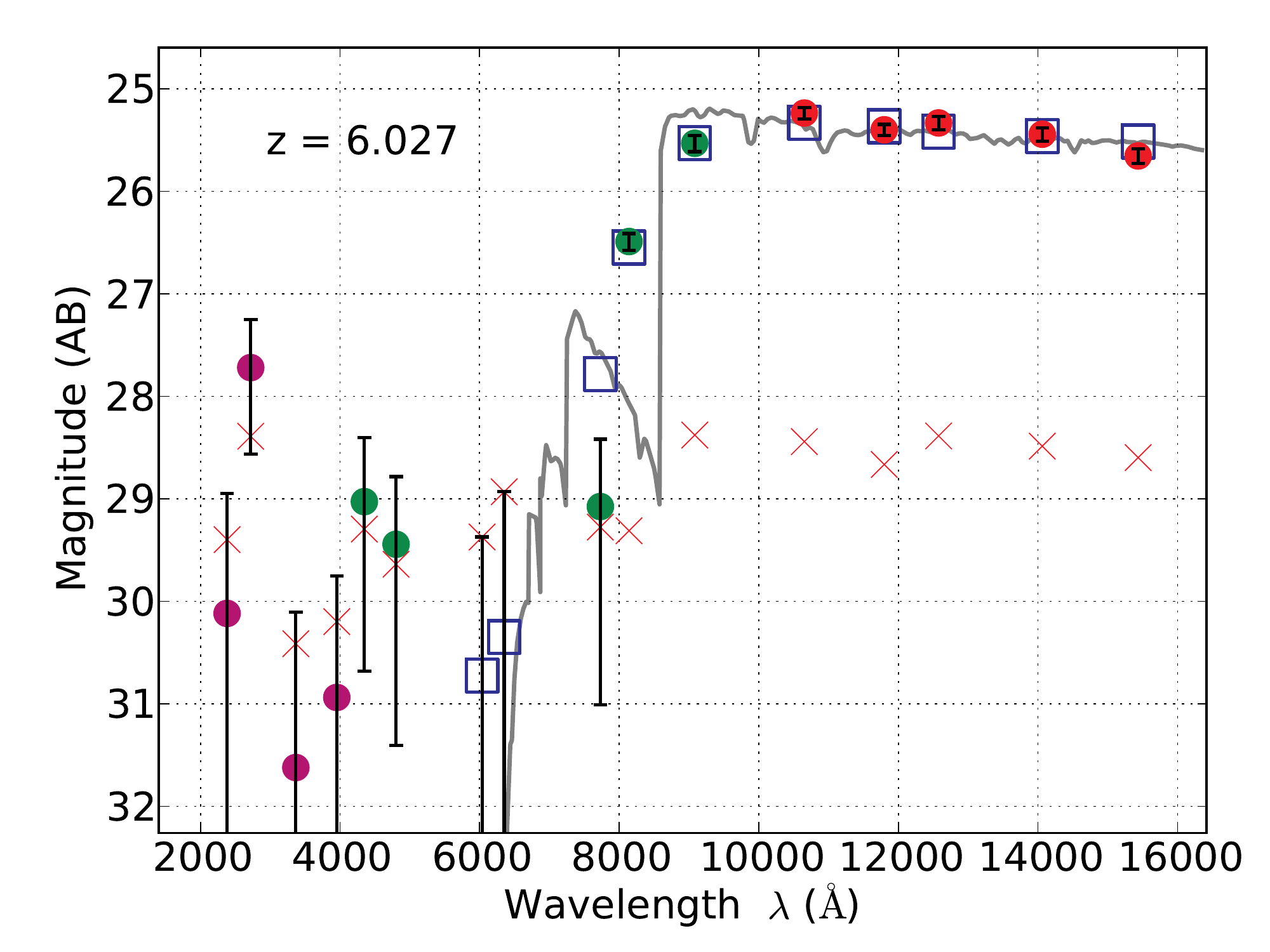}{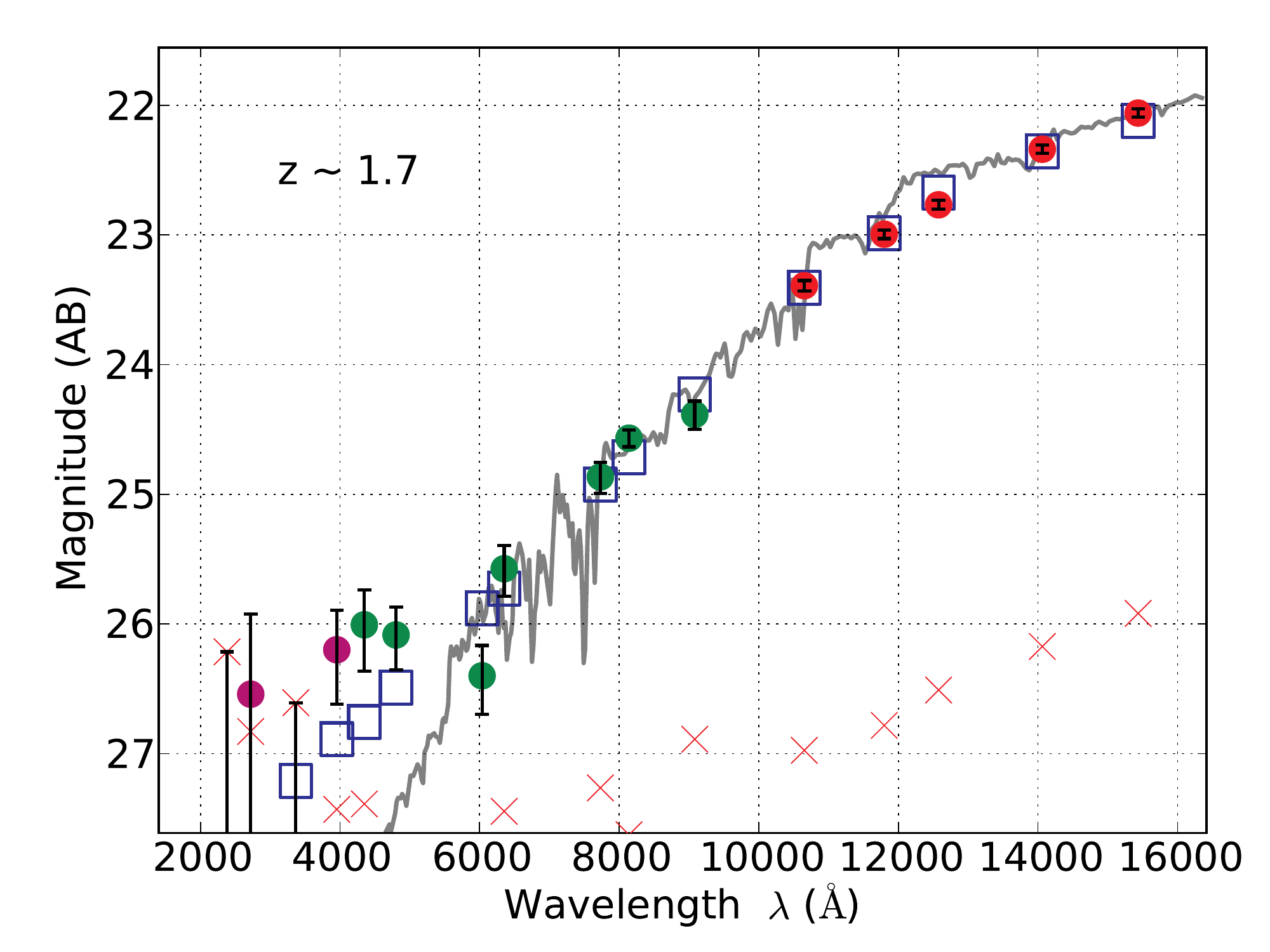}
\caption{\label{redgals}
Robustly distinguishing between high-redshift objects (left) 
and lower redshift reddened galaxies (right) with the CLASH filter set.
Photometry in the 16 CLASH filters is plotted versus wavelength for two galaxies behind CLASH clusters.
{\bf Left:} a spectroscopically confirmed $z = 6.027$ galaxy behind Abell 383 \citep{Richard11}
and best fit photometric redshift $z = 6.06$.
{\bf Right:} a reddened galaxy behind Abell 2261 with photo-$z \sim 1.7$.
Red X's show 1-$\sigma$ detection limits based on the measured flux uncertainties,
and blue boxes are SED-predicted magnitudes as integrated through each filter.
Note the high redshift galaxy SED is roughly flat in the NIR,
while the lower redshift SED continues to rise as a function of wavelength through the NIR.
This difference, highlighted clearly in the NIR photometry,
enables us to robustly distinguish high redshift galaxies from lower redshift interlopers.
}\end{figure*}

The five NIR filters provide the ability to identify $z > 7$ galaxies with high confidence.
Measured colors in these filters are essential for discriminating between
such high-redshift galaxies and lower-redshift highly-reddened galaxies
(e.g., see Figure~\ref{redgals}).
The former are generally relatively flat in $F_{\nu}$ (or AB magnitudes) in the NIR
while the latter generally increase in brightness as a function of wavelength.
Quantitatively, the following selection criteria have proven effective:
$z-J > 0.8$ (a flux decrement factor of 2.1) and $J-H < 0.5$
 \citep{Bradley08,Zheng09,Bradley11}.
To reliably measure the above colors for high-$z$ lensed galaxy detection,
our exposures are set to reach the following AB magnitude limits:
F850LP: 26.7 (5$\sigma$), F110W: 27.8 (5$\sigma$), 
F125W: 26.5 (10$\sigma$), and F160W 26.7 (10$\sigma$). 

\subsection{Dither Pattern}
\label{sec:dither}

In each orbit we use a compact $4-$point dither pattern that provides
half-pixel sampling along both detector axes.  The dither pattern
serves to both improve the spatial sampling of the point spread function (PSF), especially for
the {\em WFC3/IR} detector with its large pixel scale of 0.128 arcsec/pixel,
and to help remove hot pixels and other detector imperfections that
may be unaccounted for in the calibration reference files.  In the
first epoch of each roll angle, we use a small-scale dither pattern to
preserve, as best as possible in light of significant geometric
distortions, half-pixel sampling across the detector.  In subsequent
epochs involving {\em WFC3/IR} observations, either in prime or parallel, we
use a slightly larger dither pattern to help identify and remove
persistence artifacts from compact sources, which, if uncorrected, could
possibly be misidentified as SNe candidates.  While our small-scale dither
patterns are much smaller than is needed to cover the {\em WFC3/UVIS} and
{\em ACS/WFC} detector gaps, our cluster observations are obtained at two
orientations, leaving only two small diamond-shaped regions  ($\sim 4.4$ square arcseconds each) 
in the central cluster area without data in all 16-filters.

\subsection{Observation Cadence and Supernova Follow-Up}
\label{sec:cadence}

Parallel observations are being obtained with a primary science goal of detecting Type Ia SNe.
While the cluster core is being imaged with \ACS, we obtain parallels with \WFCiii, and vice versa.
For each of the two roll angles, 
observations are obtained over the course of four epochs spread out over $\sim 30$ to 45 days
($\sim 10$ to 14 days between observations).
Image differencing analysis enables us to detect supernovae which have gone off between observations.
As noted above, the exposure sequences are executed either sequentially or interleaved, 
depending on scheduling availability.

Upon detection of a promising high-$z$ type Ia SN in the parallel field around a cluster, 
we can reprogram the remaining parallels to provide the follow-up 
(light curve imaging and grism spectroscopy) necessary to measure its distance. 
The ability to reprogram the later of the two orientations to follow-up a SN detected early in
a cluster observing sequence sets the upper limit on the angular offset between the two orientations
-- both orientations must be accessible during the entire cluster sequence. This
constraint means that the two orientations cannot be more than $\sim 30^{o}$ apart. A lower
limit on the angular separation between the two orientations is set to $\sim 20^{o}$ to ensure
there is not excessive overlap in the \ACS\ parallel pointings (with the exception of RXJ1347 where an
orientation shift of 15$^{o}$ was needed to avoid placing a bright star in the parallel fields).
The ability to reprogram a downstream orientation shift means that a new target of opportunity (ToO) 
observation will
not be required if the SN is discovered early in the cluster observing sequence. In this regard,
the design of the CLASH SN program has part of the follow-up built into its implementation.
For the flexibility required to follow SNe found at the end of their corresponding cluster observing sequence, 
a small number of ToOs (four) and follow-up reserve orbits
are included in the CLASH orbit allocation.  
CLASH parallel fields are {\it not} observed in 16 filters. 
The \ACS\ parallels are taken in F775W and F850LP and
the \WFCiii\ parallels are taken in F350LP (UVIS), F125W (IR) and F160W (IR) (see
Table~\ref{tab:visits}).

The CLASH and CANDELS \citep{candels11,candels11a} supernova programs are tightly coordinated and, in fact, share
a common pool of reserve orbits from which both programs can draw upon for follow-up. This
common reserve has been allocated 200 orbits: 50 from the CLASH allocation and 150 from CANDELS. The
coordinated program is led by A. Riess who is a co-I in both programs.

\subsection{ACS Failure Options}
\label{sec:ACSfail}

While \ACS\ functionality was restored in SM4, it is now only a ``single-string'' instrument, 
meaning there is no redundant path if the main CCD electronics box experiences a failure. This
was a constraint imposed by the nature of the \ACS\ repair.
The primary impact of such a failure would be the loss of our parallel observations for supernova searches.
Supernovae could still be detected in our primary cluster core observations assuming we 
were to continue to distribute the exposures over multiple epochs.
However, the use of type Ia SNe for cosmology in strongly lensed regions is fraught with difficulty (see \S\ref{sec:SNe}). While a failure of \ACS\ would be a significant blow to the survey's dark energy science objectives, all other components of the CLASH science program, including our prime objective of studying the distribution of mass profile properties in clusters, would be able to be pursued with \WFCiii\ alone. 

If \ACS\ fails permanently, we would abandon the dual orientation strategy and would, most likely,
abandon the multiple epoch exposures, allowing the
observations of each cluster to be completed on a much shorter timeframe.
By observing each cluster at a single roll angle, we would  
slightly increase the area in which we obtain full 16-filter coverage.
All of the \ACS\ filters are available with {\em WFC3/UVIS},
which would allow us to continue our 16-filter observations with \WFCiii\ alone.
However there would be noticeable reductions in signal-to-noise
in the redder filters (F775W, F814W, F850LP).
In order to continue to achieve the same depths,
we would require exposure times in the redder filters to be increased by 80\%,
adding about four orbits of integration time to each cluster.

The {\em WFC3/UVIS} and {\em WFC3/IR} fields of view are large enough to contain and yield robust photometric redshifts 
for most of the strongly lensed arcs.
We would lose the extra areal coverage provided by the \ACS\ observations,
slightly reducing the area of overlap with our Subaru images.
However, we estimate the impact on our
ability to measure the mass profiles over the full radial range will be negligible.

\section{The CLASH Data Pipeline}
\label{sec:pipeline}

The bulk of our \HST\ data analyses make use of mosaics of globally aligned and co-added images. To accomplish this, we use the {\tt MosaicDrizzle} pipeline \citep{Koekdriz02}. The {\tt MosaicDrizzle} pipeline takes as its input the calibrated {\tt FLT} files which have been produced by {\tt calacs} and {\tt calwf3}. The 
ACS/WFC data, however, are first corrected for bias-striping and charge-transfer efficiency (CTE) degradation effects \citep{CTE}. 
The CTE correction approximately reverses the effects of
charge being trapped and trailed across the image during data readout.
However, a similar procedure has yet to be developed and implemented for WFC3/UVIS images.
We find that this uncorrected CTE can most significantly affect our UVIS photometry as follows.
Trails from cosmic rays can leak into photometric apertures of non-detections,
artificially boosting their observed fluxes.
Objects otherwise expected to be UVIS dropouts, or non-detections
(based on ACS+IR photometry),
may commonly have significant detections in one or more UVIS filters.
We find this can be greatly mitigated by adopting a more aggressive rejection of cosmic rays and their trails.
The result is that most expected UVIS dropouts do actually drop out, 
though with some significant (as high as $\sim 5$-$\sigma$) detections remaining.

{\tt MosaicDrizzle} then carries out a sequence of steps aimed at aligning the 
exposures in all the different camera/filter combinations within each visit and also 
across visits, using a combination of catalog matching and cross-correlation. 
The catalog-matching is done using a deep ground-based catalog (typically from 
Subaru's Suprime Cam) as an initial reference point, followed by subsequent 
matching using catalogs from the CLASH \HST\ images themselves, solving for shifts 
and rotations, which yields an accuracy of ~0.1-0.2 pixel. Shifts are further refined using 
cross-correlation which can achieve accuracies to the level of 0.02-0.05 
pixel, thus ~1-2 mas for \ACS\ images, limited essentially at that point by the stability of the 
PSF from one exposure to the next, as well as the accuracy of the 
distortion correction which is also removed from the images at this step.  

The iterative shift refinement is then used to enable cosmic ray rejection to be carried out 
for all the exposures in a given filter across multiple epochs, leading to a 
final set of cosmic ray and bad pixel masks. These are then weighted according to the sky 
level in each input exposure (as modulated by the flatfield / detected quantum efficiency 
variation across the detector for each filter), along with the readnoise and accumulated 
dark current, to form an inverse variance image for each exposure.  The 
drizzle combination is then carried out using the inverse variance images as weights, 
using a square kernel as well as a"pixfrac" parameter typically set to 0.8 
which is chosen to be appropriate to the output image pixel scale and the number of dithers per filter.  
An advantage of this approach is that it involves only a single 
geometric operation, transforming the pixel values from the {\tt FLT} files 
directly onto the output frame and avoiding any additional convolution. 
     
It is worth noting that the inverse variance describes the expected noise in the 
absence of any correlated noise; in practice the images do have some amount of 
correlated noise (due to the PSF being sampled by the detector pixels to begin with, 
as well as the single-step transformation onto the output image plane), but 
typically this is no more than $\sim10 - 15$\% of the noise that would be expected 
in the absence of any correlation, and cannot be reduced further without making 
the output image pixels larger.  For each cluster, two sets of image mosaics are 
generated --- one drizzled to a 30 mas/pixel grid and one drizzled to a 65 mas/pixel 
grid (for both \ACS\ and \WFCiii). We register the images with north up, and the 
images are aligned so that the objects are sampled by the same pixels for all the filters.

We also run a redundant image processing pipeline that is built upon the framework developed for the \ACS\ GTO 
pipeline {\tt APSIS} \citep{APSIS}. For the CLASH application, {\tt APSIS} was 
modified to handle the {\em WFC3} detectors. The use of this second pipeline provides 
a useful verification of the photometry and astrometry as well as a
redundant processing facility. The main difference between {\tt MosaicDrizzle} and {\tt APSIS}
is that {\tt APSIS} does not require an external astrometric reference catalog and 
can derive the relative image shifts from the \HST\ data alone. To date,
both pipelines appear to generate data of roughly similar quality with {\tt MosaicDrizzle} 
producing slightly better alignment precision.
 
The SN detection pipeline has been developed jointly by the CLASH and CANDELS teams. 
It will be described in more detail in a separate paper. Briefly, the SN detection 
pipeline uses an image-differencing scheme to identify SN candidates. 
Extensive work has already been devoted to minimizing false positive detections. 
IR persistence is one potential source for false positives
in both the SN and high-z galaxy detection projects. 
To address this, a dark exposure is obtained in the Earth-occultation 
just preceding the first visit for all CLASH and CANDELS observations. 
This enables a mask to be generated that can flag any suspect pixels in the
initial exposure of a CLASH visit. 
Subsequent orbits in the visit are flagged based on our actual data.

\subsection{Object Detection and Characterization}
\label{sec:objdet}

SExtractor \citep{SExtractor} is used to detect objects and measure their photometry. 
We run SExtractor (version 2.5.0) in dual image mode, 
using a detection image created from a weighted sum of the {\em ACS/WFC} and {\em WFC3/IR} images. 
The weights come from the inverse 
variance images produced by {\tt Mosaicdrizzle}. We do not use the {\em WFC3/UVIS} images in the construction of the 
detection image but, of course, run SExtractor on the {\em UVIS} data to compute source photometry. 
We also create a detection image solely from the {\em WFC3/IR} images to optimize the search for high redshift ($z > 6$) objects.

\begin{figure*}
\plotone{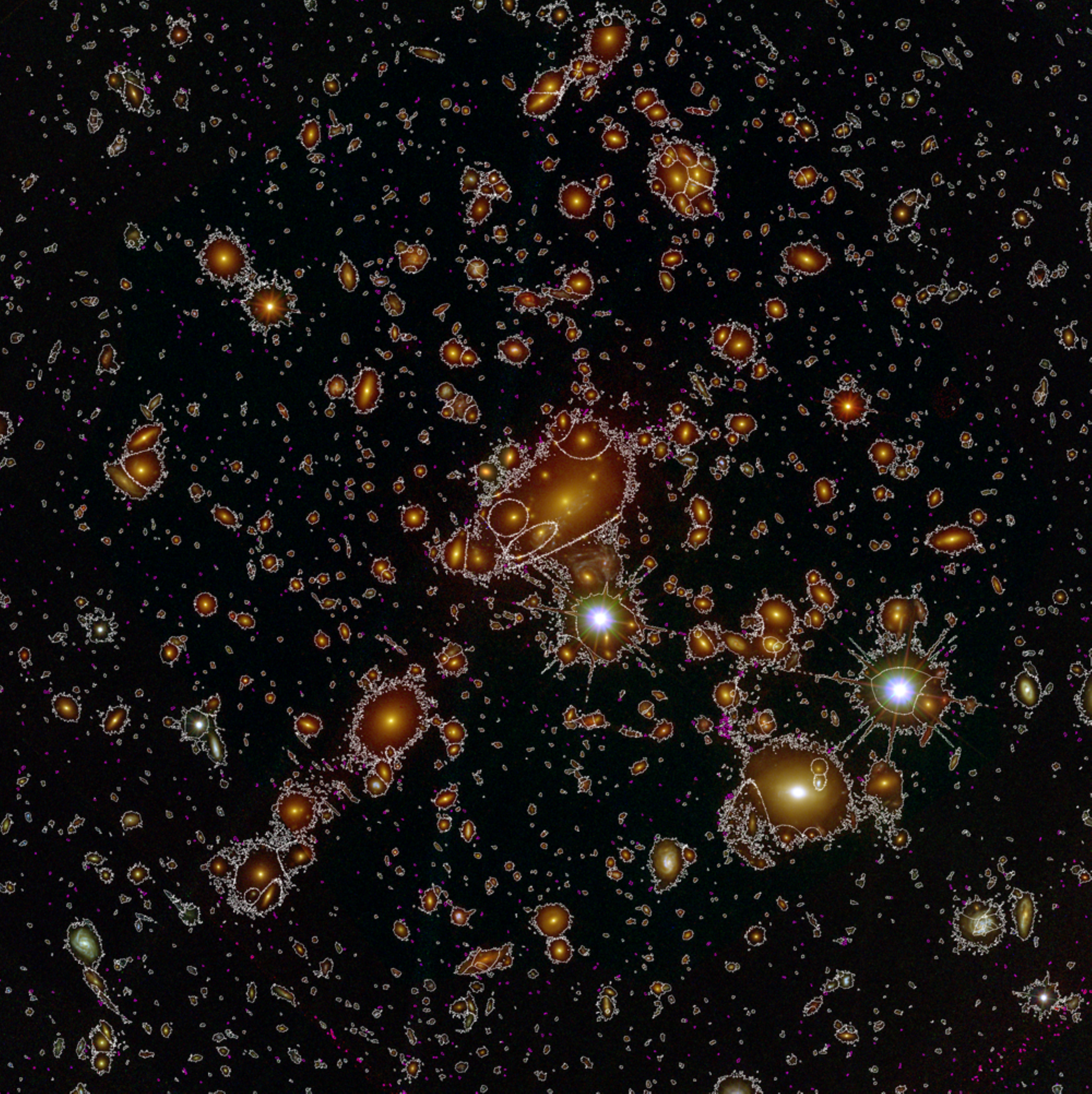}
\caption{\label{m1149segm}%
SExtractor segmentation map for the central region of the cluster MACS1149.6$+$2223. Detection of additional objects behind the cluster
 is accomplished by subtracting the best fits to the surface brightness distributions of the brightest of the cluster galaxies. 
}\end{figure*}

For object detection in the {\em ACS}$+${\em IR} images, 
we require a minimum of 9 contiguous pixels at the level of the observed background RMS or higher.
The detection phase background sky level is computed in $5 \times\ 5$ grids of cells, with each cell being $128 \times\ 128$ pixels. 
For photometry, the local sky is estimated from a 24-pixel wide rectangular annulus around each detected object. 
The deblender minimum contrast ratio and number of threshold levels are 0.0015 and 32, respectively.
These parameters were chosen after a systematic investigation of possible values 
and yields reasonable performance in minimizing spurious detections and suppressing over-deblending of bright objects, 
while achieving reasonable completeness in detecting faint sources behind bright cluster galaxies.
The segmentation map for the cluster MACS1149.6$+$2223 using the above parameters is shown in Figure~\ref{m1149segm}.

Our image processing effectively rejects cosmic rays from the central regions of our {\em UVIS} and {\em ACS} images
where we have four or more overlapping exposures.
However such rejection is not possible in the corners and edges of our images,
where we have fewer exposures due to our observing strategy (two roll angles and dithering).
As each cosmic ray only affects a single observation,
it will often only be detected in a single filter.
We prune these detections from our catalog by rejecting any object
with only a single 5-$\sigma$ detection in one {\em UVIS}/{\em ACS} filter, as measured by SExtractor.
We also reject any object without any 5-$\sigma$ detections.

We perform a separate {\em IR}-based detection 
with more aggressive deblending (64 levels of 0.0001 minimum contrast)
and background subtraction ($3 \times\ 3$ grids of $64 \times\ 64$ pixels).
This detection is slightly more sensitive to redder objects, including those at high-redshift.
It also performs slightly better at deblending these fainter objects, 
including those at high-$z$ as well as arcs (strongly-lensed galaxies),
from brighter nearby cluster galaxies.
As all {\em WFC3/IR} observations are obtained with multiple readouts and cosmic ray rejection,
and to preserve our sensitivity to faint high-redshift objects,
we do not prune this catalog based on 5-$\sigma$ detections.

Automated detection of strongly lensed galaxies is often challenging due to crowding in the cluster core,
especially for low-surface brightness galaxies stretched into long, thin arcs.
Progress has been made \citep[e.g.,][]{SeidelBartelmann07}, and we are testing this method for use with CLASH data.
Currently, manual intervention is still required for those arcs which elude detection.
We compare the initial source list with arc candidates identified both visually and based on our strong lens modeling.
For arcs that are either missing from the detection list or that are only partially detected, we construct manual photometric apertures.
We force SExtractor to adopt these apertures using the software package SExSeg \citep{Coe06}.
Arc photometry is then derived from images in which light from the BCG has been modeled and subtracted.

Isophotal apertures are used as they have been shown to yield robust colors \citep{Benitez04}.
In the source list presented in \S\ref{sec:catalog},
no aperture corrections have been applied to the magnitudes.
The PSF FWHM's span $\sim 0.07\arcsec$--$0.15\arcsec$ \citep{ACS,Sirianni03,WFC3handbook}.
The photometric corrections for PSF variation are small for extended objects 
for all but the faintest sources, in part because we use relatively large isophotal apertures. 

Extinction corrections are derived from the \cite{Schlegel98} IR dust emission maps,
and the resulting $A_\lambda$ coefficients, in AB mag per unit E(B-V), are given in Table~\ref{tab:filters}.
Flux uncertainties are derived by SExtractor using the inverse variance images. We have compared the flux errors so derived with the
predicted uncertainties from the {\em ACS} and {\em WFC3} exposure time calculator and find excellent agreement.

\subsection{Photometric Redshifts}
\label{sec:photoz}

We derive photometric redshift estimates using two independent packages: 
{\tt BPZ} \citep{Benitez00, Benitez04, Coe06} and {\tt LePHARE} \citep[hereafter {\tt LPZ}]{Ilbert06}.  Both software packages 
use $\chi^2$ minimization and template fitting but differ in their specific templates and their assumed priors. 
{\tt LPZ} uses the SED library optimized for the
COSMOS survey \citep{Ilbert09} without template interpolation. 
{\tt BPZ} currently uses PEGASE SED templates \citep{Fioc97}
which have been heavily recalibrated based on the FIREWORKS spectroscopic and photometric catalog \citet{Wuyts08}.
{\tt BPZ} allows for interpolation between adjacent templates 
and uses an empirically-derived prior on redshift and type based on observed magnitude.
A full analysis of the accuracy and precision of our photometric redshift estimates will be discussed in an upcoming paper
\citep{Jouvel11}. The photoz accuracy analyses will be based on over $1,000$ redshifts, obtained from the
facilities described in \S\ref{sec:supporting}.

\subsection{Source List for MACS1149.6$+$2223}
\label{sec:catalog}

\begin{figure}
\plotone{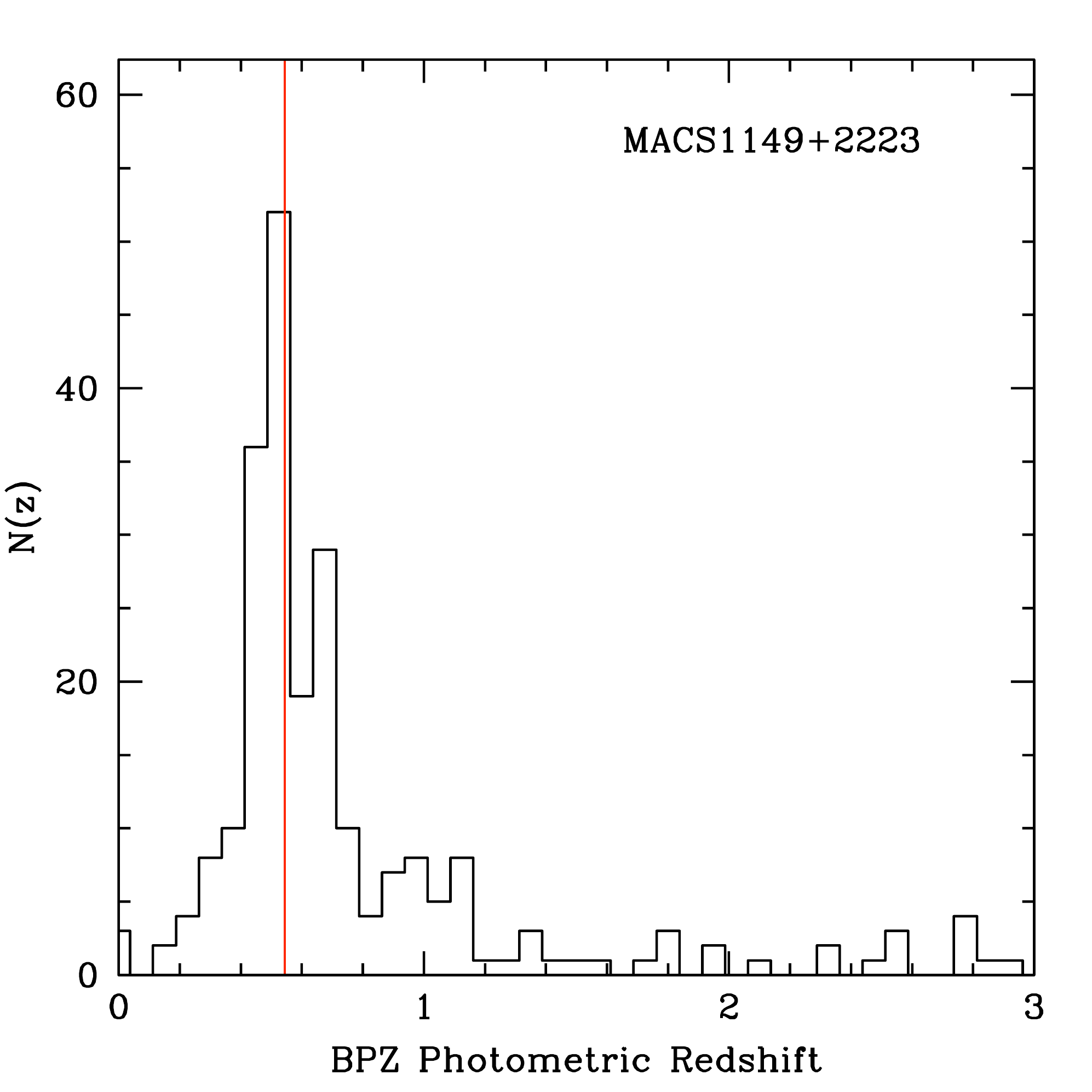}
\caption{\label{m1149photz}%
Histogram of the BPZ photometric redshifts for galaxies within 1 arcminute of the center of the cluster MACS1149.6$+$2223.
The 232 galaxies included here have
BPZ Odds $\ge 0.90$, F850LP magnitude $\le$ 26.5 and $z_{phot} \le 3$. The histogram
shows the cluster peak very clearly. The mean spectroscopic redshift of the cluster ($z = 0.544$) is
denoted by the vertical red line. The average photometric redshift in the range $0.4 \le z_{phot} \le 0.65$ is $\left<z_{ph}\right> = 0.515$.  
The accuracy of our photometric
redshifts is expected to improve further as our spectroscopic sample increases. See \cite{Jouvel11} for details.
}\end{figure}

We present here an initial version of a representative CLASH source list for the $z = 0.544$ cluster MACS1149.6$+$2223. 
The full list is available on-line (at MAST treasury program archive website) and has over 100 parameters per object. 
In Table~\ref{tab:m1149cat} we present just a small subset of the full list: 
100 galaxies within 1 arcminute of the cluster center. 
This source list is based on the ACS$+$WFC3/IR detection image (see \S\ref{sec:objdet}). 
Column 1 gives the SExtractor object ID, 
columns 2 and 3 give the equatorial coordinates (J2000), 
column 4 gives the isophotal area of the detection (in 0.065$''$ pixels), 
column 5 gives the ellipticity of the source as measured by SExtractor, 
column 6 gives the number of passbands in which the object is detected at a significance of $5\sigma$ or greater, 
and column 7 gives the total number of bands in which the object is detected. 
In the case of MACS1149.6$+$2223, there are 17 HST passbands available 
because the archival imaging included {\em ACS} F555W.
Note that even bright sources will sometimes not be detected in all bands because the footprints of
each detector on the sky are different (see Figure~\ref{pointings}). 
Columns 8 -- 10 give the BPZ photometric redshift estimate, 
the BPZ Odds parameter \citep{Benitez04},
and the $\chi^2$ measurement presented in \cite{Coe06}. The 95\% confidence limits on the range of the photometric 
redshift estimate is given in the row just below the best fit redshift value. This range is an indicator of the uncertainty 
in the redshift value.
In general, BPZ photo-z estimates with Odds $>$ 0.90 are the most reliable estimates. 
Figure~\ref{m1149photz} shows the histogram of the BPZ photometric redshifts for 232 galaxies in the cluster MACS1149.6$+$2223 with
BPZ Odds $\ge 0.90$, F850LP magnitude $\le$ 26.5 and $z_{phot} \le 3$. The histogram
shows the cluster peak very clearly. 

The remaining columns in Table~\ref{tab:m1149cat} give the the isophotal magnitudes (corrected for Galactic extinction) 
for five of the CLASH passbands: F275W, F390W, F850LP, F125W, and F160W. 
The photometric error in the isophotal magnitude is given, in parentheses, in the row directly below the magnitude value. 
Photometry has been corrected for Galactic extinction using an E(B-V) = 0.02297 and
all magnitudes given are on the AB photometric system (see Table~\ref{tab:filters} for extinction coefficients used). 
A magnitude value of 99 indicates a non-detection (negative flux) in that band.
In this case, the magnitude error value gives the $1\sigma$ detection limit in that band. 
A magnitude value of $-99$ indicates the source lies outside of the detector field of view or lies within a gap between detectors. 
Source lists for each CLASH cluster are released, via MAST, $\sim 6$ months after the final observation for that cluster is acquired.

\begin{deluxetable*}{rrrcrrrcrrrrrrr}
\tablewidth{0pt}
\tablecaption{\label{tab:m1149cat}Sample of CLASH Source List for MACS1149.6+2223}
\tablehead{
\colhead{}&
\colhead{}&
\colhead{}&
\colhead{Area}&
\colhead{}&
\colhead{}&
\colhead{}&
\colhead{zbpz}&
\colhead{}&
\colhead{}&
\colhead{F275W}&
\colhead{F390W}&
\colhead{F850LP}&
\colhead{F125W}&
\colhead{F160W}\\
\colhead{ID}&
\colhead{$\alpha_{\rm J2000}$}&
\colhead{$\delta_{\rm J2000}$}&
\colhead{(Pixels)}&
\colhead{Ell}&
\colhead{Nsig5}&
\colhead{Nfobs}&
\colhead{[min,max]}&
\colhead{odds}&
\colhead{$\chi^2$}&
\colhead{(err)}&
\colhead{(err)}&
\colhead{(err)}&
\colhead{(err)}&
\colhead{(err)}
}
\startdata
 4756&177.4090& 22.4002&    9&0.36& 4&17&   2.990   &0.63& 1.32& 29.64& 29.05& 27.18& 28.65& 28.10\\
     &        &        &     &    &  &  &[2.73,3.09]&    &     &( 1.50) &( 0.65) &( 0.22) &( 0.52) &( 0.30) \\
 4761&177.3900& 22.3999&   65&0.47&12&17&   1.701   &0.94& 2.44& 26.85& 26.73& 26.87& 25.42& 25.65\\
     &        &        &     &    &  &  &[1.62,1.75]&    &     &( 0.49) &( 0.24) &( 0.39) &( 0.08) &( 0.08) \\
 4762&177.3900& 22.4002&   10&0.45& 4&17&   1.717   &0.59& 0.63& 29.20& 28.16& 27.16& 26.82& 27.11\\
     &        &        &     &    &  &  &[1.49,1.77]&    &     &( 1.53) &( 0.48) &( 0.33) &( 0.17) &( 0.19) \\
 4763&177.3900& 22.4000&    8&0.42& 1&17&   0.765   &0.12& 0.63& 99.00& 28.19& 27.28& 27.26& 27.36\\
     &        &        &     &    &  &  &[0.76,2.22]&    &     &(28.12) &( 0.44) &( 0.32) &( 0.21) &( 0.20) \\
 4764&177.3910& 22.4002&   29&0.27&12&17&   1.222   &0.63& 1.98& 99.00& 27.69& 26.44& 26.24& 26.26\\
     &        &        &     &    &  &  &[1.12,1.32]&    &     &(28.01) &( 0.34) &( 0.18) &( 0.10) &( 0.09) \\
 4765&177.4080& 22.4002&   15&0.16& 8&17&   0.745   &0.31& 1.47& 99.00& 99.00& 26.59& 26.94& 26.85\\
     &        &        &     &    &  &  &[0.67,4.03]&    &     &(28.04) &(28.89) &( 0.19) &( 0.18) &( 0.14) \\
 4773&177.4140& 22.4002&    4&0.31& 1&17&   1.150   &0.14& 0.87& 28.89& 27.87& 27.97& 28.95& 28.03\\
     &        &        &     &    &  &  &[1.25,2.41]&    &     &( 0.95) &( 0.25) &( 0.39) &( 0.60) &( 0.27) \\
 4778&177.3880& 22.3999&  152&0.13&12&17&   0.453   &0.98& 4.00& 25.34& 26.62& 23.78& 23.48& 23.30\\
     &        &        &     &    &  &  &[0.43,0.47]&    &     &( 0.20) &( 0.30) &( 0.04) &( 0.02) &( 0.01) \\
 4780&177.4130& 22.4000&  260&0.03&13&17&   0.685   &0.91& 5.22& 26.26& 26.32& 23.35& 23.23& 23.00\\
     &        &        &     &    &  &  &[0.66,0.72]&    &     &( 0.53) &( 0.29) &( 0.03) &( 0.02) &( 0.01) \\
 4782&177.3890& 22.4001&    8&0.06& 2&17&   2.344   &0.32& 0.28& 99.00& 28.35& 28.67& 27.85& 27.07\\
     &        &        &     &    &  &  &[2.20,2.76]&    &     &(28.58) &( 0.36) &( 0.67) &( 0.25) &( 0.11) \\
 4784&177.4100& 22.4000&   10&0.51& 4&17&   0.887   &0.37& 1.86& 26.50& 27.07& 26.65& 27.06& 27.52\\
     &        &        &     &    &  &  &[0.51,1.00]&    &     &( 0.28) &( 0.22) &( 0.22) &( 0.23) &( 0.29) \\
 4786&177.3870& 22.4001&    5&0.03& 2&17&   3.898   &0.64& 1.30& 99.00& 30.42& 28.65& 29.30& 29.74\\
     &        &        &     &    &  &  &[3.65,4.08]&    &     &(28.87) &( 1.21) &( 0.53) &( 0.60) &( 0.72) \\
 4787&177.4100& 22.3998&  164&0.53&12&17&   0.507   &0.80& 3.54& 25.89& 27.03& 24.17& 24.19& 24.23\\
     &        &        &     &    &  &  &[0.46,0.54]&    &     &( 0.34) &( 0.43) &( 0.06) &( 0.04) &( 0.04) \\
 4788&177.4140& 22.4001&    7&0.20& 2&17&   2.472   &0.40& 0.83& 99.00& 27.85& 27.84& 28.24& 28.82\\
     &        &        &     &    &  &  &[1.93,2.68]&    &     &(28.45) &( 0.27) &( 0.38) &( 0.38) &( 0.51) \\
 4792&177.3820& 22.4000&    5&0.45& 6&17&   0.587   &0.31& 0.67& 99.00& 28.13& 27.33& 26.98& 26.83\\
     &        &        &     &    &  &  &[0.45,2.98]&    &     &(27.95) &( 0.49) &( 0.40) &( 0.20) &( 0.15) \\
 4793&177.3910& 22.3998&  387&0.29&14&17&   3.248   &1.00& 1.93& 25.38& 25.03& 23.05& 23.01& 22.69\\
     &        &        &     &    &  &  &[3.23,3.27]&    &     &( 0.28) &( 0.11) &( 0.03) &( 0.02) &( 0.01) \\
 4794&177.3910& 22.3999&   26&0.36&11&17&   3.274   &0.92& 1.34& 99.00& 28.92& 26.96& 26.71& 26.44\\
     &        &        &     &    &  &  &[3.25,3.42]&    &     &(28.36) &( 0.64) &( 0.20) &( 0.12) &( 0.08) \\
 4799&177.3830& 22.3998&  115&0.21&16&17&   1.140   &1.00& 2.94& 25.07& 24.97& 24.06& 23.98& 23.74\\
     &        &        &     &    &  &  &[1.13,1.17]&    &     &( 0.15) &( 0.07) &( 0.05) &( 0.03) &( 0.02) \\
 4800&177.3840& 22.3998&  132&0.20&13&17&   0.714   &0.88& 2.33& 25.64& 26.46& 24.59& 24.71& 24.78\\
     &        &        &     &    &  &  &[0.67,0.74]&    &     &( 0.22) &( 0.23) &( 0.07) &( 0.05) &( 0.05) \\
 4801&177.3840& 22.3997&  472&0.10&15&17&   0.528   &0.82& 4.85& 24.30& 24.56& 22.75& 22.72& 22.63\\
     &        &        &     &    &  &  &[0.46,0.54]&    &     &( 0.14) &( 0.09) &( 0.03) &( 0.02) &( 0.01) \\
 4805&177.3950& 22.4000&   29&0.21&12&17&   0.906   &0.95& 3.55& 27.30& 27.30& 25.84& 26.34& 26.50\\
     &        &        &     &    &  &  &[0.88,0.94]&    &     &( 0.49) &( 0.26) &( 0.12) &( 0.12) &( 0.12) \\
 4806&177.4070& 22.4000&   58&0.34&14&17&   0.776   &0.81& 2.52& 26.09& 26.48& 25.48& 24.99& 24.91\\
     &        &        &     &    &  &  &[0.74,0.84]&    &     &( 0.25) &( 0.18) &( 0.11) &( 0.05) &( 0.04) \\
 4812&177.4150& 22.3999&   69&0.19&10&17&   0.347   &0.74& 1.53& 99.00& 27.99& 25.82& 24.93& 24.81\\
     &        &        &     &    &  &  &[0.30,0.39]&    &     &(27.49) &( 0.61) &( 0.16) &( 0.05) &( 0.04) \\
 4813&177.4050& 22.3998&  321&0.04&14&17&   0.505   &1.00& 2.09& 27.06& 25.72& 22.59& 22.17& 21.85\\
     &        &        &     &    &  &  &[0.49,0.52]&    &     &( 0.94) &( 0.19) &( 0.02) &( 0.01) &( 0.01) \\
 4820&177.3840& 22.4007&   76&0.02&11&17&   0.772   &0.73& 3.08& 99.00& 26.97& 25.31& 25.03& 25.00\\
     &        &        &     &    &  &  &[0.71,0.84]&    &     &(27.36) &( 0.32) &( 0.12) &( 0.06) &( 0.05) \\
 4821&177.3920& 22.3995&   76&0.11&10&17&   0.742   &0.49& 2.22& 27.21& 99.00& 25.56& 24.99& 24.78\\
     &        &        &     &    &  &  &[0.54,0.81]&    &     &( 0.68) &(28.17) &( 0.14) &( 0.06) &( 0.04) \\
 4822&177.3920& 22.3998&   32&0.34& 9&17&   0.560   &0.11& 0.89& 27.40& 27.48& 26.13& 26.52& 26.01\\
     &        &        &     &    &  &  &[0.34,3.09]&    &     &( 0.69) &( 0.41) &( 0.21) &( 0.19) &( 0.11) \\
 4827&177.3860& 22.3998&    5&0.09& 1&17&   0.650   &0.40& 0.71& 27.72& 28.66& 28.34& 28.44& 28.33\\
     &        &        &     &    &  &  &[0.53,0.82]&    &     &( 0.37) &( 0.42) &( 0.49) &( 0.36) &( 0.29) \\
 4832&177.4120& 22.3996&   66&0.54&14&17&   2.739   &0.96& 1.41& 99.00& 26.10& 25.33& 25.31& 25.03\\
     &        &        &     &    &  &  &[2.64,2.75]&    &     &(27.56) &( 0.13) &( 0.10) &( 0.07) &( 0.05) \\
 4841&177.4040& 22.3996&   74&0.37&14&17&   0.371   &0.70& 5.95& 26.80& 27.16& 24.63& 25.42& 25.07\\
     &        &        &     &    &  &  &[0.34,0.47]&    &     &( 0.57) &( 0.41) &( 0.07) &( 0.10) &( 0.06) \\
 4843&177.3890& 22.3996&   21&0.55& 6&17&   0.768   &0.37& 0.92& 99.00& 27.84& 26.57& 26.19& 26.03\\
     &        &        &     &    &  &  &[0.64,1.03]&    &     &(27.85) &( 0.42) &( 0.23) &( 0.11) &( 0.08) \\
 4853&177.4140& 22.3994&   56&0.28&13&17&   2.201   &0.82& 2.38& 27.57& 25.83& 26.66& 25.94& 25.27\\
     &        &        &     &    &  &  &[2.19,2.33]&    &     &( 0.75) &( 0.10) &( 0.29) &( 0.11) &( 0.05) \\
 4856&177.4050& 22.4009&   21&0.33& 8&17&   0.767   &0.41& 2.30& 26.01& 27.68& 26.15& 26.34& 26.01\\
     &        &        &     &    &  &  &[0.63,0.92]&    &     &( 0.20) &( 0.40) &( 0.17) &( 0.14) &( 0.09) \\
 4868&177.4060& 22.3994&   21&0.17& 7&17&   0.558   &0.27& 1.64& 27.33& 27.98& 26.42& 26.68& 26.26\\
     &        &        &     &    &  &  &[0.51,0.97]&    &     &( 0.53) &( 0.47) &( 0.19) &( 0.17) &( 0.10) \\
 4869&177.4060& 22.3991&  215&0.03&12&17&   0.455   &0.99& 2.13& 25.70& 26.90& 23.48& 23.16& 22.90\\
     &        &        &     &    &  &  &[0.43,0.47]&    &     &( 0.33) &( 0.44) &( 0.04) &( 0.02) &( 0.01) \\
 \enddata
\tablecomments{The full source list with over 100 parameters per source 
is available on-line from IOP and from the MAST CLASH Treasury Archive 
webpage at http://archive.stsci.edu/prepds/clash/ \\
\\
All magnitudes are on AB system.}
\end{deluxetable*}

\section{Supporting Observations}
\label{sec:supporting}

As discussed above, having both weak and strong lensing information as well as information about the cluster
baryonic mass distribution are critical for deriving robust mass profiles and concentrations. A partial list of the 
supporting multi-wavelength imaging being used in the CLASH program is
summarized in Table~\ref{tab:OtherObs}.
All CLASH clusters have X-ray imaging as well as wide-field multi-band ground-based optical imaging. The X-ray imaging
is from Chandra/ACIS \citep{CXOACIS} and some of the clusters have XMM/EPIC \citep{XMMEPIC,XMMEPIC2} imaging as well. 

The ground-based optical imaging is, for 24 out of the 25 clusters, from the Subaru Observatory
(and, in particular, performed with the SuprimeCam instrument \citep{SuprimeCam}).  
The Subaru SuprimeCam imaging is currently available in three or more filters for 22 of the 24 clusters reachable from 
Mauna Kea and we are pursuing additional optical filter coverage for the two remaining clusters. Wide-field optical
imaging for Abell 2261 from Subaru, for instance, 
was supplemented with SDSS {\em iz} band imaging from the Kitt Peak 4-m Mayall telescope
using the Mosaic1 camera.
RXJ2248.7-4431, is too far south to reach with Subaru but has multi-band imaging from the wide-field imager (WFI) 
on the European Southern Observatory's 2.2m telescope.
The multiband optical imaging enables robust selection of background galaxies for weak lensing analysis 
\citep[e.g.,][]{Medezinski07}. 

We have also acquired mm-wave imaging to map the Sunyaev-Zel'dovich Effect (SZE) in order to 
measure the integrated line-of-sight gas pressure towards each cluster.
The SZE data come primarily from two facilities -- the Bolocam instrument at the Caltech Submillimeter Observatory
\citep{Golwala09,Sayers11} 
and Multiplexed SQUID/TES Array at Ninety 
Gigahertz (MUSTANG; \citealt{Dicker08,Dicker09}) at the Robert C. Byrd Green Bank Telescope. All 25 CLASH clusters have a detected SZE signal with 
the lowest SNR$\sim 7$ and most with SNR $>10$. 
The combination of the X-ray and mm-wave observations allow
the mass scaling relations to be accurately calibrated for use in cosmological surveys \citep[e.g.,][]{Okabe10b}.
The X-ray and SZE data will be used to measure gas density profiles
for subtraction from lensing-derived total mass profiles
to yield DM-only mass profiles \citep[e.g.,][]{Lemze08}. These data also allow
assumptions of hydrostatic equilibrium to be tested and constraints on thermal motion to be made from 
measurements of excess pressure \citep{Umetsu09,Zhang10,Molnar10}. 

Spitzer Space Telescope \citep{SST2004} data is available for most of the CLASH clusters from 
the IRAC Lensing Survey (E. Egami, P.I.). 
The few remaining clusters that lacked sufficiently deep Spitzer data will be observed in our cycle 8 ICLASH program
(R. Bouwens, P.I.). The Spitzer data
are critical for characterizing the high-$z$ galaxy population (e.g., see \cite{Richard11}). Deep IRAC \citep{IRAC2004} imaging
samples the rest-frame optical in high-$z$ galaxies, allowing us to perform stellar population modeling and derive
stellar masses, and enables robust discrimination between star-forming $z > 7$ galaxies and dusty/evolved galaxies at $z \sim 2$.

We have also been awarded 225 hours of time on the VLT as part of a large program (P. Rosati, P.I.)
to obtain VIMOS spectroscopy of the 14 southern clusters ($\delta < +1.5^{\circ}$). In addition, we are
acquiring cluster and lensed galaxy redshifts from spectrographs on {\em Magellan (GISMO; LDSS), LBT (MODS), MMT (Hectospec),} and 
{\em Palomar} (double spectrograph). 

\begin{deluxetable*}{lccrrcc}
\tabletypesize{\scriptsize}
\tablewidth{0pt}
\tablecaption{\label{tab:OtherObs}Partial List of Supporting Observations for CLASH at Time of Publication}
\tablehead{
\colhead{}&
\colhead{Proj. FOV of}&
\colhead{}&
\colhead{}&
\colhead{}&
\colhead{}&
\colhead{}\\
\colhead{}&
\colhead{Suprime Cam}&
\colhead{Suprime Cam}&
\colhead{Chandra}&
\colhead{XMM}&
\colhead{Spitzer$^a$}&
\colhead{Bolocam}\\
\colhead{Cluster}&
\colhead{(Mpc)}&
\colhead{Filters}&
\colhead{(ksec)}&
\colhead{(ksec)}&
\colhead{(ksec)}&
\colhead{(ksec)}
}
\startdata
X-ray Selected Clusters:  \\
   ~~Abell 209                      	&~5.68		& BVRIZ 		& 20		&\nodata		&	(31.2)$^b$	&	66	      	  \\
   ~~Abell 383                      	&~5.26		& BVRIZ 		& 50		&\nodata		&	20.4		&	91	      	  \\
   ~~MACS0329.7-0211         	&~9.68		& BVRZ   		& 70		&\nodata		&	(31.2)$^b$&	39	      	  \\
   ~~MACS0429.6-0253         	&~9.30		& VRI       		& 24		&\nodata		&	(31.2)$^b$&	62	      	  \\
   ~~MACS0744.9+3927        	&11.90		& BVRIZ  	& 90		&138		&	18.0		&	62	      	  \\
   ~~Abell 611                      		&~7.28		& BVRIZ  	& 40		&28			&	18.0		&	79	      	  \\
   ~~MACS1115.9+0129         	&~8.34		& BVRIZ  	& 56		&\nodata		&	(31.2)$^b$&	56	      	  \\
   ~~Abell 1423                         &~5.82          & VI                    & 36            &31                     &       \nodata &   TBA$^d$     \\
   ~~MACS1206.2-0847           	&~9.56		& BVRIZ 		& 24		&\nodata		&	(31.2)$^b$&	41	      	 \\
   ~~CLJ1226.9+3332              	&13.04		& BVZ      	& 75		&120		&	 17.0		&	68	      	   \\ 
   ~~MACS1311.0-0310          	&10.20		& R            	& 85		&\nodata		&	(31.2)$^b$&	41	      	   \\
   ~~RXJ1347.5-1145               	&~9.68		& BVRIZ  	&185	&36			&	19.2		&	59	      	   \\
   ~~MACS1423.8+2404         	&10.72		& BVRIZ   	&140	&\nodata		&	21.4		&	85	      	   \\
   ~~RXJ1532.9+3021             	&~8.22		& BVRIZ   	& 40		&\nodata		&	20.2		&	53	      	   \\
   ~~MACS1720.3+3536         	&~8.90		& BVRIZ   	&60		&\nodata		&	20.0		&	68	      	   \\
   ~~Abell 2261                     	&~6.06		& BVR      	&35		&33			&	22.4		&	46	      	\\
   ~~MACS1931.8-2635          	&~8.34		& BVRIZ  	&114		&\nodata		&	(31.2)$^b$&	63	      	   \\
   ~~RXJ2129.7+0005             	&~6.26		& BVRIZ  	&40		&\nodata		&	18.2		&	49	      	   \\
   ~~MS2137-2353                	&~7.70		& BVRIZ 		&150	&\nodata		&	20.4		&	47	      	\\
   ~~RXJ2248.7-4431 (Abell 1063S) &\nodata	& {\it Not accessible}&26	&\nodata		&	20.4		&	20	      \\
                                                       &  		&  See note {\it c} & & & & \\
 ~~\\
   High Magnification Clusters:  \\
   ~~MACS0416.1-2403          	&~9.30		& BRZ      		&16		&\nodata		&	(31.2)$^b$&	29	      	   \\
   ~~MACS0647.8+7015         	&11.10		& BVRIZ  	&40		&\nodata		&	18.0		&	43	      	    \\
   ~~MACS0717.5+3745         	&10.76		& BVRIZ  	&104	&\nodata		&	21.4		&	47	      	    \\
   ~~MACS1149.6+2223          	&10.72		& BVRIZ 		&20		&\nodata		&	18.0		&	45	      	   \\
   ~~MACS2129.4-0741           	&10.96		& BVRIZ 		&40		&\nodata		&	19.5		&	47		\\
\enddata
\tablecomments{Observations with MUSTANG (NRAO/GBT) pending. Spectroscopic observations summarized
in text.}
\tablenotetext{a}{Times shown are the total for both IRAC channels 1,2 (3.6 and 4.5 microns).}
\tablenotetext{b}{These 8 clusters will be observed with the indicated exposures in Spitzer cycle 8.}
\tablenotetext{c}{RXJ2248.7-4431 has UBVRIZ imaging with the WFI on the 2.2-meter telescope at La Silla. Total exposure times in each band are 10-11 hours.}
\tablenotetext{d}{SZE data for A1423 will be acquired in spring 2012.}
\end{deluxetable*}

\section{Summary}
\label{sec:summary}

The Cluster Lensing And Supernova survey with Hubble will produce a major advance in our understanding of 
the DM power spectrum and 
the internal structure of cluster halos on scales from 10 kpc to 2 Mpc. The precision to which these measurement are being 
made will provide an unprecedented foil against which we will challenge and ultimately expand our 
current ideas about structure formation and the nature of dark energy. 
In November 2010, the CLASH MCT program initiated its three year observing plan
to obtain deep (20-orbit) 16-band \HST\ imaging for each of 25 galaxy cluster cores.
The clusters in our sample are
massive ($5\times 10^{14} < M_{vir}/M_\odot < 3\times 10^{15}$),
span a range of (intermediate) redshifts ($0.18 < z < 0.9$), and most (20 of the 25) were
selected based on their X-ray properties (reasonably relaxed and $T_x > 5$ keV).
The 16-band \HST\ imaging yields precise ($2\% (1+z)$) photometric redshifts
for all galaxies brighter that F775W AB mag 26, including hundreds of strongly lensed galaxies. 
When combined with weak lensing maps from wide field Subaru imaging and constraints on the baryonic gas distribution
from mm-wave and X-ray imaging, CLASH data will allow the cluster mass profiles
to be tightly constrained over a large range of scales, yielding
robust measurements of their central density concentrations. We also expect to 
provide important constraints on the degree of substructure in the DM distribution
by studying galaxy-scale lensing. 
The strong lens magnifying power of our clusters should also enable detection of 
dozens of relatively bright ($m < 26$ AB) $z > 7$ galaxies, including some bright enough
for spectroscopic follow-up with ground-based telescopes and, certainly, many that can be
studied with the spectrographs on {\em JWST}.
Parallel observations may detect up to $\sim30\ z > 1$ Type Ia supernovae that, when combined
with $z  > 1$ SNe Ia detected by the CANDELS program, will provide new constraints
on the time-dependence of $w$ and a dramatic improvement in our understanding of SNe Ia evolution
in the matter-dominated universe ($z > 1.5$).

CLASH data will also provide the mass-calibrators for the next generation of big 
cosmological surveys such as the Dark Energy Survey (DES), Sunayev-Zeldovich surveys (e.g., South Pole Telescope) 
and next generation X-ray cluster surveys. The DES will measure the evolution of the space density 
of clusters (among other probes) as a way to measure Dark Energy. These cluster counts tell us 
much about cosmological parameters through their impact on both the volume and the growth of perturbations. 
To do so, however, requires that the cluster mass scaling relations be well calibrated. The CLASH measurements
of the enclosed mass in each cluster will be invaluable in calibrating the relation between the 
cluster mass and a number of observable mass proxies.

MCT programs represent a very large investment of \HST\ observing time.
These data are intended to be a community resource, and as such they have no proprietary period.
The zero proprietary period policy also applies to our cycle 8 Spitzer observations.
All \HST\ and Spitzer images may be downloaded immediately after they are obtained 
via the MAST and IRSA archives, respectively.
The CLASH collaboration is producing high-level science products that will
be publicly distributed via MAST's Treasury Program archive site\footnote{http://archive.stsci.edu/hst/tall.html}. This paper
includes the first public release of a CLASH source list for the cluster MACS1149.6$+$2223 as well as uniformly re-derived X-ray properties for the 25 clusters in the sample.
The CLASH HLSP include co-added and mosaicked 
images, weight maps, inverse variance maps, high-resolution color images,
and source lists. At the end of the survey we will also release the corresponding registered X-ray surface brightness images, 
co-added Spitzer Space Telescope images, co-added SZE data, the calibrated lensing-derived mass and source magnification maps, the source lists with all photometric 
redshift information, SN coordinates and their light curves and grism spectra, and the associated spectroscopic redshift catalogs.
The CLASH data provide a vast legacy 
archive for studies of the formation and evolution of cosmic structure. 

\acknowledgements{
We are especially grateful to our program coordinator Beth Perrillo 
for her expert assistance in implementing the \HST\ observations in this program.  
We thank Jay Anderson and Norman Grogin 
for providing the \ACS\ CTE and bias striping correction algorithms used in our data pipeline. Finally,
we are indebted to the hundreds of people 
who have labored many years to plan, develop, manufacture, install, repair, and calibrate
the \WFCiii\ and \ACS\ instruments as well as to all those who maintain and operate the 
{\em Hubble Space Telescope}.

The CLASH Multi-Cycle Treasury Program (GO-12065) is based on observations made with the NASA/ESA Hubble Space Telescope.
The Space Telescope Science Institute is operated by the Association of Universities for Research in Astronomy, Inc. under NASA contract NAS 5-26555.
ACS was developed under NASA contract NAS 5-32864.

This research is supported in part by NASA grant HST-GO-12065.01-A, 
the Israel Science Foundation, 
the Baden-Wuerttemberg Foundation, 
the German Science Foundation (Transregio TR 33), 
Spanish MICINN grant AYA2010-22111-C03-00, 
funding from the Junta de Andaluc'a Proyecto de Excelencia NBL2003, 
INAF contracts  ASI-INAF I/009/10/0, ASI-INAF I/023/05/0, ASI-INAF I/088/06/0, PRIN INAF 2009, and PRIN INAF 2010, 
NSF CAREER grant AST-0847157, 
the UK's STFC, the Royal Society, the Wolfson Foundation, 
and National Science Council of Taiwan grant NSC97-2112-M-001-020-MY3. 
AZ acknowledges support by the John Bahcall excellence prize. LI acknowledges support from a Conicyt FONDAP/BASAL grant.
PR and SS acknowledge support from the DFG cluster of excellence Origin and Structure of the Universe program.
}

\bibliographystyle{astroads}
\bibliography{CLASH2}

\end{document}